\documentclass[10pt]{article}
\usepackage[utf8]{inputenc}
\usepackage{cite}
\usepackage{hyperref}
\hypersetup{colorlinks=true,linkcolor=bluecyan,citecolor=orange3,urlcolor=green3,pdfencoding=auto,linktocpage}
\usepackage{amsmath,amssymb,amsbsy,amstext,amsthm,simplewick,amsfonts}
\usepackage{mathrsfs}
\usepackage{graphicx}
\usepackage{wrapfig}
\usepackage{upgreek}
\usepackage{bm} 
\usepackage{framed}
\usepackage{bbm}
\usepackage{textcomp}
\usepackage{adjustbox}
\usepackage{makecell}
\usepackage{tcolorbox}
\usepackage{empheq}
\usepackage[normalem]{ulem}
\usepackage{enumitem}
\usepackage{braket} 
\usepackage{array}
\usepackage{dsfont}
\usepackage{physics}
\usepackage{ulem}
\usepackage{xcolor}
\usepackage{caption}
\usepackage{subcaption}
\usepackage{tocloft}
\usepackage{tikz}
\usepackage{float}
\usepackage[compat=1.1.0]{tikz-feynman}
\tikzfeynmanset{
      every blob={/tikz/fill=gray!20}
    }
\usepackage{appendix}
\usepackage[framemethod=default]{mdframed}
\usepackage[makeroom]{cancel}
\usepackage{diagbox}
\usepackage{tabularray}
\usepackage{multirow}

\newmdenv[backgroundcolor=gray!15,%
skipabove=5pt,%
skipbelow=5pt,%
leftmargin=2pt,%
rightmargin=2pt,%
innertopmargin=-6pt,%
innerbottommargin=5pt,%
innerleftmargin=5pt,%
innerrightmargin=5pt,%
splittopskip=0pt,%
splitbottomskip=0pt,%
linewidth=0pt,%
nobreak=true]%
{keyeqn}

\newmdenv[backgroundcolor=gray!15,%
skipabove=5pt,%
skipbelow=5pt,%
leftmargin=2pt,%
rightmargin=2pt,%
innertopmargin=-2pt,%
innerbottommargin=5pt,%
innerleftmargin=5pt,%
innerrightmargin=5pt,%
splittopskip=0pt,%
splitbottomskip=0pt,%
linewidth=0pt,%
nobreak=true]%
{keythrm}

\graphicspath{{Fig/}}

\definecolor{lightgreen}{cmyk}{0.2, 0, 0.2, 0.2}
\definecolor{lightgray}{cmyk}{0.1,0.2,0,0.1}
\definecolor{lightgray2}{cmyk}{0.1,0.1,0,0.1}
\definecolor{bluecyan}{RGB}{0, 100, 200}
\definecolor{blue3}{RGB}{31,119,180}
\definecolor{red3}{RGB}{214,39,40}
\definecolor{orange3}{RGB}{255,127,14}
\definecolor{green3}{RGB}{44,160,44}

\definecolor{red2}{RGB}{255,0,0}
\definecolor{green2}{RGB}{0,170,0}
\definecolor{blue2}{RGB}{0,128,255}
\definecolor{magenta2}{RGB}{191,64,191}
\definecolor{purple2}{RGB}{112,48,160}
\definecolor{orange2}{RGB}{255,192,0}

\def\bfk{\textbf{k}}
\def\bfx{\textbf{x}}

\def\Re{\mathrm{Re}\,}

\numberwithin{equation}{section}

\allowdisplaybreaks[1]
\begin{document}

\begin{titlepage}
	\setcounter{page}{1} \baselineskip=15.5pt 
	\thispagestyle{empty}

    \begin{center}
		{\fontsize{18}{18}\centering {{\bf{Massive Spinning Fields During Inflation:}} \\ \vspace{0.2cm} \textit{\Large Feynman rules and correlator comparison}}}\\
	\end{center}
 
	\vskip 18pt
	\begin{center}
		\noindent
		{\fontsize{12}{18}\selectfont Trevor Cheung\footnote{\tt kaiheitrevor.cheung@nottingham.ac.uk}$^{,a}$ and David Stefanyszyn\footnote{\tt david.stefanyszyn@nottingham.ac.uk}$^{,a,b}$}
	\end{center}
	
	\begin{center}
		\vskip 8pt
		$a$ \textit{School of Physics and Astronomy,
			University of Nottingham, University Park, \\ Nottingham, NG7 2RD, United Kingdom} \\
            $b$ \textit{School of Mathematical Sciences,
			University of Nottingham, University Park, \\ Nottingham, NG7 2RD, United Kingdom} \\
	\end{center}
	
	
	\noindent\rule{\textwidth}{0.4pt}
	\noindent \textbf{Abstract} ~~ We consider the dynamics of massive spinning fields during inflation and the resulting signatures in the cosmological correlators of inflaton perturbations computed in the Poincar\'{e} patch of de Sitter space. There are (at least) two ways to describe the fluctuations of such new spinning degrees of freedom and these are distinguished by the symmetries of the de Sitter group that they linearly realise. The primary question we ask is: do these two set-ups yield distinct signatures in cosmological observables? After systematically deriving the Feynman rules for exchange diagrams consisting of massive spinning fields, where we discover the necessity of \textit{effective propagators} that augment the naive Schwinger-Keldysh ones by delta functions corresponding to instantaneous propagation, we show that the two set-ups are indistinguishable at the level of the inflaton bispectrum but distinguishable at the level of the trispectrum and other higher-point correlation functions. The bispectrum is special since in the corresponding tree-level Feynman diagrams, only the helicity-zero modes of the spinning fields can propagate. The bispectrum correspondence holds up to the addition of contact diagrams arising from the self-interactions of the inflaton, and is consistent with the symmetries of the effective field theory of inflation. Our results suggest that the cosmological collider signals in the bispectrum are universal and do not depend on the detailed description of the massive spinning field.     
	
	\noindent\rule{\textwidth}{0.4pt}
	
	
\end{titlepage} 


\newpage
\setcounter{page}{2}
{
	\tableofcontents
}


\section{Introduction}
The incredibly high energies associated with inflationary cosmology presents us with an unrivalled opportunity to probe particle physics in regimes not accessible to terrestrial colliders \cite{Chen:2009zp,Chen:2009we,Arkani-Hamed:2015bza,Lee:2016vti,Noumi:2012vr}. The early universe is a natural laboratory where very heavy states can be produced thanks to the highly-energetic background of the accelerating spacetime. Such states quickly decay, and if the underlying symmetries allow, they decay into light states such as the inflaton and graviton, and ultimately leave imprints of their existence in the shape dependence of cosmological correlators. Such spatial correlators are evaluated at the end of inflation and provide the initial conditions for the Hot Big Bang model of cosmology. Observations of the Cosmic Microwave Background (CMB) and Large Scale Structure (LSS) probe these spatial correlators thereby providing us with the tantalising prospect of searching for signatures of such heavy degrees of freedom in cosmological data sets. 

Assuming scale invariance of inflaton correlators, the effects of additional degrees of freedom cannot be discerned from the power spectrum alone since, up to an overall constant, this two-point function is completely fixed by symmetry and momentum conservation. The first observable where such effects can be seen is the three-point function, or bispectrum $B_3$, and in perturbation theory new heavy states affect the bispectrum via exchange diagrams \cite{Pimentel:2022fsc,Jazayeri:2022kjy,Jazayeri:2023xcj,Kumar:2017ecc}. A necessary ``interaction" for such processes to exist is a linear-mixing vertex where the inflaton equation of motion is linearly sourced by the new state. In the presence of such a linear-mixing, the relevant tree-level Feynman diagrams that compute the bispectrum are 
 \begin{equation} \label{generalcorrespondence}
 B_3 \supset    \begin{gathered}
       \begin{tikzpicture}
  \begin{feynman}
    \vertex[dot] (a) at (0,0){};
    \vertex[dot] (b) at (2,0){};
    \vertex (i1) at (-0.5,1) {};
    \vertex (i2) at (-0.5,-1) {};
    \vertex (i3) at (3,0) {};
   
       \diagram* {
      (i1) -- (a) [blob] -- (i2),
      (b) --  (i3),
      (a) -- [boson] (b),
    };
  \end{feynman}
\end{tikzpicture}
\end{gathered} + 
\begin{gathered}
  \begin{tikzpicture}
  \begin{feynman}
    \vertex (a) at (0,0){};
    \vertex[dot] (b) at (-1,0){};
    \vertex [dot] (c) at (-1.5,0.866025403784){};
    \vertex (i2) at (-2,1.73205080757){};
    \vertex [dot](d) at (-1.5,-0.866025403784){};
     \vertex (i3) at (-2,-1.73205080757){};
       \diagram* {
 (a) -- [plain] (b) -- [boson] (c) -- [plain] (i2),
      (b) -- [boson] (d) -- [plain] (i3),
    };
  \end{feynman}
\end{tikzpicture}
\end{gathered}
+
\begin{gathered}
  \begin{tikzpicture}
  \begin{feynman}
    \vertex[dot] (a) at (0,0){};
    \vertex[dot] (b) at (-1,0){};
    \vertex (i1) at (1,0){};
    \vertex [dot] (c) at (-1.5,0.866025403784){};
    \vertex (i2) at (-2,1.73205080757){};
    \vertex [dot](d) at (-1.5,-0.866025403784){};
     \vertex (i3) at (-2,-1.73205080757){};
       \diagram* {
      (i1) -- [plain] (a) -- [boson] (b) -- [boson] (c) -- [plain] (i2),
      (b) -- [boson] (d) -- [plain] (i3),
    };
  \end{feynman}
\end{tikzpicture}
\end{gathered}
+ \ldots
\end{equation}
Evaluating these diagrams is tricky, however. Indeed, we are obliged to evaluate nested time integrals that are non-trivial thanks to the breaking of time translation symmetry, and \textit{very} non-trivial since the mode functions of the heavy fields are Hankel functions (or Whittaker functions for higher-point correlators where there is parity-violation \cite{Stefanyszyn:2023qov,Tong:2022cdz,Qin:2025xct,Agrawal:2017awz}). In recent years, much effort has been focused on evaluating diagrams of this form bypassing the traditional bulk computations and instead utilising ideas of the cosmological bootstrap programme where one works directly with the observable of interest and imposes constraints from symmetries, locality and unitarity \cite{Chen:2009zp,Chen:2009we,Arkani-Hamed:2015bza,Lee:2016vti,Bonifacio:2022vwa,Cabass:2021fnw,Jazayeri:2021fvk,Stefanyszyn:2023qov,Tong:2022cdz,Qin:2025xct,Agrawal:2017awz,Goodhew:2020hob,Baumann:2022jpr,Pajer:2020wxk,Werth:2024mjg,Arkani-Hamed:2018kmz,Baumann:2019oyu,Baumann:2020dch,Baumann:2021fxj,Arkani-Hamed:2023kig,Pimentel:2022fsc,Jazayeri:2022kjy,Jazayeri:2023xcj,Sleight:2019hfp,Sleight:2019mgd,Sleight:2020obc,Bzowski:2013sza,Grimm:2025zhv,Liu:2024xyi,Xianyu:2023ytd,Qin:2023ejc,Qin:2022fbv,DiPietro:2021sjt,Aoki:2024uyi,Hogervorst:2021uvp,Arkani-Hamed:2017fdk,Albayrak:2023hie,Noumi:2012vr,Chakraborty:2023eoq,Chakraborty:2023qbp,Kumar:2019ebj,deRham:2025jle,Wang:2022eop}. The output is an oscillatory shape dependence of the bispectrum, and of higher-order correlators, that contains information on the mass and spin of the exchanged state. This area of research is often referred to as \textit{cosmological collider physics} with the imprints of new heavy degrees of freedom referred to as the \textit{cosmological collider signals}, which are being searched for in the data with additional prospects for future searches  \cite{MoradinezhadDizgah:2018ssw,Sohn:2024xzd,Cabass:2024wob,Meerburg:2016zdz}.

During inflation, massive spinning fields do not necessarily adhere to a unique interacting Lagrangian. Indeed, the background vev of the inflaton breaks de Sitter boost symmetry and therefore the theory of fluctuations is not expected to linearly realise the full de Sitter group. The way massive spinning fields are usually described during inflation is outlined in \cite{Lee:2016vti} where the free theory is de Sitter invariant with the interactions with the inflaton breaking boosts. However, another perspective was offered in \cite{Bordin:2018pca} where not even the free theory is assumed to be de Sitter invariant. In this latter case, the spinning fields are representations of the unbroken $SO(3)$ symmetry of FRW cosmologies. Both approaches are admissible, and ultimately describe the propagation of $2S+1$ degrees of freedom on top of an inflationary background with couplings to the inflaton fixed by non-linearly realised symmetries. We refer to the set-up of \cite{Lee:2016vti} as \textit{cosmological collider (CC) physics} and that of \cite{Bordin:2018pca} as \textit{cosmological condensed matter (CCM) physics}.  
In this paper we ask the following question:
\begin{center}
    \textit{Do these two set-ups, CC and CCM, produce distinct signatures in the \\ cosmological correlators of inflaton perturbations?}
\end{center}
In order to answer this question, we consider the computation of correlators in the two set-ups and see if they can be matched while maintaining local interactions in both cases and without altering the total number of degrees of freedom. While the rules for computing correlators in the CCM case are well understood \cite{Bordin:2018pca,Stefanyszyn:2023qov,Goodhew:2021oqg}, those of the CC set-up are complicated by the non-dynamical modes that are required to form multiplets of the full de Sitter group. We find that the naive method of constructing Schwinger-Keldysh propagators out of the mode functions is not the full story, and that the true propagators may also contain delta function terms that correspond to instantaneous propagation. These additional delta function pieces yield additional diagrams. For example, when computing the three-point function due to single exchange, the delta function terms yield additional contact diagrams that are of the same order in couplings as the exchange. Such delta function terms also appear in flat-space in the canonical quantisation of e.g. a massive spin-$1$ field \cite{Weinberg:1995mt}. However, in that case one can reformulate the Feynman rules in a manifestly Lorentz covariant way where the delta functions cancel out. In our case, we are considering correlators that arise from boost-breaking theories and therefore such a covariant reformulation is not useful or necessary, and the delta functions we have must be included to yield the correct perturbative results. After reviewing the CCM rules in Section \ref{FeynmanRulesCCM}, we derive the CC rules in Section \ref{FeynmanRulesCC} concentrating on the spin-$1$ and spin-$2$ cases. We initially work with the wavefunction of the universe and derive expressions for the bulk-bulk propagations followed by converting to correlators using the Born rule.

With the Feynman rules for both the CCM and CC set-ups at hand, we go ahead and compare the resulting correlators in Section \ref{CorrelatorComparison}. There are two zeroth-order obstacles to finding a correspondence between the two set-ups: 
\begin{enumerate}
    \item The CC fields are forced to propagate with unit sound speed by symmetry, whereas the CCM fields can propagate with a reduced sound speed and different helicity degrees of freedom can propagate with different speeds.
    \item The mass spectrum of the CC fields is constrained by the Higuchi bound \cite{Higuchi:1986py} whereas there are no such bounds for the CCM fields \cite{Bordin:2018pca}.
\end{enumerate}
These two obstacles show that the CCM set-up contains more freedom than the CC set-up, so the question we are really asking is:
\begin{center}
    \textit{Can the correlators produced in the CC set-up be mimicked by those of the CCM set-up?}
\end{center}
This is the question we set out to answer. We find that the bispectra computed in the CC set-up can indeed be mimicked by that of the CCM set-up with local interactions on each side meaning that the cosmological collider signals being searched for in the bispectrum \cite{Sohn:2024xzd,Cabass:2024wob} are universal. However, for the trispectrum and other higher-point correlators, the CC correlators cannot be mimicked by the CCM ones while maintaining local interactions and the number of degrees of freedom. The bispectrum is special since the relevant exchange diagrams at tree-level only allow for the exchange of modes with zero helicity since such diagrams rely on the linear mixing vertex and the non-zero helicity modes decouple from such a vertex by rotational invariance (see e.g. \cite{Pimentel:2022fsc}). As soon as multiple helicity exchanges become relevant, the correspondence fails. We initially only assume the linear realisation of $SO(3)$ symmetries when we build interactions between the inflaton and new states, but in Section \ref{CCvCCMEFToI} we restrict our analysis to interactions that adhere to the symmetries of the \textit{effective field theory of inflation} (EFToI) \cite{Cheung:2007st}. There we find that any bispectrum computed in the CC set-up with interactions constrained by the symmetries of the EFToI, can be mimicked by CCM interactions that also adhere to those symmetries. The EFToI symmetries are therefore preserved in the correspondence. We conclude and discuss avenues for future work in Section \ref{Conclusion}, and we provide a set of appendices with further calculational details, in particular on the derivation of Feynman rules for the exchange of massive spin-$2$ modes. 

\paragraph{Summary of main results} For the reader's convenience, here we summarise our main results:
\begin{itemize}
    \item We derived the Feynman rules for inflaton correlators when exchanging massive spinning fields described by the cosmological collider (CC) physics set-up where the free theory of these fields is fully de Sitter invariant. We discovered \textit{mixed propagators} that connect CC modes with the same helicity but coming from different components of the full $\Phi_{\mu_1 \ldots \mu_S}$, and the addition of delta functions in some propagators that indicate that a non-dynamical mode is being exchanged. We first derived the rules for wavefunction coefficients and then extracted rules for correlators by applying the Born rule. As an example, consider the quartic wavefunction coefficient arising from the exchange of a massive spin-$1$ field $\Phi_\mu$ (we treat $\Phi_0$ and $\Phi_i$ separately as in the EFToI they do not need to form a single multiplet \cite{Lee:2016vti}):
    \begin{itemize}
    \item when exchanging the temporal mode $\Phi_0$ (denoted as $\Phi^0_{0,1}$ in the main text), there is an extra contact contribution to the wavefunction coefficient compared to what one might naively expect, diagrammatically represented as  \begin{equation}
    \underbrace{\begin{gathered}
        \begin{tikzpicture}
  \begin{feynman}
    \vertex (a) at (0.5,0);
    \vertex (i1) at (1,0);
    \vertex (i2) at (2,0);
    \vertex (i3) at (3,0);
    \vertex (i4) at (4,0);
    \vertex (b) at (4.5,0);
    \vertex (c1) at (3/2,-2);
    \vertex (c2) at (7/2,-2);
        \vertex (i11) at (1,0.25) {\(\bfk_1\)};
    \vertex (i21) at (2,0.25) {\(\bfk_2\)};
    \vertex (i31) at (3,0.25) {\(\bfk_3\)};
    \vertex (i41) at (4,0.25) {\(\bfk_4\)};
       \diagram* {
      (a) -- [plain] (i1) -- [plain] (i2) -- [plain] (i3) -- [plain] (i4) -- [plain] (b),
      (c1) -- [plain] (i1),
      (c1) -- [plain] (i2),
      (c2) -- [plain] (i3),
      (c2) -- [plain] (i4),
      (c1) -- [double, edge label'=\(\Phi^0_{0,1}\)] (c2),
    };
  \end{feynman}
\end{tikzpicture}
\end{gathered}}_{\text{correct Feynman rules}}
= \underbrace{\begin{gathered}
\begin{tikzpicture}
  \begin{feynman}
   \vertex (a) at (0.5,0);
    \vertex (i1) at (1,0);
    \vertex (i2) at (2,0);
    \vertex (i3) at (3,0);
    \vertex (i4) at (4,0);
    \vertex (b) at (4.5,0);
    \vertex (c1) at (3/2,-2);
    \vertex (c2) at (7/2,-2);
        \vertex (i11) at (1,0.25) {\(\bfk_1\)};
    \vertex (i21) at (2,0.25) {\(\bfk_2\)};
    \vertex (i31) at (3,0.25) {\(\bfk_3\)};
    \vertex (i41) at (4,0.25) {\(\bfk_4\)};
       \diagram* {
      (a) -- [plain] (i1) -- [plain] (i2) -- [plain] (i3) -- [plain] (i4) -- [plain] (b),
      (c1) -- [plain] (i1),
      (c1) -- [plain] (i2),
      (c2) -- [plain] (i3),
      (c2) -- [plain] (i4),
      (c1) -- [ghost, edge label'=\(\Phi^0_{0,1}\)] (c2),
    };
  \end{feynman}
\end{tikzpicture}
    \end{gathered}}_{\text{naive Feynman rules}}
    + \; \underbrace{\begin{gathered}
\begin{tikzpicture}
  \begin{feynman}
    \vertex (a) at (0.5,0);
    \vertex (i1) at (1,0);
    \vertex (i2) at (2,0);
    \vertex (i3) at (3,0);
    \vertex (i4) at (4,0);
    \vertex (b) at (4.5,0);
    \vertex (c) at (5/2,-2);
    \vertex (i11) at (1,0.25) {\(\bfk_1\)};
    \vertex (i21) at (2,0.25) {\(\bfk_2\)};
    \vertex (i31) at (3,0.25) {\(\bfk_3\)};
    \vertex (i41) at (4,0.25) {\(\bfk_4\)};
    \vertex (d) at (5/2,-2.55) {\(\)};
       \diagram* {
      (a) -- [plain] (i1) -- [plain] (i2) -- [plain] (i3) -- [plain] (i4) -- [plain] (b),
      (c) -- [plain] (i1),
      (c) -- [plain] (i2),
      (c) -- [plain] (i3),
      (c) -- [plain] (i4),
    };
  \end{feynman}
\end{tikzpicture}
\end{gathered}}_{\text{new contact diagram}}.
\end{equation}
Effectively, the bulk-bulk propagator acquires a delta function correction:
\begin{keyeqn}
    \begin{align}
    \left (G^0_{0,1\to 0,1}\right )_{\text{eff}}(\eta,\eta',k) = G^0_{0,1}(\eta,\eta',k) + \frac i{m^2a^2}\delta(\eta-\eta')\,, \label{Gspin1effective}
\end{align}
\end{keyeqn}
where here $G^0_{0,1}(\eta,\eta',k)$ is the naive bulk-bulk propagator constructed out of the mode functions. Such a delta function also appears in the corresponding flat-space propagator when we work in the mixed domain where we only Fourier transform in the spatial directions (we discuss this in Section \ref{DeltaSection}) \cite{Weinberg:1995mt}.
\item A vertex linear in $\Phi_0$ can be connected to one linear in $\Phi_i$ thanks to a mixed propagator. Indeed, there is only a single dynamical helicity-zero mode in the massive spin-$1$ field (denoted by $\Phi^0_{0,1}$ in $\Phi_0$ and $\Phi^0_{1,1}$ in $\Phi_i$) and therefore in canonical quantisation the creation and annihilation operators for the helicity-zero modes in $\Phi_0$ and $\Phi_i$ are identical \cite{Higuchi:1986py}. Diagrammatically, the following contribution is therefore non-zero:
\begin{equation}
    \begin{gathered}
        \begin{tikzpicture}
  \begin{feynman}
    \vertex (a) at (0.5,0);
    \vertex (i1) at (1,0);
    \vertex (i2) at (2,0);
    \vertex (i3) at (3,0);
    \vertex (i4) at (4,0);
    \vertex (b) at (4.5,0);
    \vertex (c1) at (3/2,-2);
    \vertex (c) at (5/2,-2);
    \vertex (c2) at (7/2,-2);
        \vertex (i11) at (1,0.25) {\(\bfk_1\)};
    \vertex (i21) at (2,0.25) {\(\bfk_2\)};
    \vertex (i31) at (3,0.25) {\(\bfk_3\)};
    \vertex (i41) at (4,0.25) {\(\bfk_4\)};
       \diagram* {
      (a) -- [plain] (i1) -- [plain] (i2) -- [plain] (i3) -- [plain] (i4) -- [plain] (b),
      (c1) -- [plain] (i1),
      (c1) -- [plain] (i2),
      (c2) -- [plain] (i3),
      (c2) -- [plain] (i4),
      (c1) -- [double, edge label'=\(\Phi^0_{0,1}\)] (c) -- [scalar, edge label'=\(\Phi^0_{1,1}\)] (c2)
    };
  \end{feynman}
\end{tikzpicture}
    \end{gathered}
    .
\end{equation}
This mixed propagator was used in \cite{Kumar:2017ecc}.
\end{itemize}
A similar story applies for the exchange of a massive spin-$2$ field $\Phi_{\mu\nu}$ with the full set of propagators given in Section \ref{FeynmanRulesCC}.
\item We then compared the correlators computed using the CCM and CC descriptions and checked if the CC ones could be mimicked by the CCM ones while maintaining local interactions and the number of derivatives. We found that this was possible for the bispectrum but not for the trispectrum and other higher-point correlators. For the bispectrum, any diagram on the CC side can be mimicked by a linear combination of diagrams on the CCM side. For example, the single-exchange correspondence reads:
    \begin{equation} 
    \begin{gathered}
        \begin{tikzpicture}
  \begin{feynman}
    \vertex[dot] (a) at (0,0){};
    \vertex[dot] (b) at (2,0){};
    \vertex (i1) at (-0.5,1) {};
    \vertex (i2) at (-0.5,-1) {};
    \vertex (i3) at (3,0) {};
    \vertex (c) at (1,-0.3) {\(\text{CC}\)};
   
       \diagram* {
      (i1) -- (a) [blob] -- (i2),
      (b) --  (i3),
      (a) -- [gluon] (b),
    };
  \end{feynman}
\end{tikzpicture}
\end{gathered}
= \begin{gathered}
\begin{tikzpicture}
  \begin{feynman}
    \vertex[dot] (a) at (0,0){};
    \vertex[dot] (b) at (2,0){};
    \vertex (i1) at (-0.5,1) {};
    \vertex (i2) at (-0.5,-1) {};
    \vertex (i3) at (3,0) {};
    \vertex (c) at (1,-0.3){\(\text{CCM}\)};
   
       \diagram* {
      (i1) -- (a) [blob] -- (i2),
      (b) --  (i3),
      (a) -- [gluon] (b),
    };
  \end{feynman}
\end{tikzpicture}
    \end{gathered}
    +     \begin{gathered}
    \begin{tikzpicture}
  \begin{feynman}
    \vertex[dot] (a) at (0,0){};;
    \vertex (i1) at (-0.5,1) {};
    \vertex (i2) at (-0.5,-1) {};
    \vertex (i3) at (1,0) {};
   
       \diagram* {
      (i1) -- (a) -- (i3),
      (a) --  (i2),
    };
  \end{feynman}
\end{tikzpicture}
\end{gathered}.
\end{equation}
\end{itemize}

\paragraph{Notations and conventions}
We denote $n$-point inflaton correlators as 
\begin{align}
\langle \pi(\bfk_1) \ldots \pi(\bfk_n) \rangle =  \hat{\delta}^{(3)}(\bfk) B_n(\bfk_1, \ldots, \bfk_n) \,,
\end{align}
where $\hat{\delta}^{(3)}(\bfk) = (2\pi)^3 \delta^{(3)}(\bfk)$. For a set of fields $\phi_1, \ldots, \phi_n$, with any indices suppressed, we parametrise the wavefunction of the universe as 
\begin{align}
\Psi =\exp & \left [- \frac 1{2!} \sum_{i=1}^n\int_{\bfk} \psi_2^{\phi_i\phi_i}(\bfk) + \sum_{i\neq j} \int_{\bfk} \psi_2^{\phi_i\phi_j}(\bfk) + \frac{1}{3!} \int_{\bfk_1, \bfk_2, \bfk_3} \psi_3^{\phi_1 \phi_1 \phi_1}(\{ \bfk\})\hat{\delta}^{(3)}\left(\sum \bfk \right) \right .\nonumber \\
& \quad \left . + \frac{1}{2!} \int_{\bfk_1, \bfk_2, \bfk_3} \psi_3^{\phi_1 \phi_1 \phi_2}(\{ \bfk\})\hat{\delta}^{(3)}\left(\sum \bfk \right) +  \int_{\bfk_1, \bfk_2, \bfk_3} \psi_3^{\phi_1 \phi_2 \phi_3}(\{ \bfk\})\hat{\delta}^{(3)}\left(\sum \bfk \right)  + \ldots \right]\,, \label{WavefunctionParam}
\end{align}
where $\{ \bfk \}$ collectively denotes the external momenta. 

We require the polarisation tensors $e^{(h)}_{i_1...i_S}$ to satisfy
\begin{align}
    \left [e^{(h)}_{i_1...i_S}(\bfk)\right ]^* & = e^{(h)}_{i_1...i_S}(-\bfk)\tag{reality}\\
    e^{(h)}_{i_1...i_S}(\bfk) e^{(h')}_{i_1...i_S}(-\bfk) & = S! \delta_{hh'}\tag{normalisation}\\
    e^{(h)}_{i_1...i_S}(\bfk) & = e^{(h)}_{(i_1...i_S)}(\bfk) \tag{symmetry} \\
    e^{(h)}_{jj i_3 ...i_S}(\bfk) & = 0 \tag{traceless} \\
    k_{i_1}...k_{i_n} e^{(h)}_{i_1...i_S}(\bfk) & = 0 \text{ for }n> S - |h|\,. \tag{transverse}
\end{align}
For example, when $S = 1$, 
\begin{equation}
    e^{(0)}_{i}(\bfk) = i\hat{k}_i, \qquad e^{(\pm 1)}_i (\bfk) = \frac {\hat{n}_i - \hat{n}_j \hat{k}_j\hat{k}_i \pm i \epsilon_{ijk} \hat{k}_j \hat{n}_k}{\sqrt{2(1- (\hat{n}_i \hat{k}_i)^2)}}\,,
\end{equation}
where $\hat{\bf n}$ is any unit vector not parallel to $\bfk$, and when $S = 2$,
\begin{equation}
    e_{ij}^{(0)}(\bfk) = \sqrt{3} \left ( \hat{k}_i \hat{k}_j - \frac 13 \delta_{ij}\right ), \quad e^{\pm 1}_{ij}(\bfk) = i(\hat{k}_i e^{(\pm 1)}_j (\bfk) + \hat{k}_j e^{(\pm 1)}_i (\bfk)), \quad e^{(\pm 2)}_{ij} = \sqrt 2 e^{(\pm 1)}_i (\bfk) e^{(\pm 1)}_j (\bfk)\,.
\end{equation}
Throughout we model the inflationary spacetime as de Sitter space and we work in the Poincar\'{e} patch with metric
\begin{align}
ds^2 = a^2(\eta)(- d \eta^2 + dx^2), \qquad a(\eta) = -\frac{1}{H \eta} \,,
\end{align}
where $H$ is the approximately constant Hubble parameter. 
\section{Feynman rules for CCM}\label{FeynmanRulesCCM}
In this section we review the Feynman rules for computing inflationary correlators where additional massive spinning fields are described by the cosmological condensed matter (CCM) physics set-up. Further details can be found in \cite{Bordin:2018pca,Stefanyszyn:2023qov}.
\subsection{Free theory}
It was argued in \cite{Bordin:2018pca} that massive spinning fields during inflation could be described as irreducible representations of the unbroken $SO(3)$ symmetries, rather than the full de Sitter $SO(1,4)$ symmetries. In this sense, one works with a theory of fluctuations on an FRW background without worrying about how the theory extends to other backgrounds. An interesting outcome of taking this point of view is that one avoids the Higuchi bound associated with representations of $SO(1,4)$ \cite{Higuchi:1986py} and therefore these states make sense even when $m^2 \rightarrow 0$. In this CCM case, fields only have spatial indices (since we are working with representations of $SO(3)$), and a spin-$S$ field will be denoted by $\Sigma^{i_1...i_S}$. These fields live on constant inflaton slices $\phi:= t+ \pi(t,\bfx) = c$, and are symmetric and traceless\footnote{The trace should be taken with respect to the induced metric $h_{ij}$ on the constant inflaton slices, with $$h_{ij} = g_{\mu\nu} \left .\frac {dx^{\mu}}{dx^i} \right |_{\phi = c} \left .\frac {dx^{\nu}}{dx^j}\right |_{\phi = c}.$$}, so the number of degrees of freedom is $2S+1$. To couple with other 4-dimensional quantities, we pushforward $\Sigma^{i_1...i_S}$ to $\Sigma^{\mu_1...\mu_S}$. For example, when $S=1$, we have \cite{Bordin:2018pca}
\begin{equation}
    \Sigma^{\mu} := \Sigma^i \left .\frac {\partial x^{\mu}}{\partial x^i}\right |_{\phi = c}= \left (-\frac {\partial_i\pi \Sigma^i}{1+\dot{\pi}} , \Sigma^i\right ) \,,
\end{equation}
and when $S=2$, we have
\begin{align}
    \Sigma^{\mu\nu} &: = \Sigma^{ij}  \left .\frac {\partial x^{\mu}}{\partial x^i}\right |_{\phi = c} \left .\frac {\partial x^{\nu}}{\partial x^j}\right |_{\phi = c} \\
    & = \begin{pmatrix}
        \dfrac {\partial_i\pi \partial_j \pi \Sigma^{ij}}{(1+\dot{\pi})^2} & -\dfrac {\partial_i \pi \Sigma^{ij}}{1+\dot{\pi}} \\
        & \\
        -\dfrac {\partial_j \pi \Sigma^{ij}}{1+\dot{\pi}} & \Sigma^{ij}
    \end{pmatrix}
    \,.
\end{align}
In general,
\begin{equation} \label{generalpushforward}
    \Sigma^{\mu_1...\mu_S} = \Sigma^{i_1...i_S} \left .\frac {\partial x^{\mu_1}}{\partial x^{i_1}}\right |_{\phi = c} \cdots \left .\frac {\partial x^{\mu_S}}{\partial x^{i_S}}\right |_{\phi = c}\,.
\end{equation}
Note that all the components of $\Sigma^{\mu_1...\mu_S}$ depend linearly and {\it algebraically} on the spatial-indexed components $\Sigma^{i_1...i_S}$ which means that there are no non-dynamical modes contained in $\Sigma^{\mu_1...\mu_S}$ (in contrast to the CC case which we will discuss below). 

We can now construct the most general quadratic action for this field up to two derivatives. Such an action will yield a kinetic term for $\Sigma^{i_1...i_S}$, but will also introduce interactions between $\Sigma^{i_1...i_S}$ and the inflaton. The general action is \cite{Bordin:2018pca}
\begin{equation} \label{quadraticactionCCM4tensor}
\begin{split}
    \hat{S}_2 = \frac 1{2S!} \int d^4 x \sqrt{-g} & \left [(1-c^2) n^{\mu} n^{\lambda} \nabla_{\mu} \Sigma^{\nu_1...\nu_S} \nabla_{\lambda} \Sigma_{\nu_1...\nu_S} - c^2 \nabla_{\mu} \Sigma^{\nu_1...\nu_S} {\nabla^{\mu}}\Sigma_{\nu_1...\nu_S}\right .\\
    &  -\, \delta c^2 \nabla_{\mu} \Sigma^{\mu \nu_2...\nu_S}\nabla_{\lambda}{\Sigma^{\lambda}}_{\nu_2...\nu_S} - (m^2 + Sc^2H^2)\Sigma^{\nu_1...\nu_S}\Sigma_{\nu_1...\nu_S} \\
    &\left . - \, 2S\kappa n^{\mu} \epsilon_{\mu\rho\gamma\lambda}\Sigma^{\rho \nu_2...\nu_S} \nabla^{\gamma} {\Sigma^{\lambda}}_{\nu_2...\nu_S} \right ]\,,
\end{split}
\end{equation}
where $n^{\mu}n_{\mu} = -1 $ is a timelike unit vector orthogonal to the constant inflaton slice. In principle there should be 5 parameters corresponding to 5 terms, but we chose to normalise $\Sigma^{\mu_1...\mu_S}$ so the coefficients of the first two terms add up to unity. Hence, there are 4 parameters, $c^2, \delta c^2, m^2, \kappa$, which control the speeds of sound of different helicity modes, the mass, and the chemical potential in the last parity-violating term. The free theory for $\Sigma^{i_1...i_S}$ has both a parity-even and parity-odd part. In conformal time they are 
\begin{align} \label{freetheoryCCM3tensor}
        S^{\text{PE}}_2 &= \frac 1{2S!} \int d^3 x d\eta \;a^2(\eta)  [(\sigma'_{i_1...i_S})^2 - c^2 (\partial_j \sigma_{i_1...i_S})^2 - \delta c^2 (\partial_j \sigma_{ji_2...i_S})^2 - m^2a^2(\eta)(\sigma_{i_1...i_S})^2] \,, \\
        S^{\text{PO}}_2 &=- \frac 1{(S-1)!} \int d^3 x d\eta \;a^3(\eta)  \kappa \epsilon_{ijk} \sigma_{i l_2...l_S}  \partial_j \sigma_{kl_2...l_S} \,,
\end{align}
where we normalised $\sigma_{i_1...i_S} = a^{-S}\Sigma_{i_1...i_S}$ so that the kinetic term is the same as a massive scalar in dS. This will ensure that we find familiar mode functions. As usual in cosmology, in this work we will make use of the background spatial translations and work in momentum space and decompose $\sigma_{i_1...i_S}$ into helicity modes, i.e.
\begin{align}
 \sigma_{i_1...i_S}(\eta, \bfk) = \sum_{h=-S}^S \sigma_{h,S} (\eta,k) e^{(h)}_{i_1...i_S}(\bfk)\,,
\end{align}
where the polarisation tensors $e^{(h)}_{i_1...i_S} (\bfk)$ satisfy various properties that are listed in the above Notations and Conventions subsection. The full free theory for the different $2S+1$ modes is now completely decoupled and can be written as 
\begin{equation} \label{freetheoryCCMhelicitymodes}
    S_2  = \frac 12 \sum_{h=-S}^S \int d\eta \int_{\bfk} a^2(\eta) \left [(\sigma'_{h,S})^2 - \left (c_{h,S}^2k^2+ m^2a^2(\eta) + 2h \kappa ka(\eta)\right )\sigma_{h,S}^2 \right ]\,,
\end{equation}
where 
\begin{equation} \label{CCMspeeds}
    c_{h,S}^2 = c^2+ \frac {S^2 - h^2}{S(2S-1)} \delta c^2\,.
\end{equation}
We now see that each mode has the quadratic action of a massive scalar field but with possibly different sound speeds. The equation of motion for each mode in the free theory is then
\begin{align}
    \mathcal{O}_{h,S}(\eta,k) \sigma_{h,S}(\eta,k) &:= \frac {\delta S_2}{\delta \sigma_{h,S}} \nonumber \\
    &= -a^2(\eta) \left (\partial_{\eta}^2 - \frac 2{\eta} \partial_{\eta} + c_{h,S}^2k^2 + m^2a^2(\eta)+ 2h\kappa ka(\eta)\right ) \sigma_{h,S}(\eta, k) = 0 \,, \label{CCMeomhelicity}
\end{align}
and the solutions to these equations that satisfy Bunch-Davies vacuum conditions are 
\begin{equation} \label{Spin2OrderCCM}
    \sigma_{h,S}(\eta,k) = -\frac {H\eta e^{-\pi \widetilde{\kappa}/2}}{\sqrt{2c_{h,S}k}}W_{i\widetilde{\kappa},\nu} (2ic_{h,S}k\eta), \qquad \nu = \sqrt{\frac 94 - \frac {m^2}{H^2}}, \qquad \widetilde{\kappa} = \frac {h\kappa}{c_{h,S}H} \,,
\end{equation}
where $W_{a,b}(z)$ is the Whittaker W-function. Note that imposing Bunch-Davies vacuum conditions is essentially a two-step procedure: $(i)$ we demand that at very early times the modes fluctuate as $\sim e^{- i k \eta}$ i.e. as they do in flat-space and $(ii)$ the mode functions satisfy the Wronskian condition:
\begin{align} \label{CCMWronskian}
     a^2 \left (\sigma_{h,S} \partial_{\eta} \sigma^*_{h,S} - \sigma^*_{h,S} \partial_{\eta} \sigma_{h,S} \right ) = i \,,
\end{align}
which ensures we have the correctly normalised canonical commutators. 
In the absence of a chemical potential, i.e. where the theory is parity-even, the mode functions are of a familiar form:
\begin{equation} \label{sigmahsCCMfreesolution}
    \sigma_{h,S} (\eta, k) = \frac {\sqrt{\pi}H}2 (-\eta)^{3/2} e^{i\pi(\nu+1/2)/2} H_{\nu}^{(1)}(-c_{h,S}k\eta) \,,
\end{equation}
where $H_{\nu}^{(1)}(z)$ is the Hankel function of the first kind. In this work we will work within the parity-even sector so will work with the Hankel form of the mode function. We refer the reader to \cite{Stefanyszyn:2023qov,Cabass:2022rhr,Liu:2019fag,Qin:2025xct,Stefanyszyn:2025yhq,Stefanyszyn:2024msm,Jazayeri:2023kji,Thavanesan:2025kyc} for discussions on the parity-odd sector. 
\subsection{Interacting theory}
We are primarily interested in how these CCM fields affect inflationary correlators of the inflaton, and possibly the graviton, via their production and decay during inflation. We therefore need to understand how to couple these fields to the inflaton and then how to compute cosmological correlators. This has been well-understood in \cite{Bordin:2018pca,Stefanyszyn:2023qov} so here we provide only a brief review. 

The interactions with $\pi(t,\bfx)$ can come from both the temporal indices on $\Sigma$ in the quadratic action \eqref{quadraticactionCCM4tensor}, and other terms permitted by the symmetries of inflation. We work within the framework of the EFT of inflation (EFToI) \cite{Cheung:2007st}, where interactions with the inflaton can enter via $\delta g^{00}$ and the extrinsic curvature $\delta K_{\mu\nu}$ of the constant inflaton slices. We will present a number of examples of such couplings throughout this work. 

Since the different helicity modes decouple in the free theory, the Feynman rules for computing wavefunction coefficients are straightforward and closely mimic those of scalar field exchange (see e.g. \cite{Goodhew:2020hob} for details). The bulk-boundary propagator for each helicity mode is
\begin{equation}
    K_{\sigma_{h,S}} (\eta,k) = \frac {\sigma_{h,S}^*(\eta,k)}{\sigma_{h,S}^*(\eta_0,k)}\,,
\end{equation}
which satisfies
\begin{equation}
\mathcal{O}_{h,S}(\eta,k) K_{\sigma_{h,S}}(\eta,k) = 0\,, \qquad K_{\sigma_{h,S}}(\eta_0,k) = 1 \,, \qquad \lim_{\eta \to -\infty(1-i\epsilon)}K_{\sigma_{h,S}}(\eta,k) = 0 \,,
\end{equation}
and allows us to write the indexed bulk-boundary propagator, with all helicity modes combined, as
\begin{equation}
K^{\sigma}_{i_1...i_Sj_1...j_S}(\eta,\bfk) = \frac 1{S!}\sum_{h=-S}^S K_{\sigma_{h,S}}(\eta,k) e^{(h)}_{i_1...i_S}(\bfk)e^{(h)}_{j_1...j_S}(-\bfk) \,.
\end{equation}
The bulk-bulk propagator for each helicity mode is\footnote{Note that the final term relies on the ratio of $\sigma_{h,S}(\eta_0,k)$ to $\sigma^*_{h,S}(\eta_0,k)$, which may not tend to a limit as $\eta_0 \to 0$ if the field is heavy (i.e. $\Re \nu = 0$). However, this can still be defined for heavy fields for any finite $\eta_0$. This is a general problem for wavefunction coefficients, but when computing the correlators, this boundary ratio will cancel out.}
\begin{equation} \label{CCMpropagators}
G_{\sigma_{h,S}}(\eta,\eta',k) = \sigma_{h,S}(\eta,k) \sigma_{h,S}^*(\eta',k) \theta (\eta - \eta') + (\eta \leftrightarrow \eta') - \frac {\sigma_{h,S}(\eta_0,k)}{\sigma_{h,S}^*(\eta_0,k)} \sigma_{h,S}^*(\eta,k) \sigma_{h,S}^*(\eta',k)\,,
\end{equation}
which satisfies the equation
\begin{equation}
\mathcal{O}_{h,S}(\eta,k) G_{\sigma_{h,S}} (\eta,\eta',k) = i \delta (\eta - \eta') \,,
\end{equation}
subject to the boundary conditions
\begin{equation}
\lim_{\eta,\eta'\to \eta_0}G_{\sigma_{h,S}}(\eta,\eta',k) = 0\,, \qquad \lim_{\eta,\eta'\to -\infty(1-i\epsilon)}G_{\sigma_{h,S}}(\eta,\eta',k) = 0\,.
\end{equation}
Again we can write the indexed bulk-bulk propagator as
\begin{align} \label{CCMbulk-bulkindexed}
G^{\sigma}_{i_1...i_Sj_1...j_S}(\eta,\eta',\bfk)=\sum_{h=-S}^S G_{\sigma_{h,S}}(\eta,\eta',k) e^{(h)}_{i_1...i_S}(\bfk) e^{(h)}_{j_1...j_S}(-\bfk) \,.
\end{align}
Note that the in above boundary conditions we included some evolution in Euclidean time when going to very early times. This deformation projects onto the vacuum at early times and ensures convergence of all time integrals in the UV.

The inflaton propagators are derived very similarly. The inflaton mode function is given by 
\begin{equation}
    \pi(\eta,k) = \frac {H}{\sqrt{2c_{\pi}^3k^3}} (1 + ic_{\pi} k\eta) e^{-ic_{\pi}k\eta} \,,
\end{equation}
where again we allow for a general speed of sound $c_{\pi}$ and have imposed Bunch-Davies vacuum conditions. The bulk-boundary propagator is 
\begin{equation}
K_{\pi}(\eta,k) = \frac {\pi^*(\eta,k)}{\pi^*(\eta_0,k)} = (1 - ic_{\pi}k\eta) e^{ic_{\pi}k\eta} \,,
\end{equation}
and the bulk-bulk propagator is
\begin{equation}
    G_{\pi}(\eta,\eta,k) = \pi(\eta,k) \pi^*(\eta',k) \theta(\eta -\eta') + (\eta \leftrightarrow \eta') - \frac {\pi(\eta_0,k)}{\pi^*(\eta_0,k)}\pi^*(\eta,k)\pi^*(\eta',k) \,.
\end{equation}
These propagators satisfy the same boundary conditions as the spinning field. 

Equipped with these propagators, we compute general wavefunction coefficients $\psi_n^{\pi...\pi\sigma...\sigma}$ (with any indices suppressed) using the following set of rules:
\begin{mdframed}
\underline{\bf General Feynman rules for computing wavefunction coefficients with CCM exchanges}
\begin{enumerate} [label=(\arabic*)]
\item Construct Feynman-Witten diagrams as follows:
\begin{enumerate} [label=(\roman*)]
    \item For the wavefunction coefficient $\psi_n^{\pi...\pi\sigma...\sigma}$, where there are $m$ $\pi$'s and $(n-m)$ $\sigma$'s in the superscript, draw $m$ external legs for $\pi$ and $(n-m)$ external legs for $\sigma$.
    \item Draw diagrams to connect the external legs, possibly with internal lines, where the kinds of vertices are dictated by the interaction Lagrangian $\mathcal{L}_{\text{int}}$, as with normal Feynman rules.
\end{enumerate}
\item Compute the contributions of these diagrams as follows: 
\begin{enumerate} [label=(\roman*)]
\item Associate each vertex with a time $\eta_1,...,\eta_V$.
\item The external legs are assigned a bulk-boundary propagator $K_{\pi}(\eta_a,k)$ or $K^{\sigma}_{i_1...i_Sj_1...j_S}(\eta_a,k)$ depending on the type of external leg, and $\eta_a$ is the time associated with that vertex.
    \item Any line that connects two vertices is assigned a bulk-bulk propagator $G_{\pi}(\eta_a,\eta_b, k)$ for the inflaton or $G^{\sigma}_{i_1...i_Sj_1...j_S}(\eta_a,\eta_b,\bfk)$ for the spinning field, where $\bfk$ is the momentum that flows from vertex $\eta_a$ to vertex $\eta_b$.
    \item Assign appropriate factors for each vertex. This includes a factor of $i$ for each vertex, spatial derivatives (with spatial indices appropriately contracted with the propagators), and temporal derivatives that then act on the propagators; as well as symmetry factors.
    \item Integrate over each time variable $\eta_1,...,\eta_V$ and add up all the relevant permutations of $\bfk_1,...,\bfk_n$, as dictated by the external states.
\end{enumerate}
\end{enumerate}
\end{mdframed}
As an example, consider the following theory for a spin-$1$ CCM field coupled to the inflaton:
\begin{align}
  S_{\text{int}} = \int d^3x dt\sqrt{-g} \left (\lambda_1\delta g^{00} \Sigma^0 + \lambda_2 \nabla_{\mu} \delta g^{00} \Sigma^{\mu}\right )& \supset \int d^3 x d\eta \left (-2\lambda_1 a^2 \pi' \partial_i\pi \sigma_i-2\lambda_2 a^2\partial_i\pi' \sigma_i \right ).
\end{align}
The number of scale factors is fixed by scale invariance and since both fields have the kinetic term of a scalar field, the number of scale factors is simply $4 - N_{\partial}$ where $N_{\partial}$ is the total number of derivatives. There are additional interactions at higher-orders in $\pi$ that will not contribute to the cubic wavefunction coefficient of $\pi$ at tree-level, which is what we will concentrate on for this example. We construct and compute the contributions from the following diagram:
\begin{equation}
\psi_3^{\pi\pi\pi}(\bfk_1,\bfk_2,\bfk_3)=
\begin{gathered}
\begin{tikzpicture}
  \begin{feynman}
    \vertex (a) at (0.5,0);
    \vertex (i1) at (1,0);
    \vertex (i2) at (2,0);
    \vertex (i3) at (3,0);
    \vertex (i4) at (4,0);
    \vertex (b) at (4.5,0);
    \vertex (c1) at (3/2,-2);
    \vertex (c2) at (7/2,-2);
        \vertex (i11) at (1,0.25) {\(\bfk_1\)};
    \vertex (i21) at (2,0.25) {\(\bfk_2\)};
    \vertex (i41) at (4,0.25) {\(\bfk_3\)};
       \diagram* {
      (a) -- [plain] (i1) -- [plain] (i2) -- [plain] (i3) -- [plain] (i4) -- [plain] (b),
      (c1) -- [plain, edge label=\(\pi'\)] (i1),
      (c1) -- [plain, edge label'=\(\partial_i\pi \)] (i2),
      (c2) -- [plain, edge label'=\(\partial_i\pi'\)] (i4),
      (c1) -- [gluon, edge label'=\(\sigma_i\)] (c2),
    };
  \end{feynman}
\end{tikzpicture}
\label{trispectrumquadraticexchange}
\end{gathered}
=\begin{gathered}
4\lambda_1 \lambda_2 \int d\eta d\eta' a^2(\eta) K'_{\pi}(\eta,k_1) K_{\pi}(\eta,k_2) k_2^i  \times \\
G^{\sigma}_{ij}(\eta,\eta',\bfk_3) a^2(\eta')  K'_{\pi}(\eta', k_3) k_3^j  \\
+\; \text{perms of }\{\bfk_1,\bfk_2,\bfk_3\} \,,
\end{gathered}
\end{equation}
with only the $h=0$ mode of the spin-$1$ field contributing since the transverse modes cannot couple to the inflaton at quadratic order. Note that the net symmetry factor is unity: we get a factor of $1/2$ as required for a single-exchange diagram, as pointed out in \cite{Goodhew:2020hob},\footnote{To arrive at this rule for computing the wavefunction we compute the on-shell action. The on-shell solution can be plugged into the free theory and the interacting part of the theory. They turn out to give the same contribution up to a factor of $-\frac{1}{2}$ and therefore sum into what we have in the main text.} and an additional factor of 2 since the two vertices are different. We are then required to go ahead and compute these integrals. For general masses such a computation is tricky due to the nested time integrals and the Hankel functions that enter through the bulk-bulk propagator. Much effort has been focused in recent years on computing such diagrams using various techniques and the general expressions are now known \cite{Chen:2009zp, Pimentel:2022fsc,Jazayeri:2022kjy,Qin:2025xct,Werth:2024mjg,Arkani-Hamed:2018kmz}, and are being put to the test with CMB and LSS observations \cite{Sohn:2024xzd,Cabass:2024wob}. 

Our ultimate interest is in cosmological correlators rather than wavefunction coefficients, and in perturbation theory the tree-level objects are algebraically related to each other through the Born rule. For inflaton correlators we have
\begin{align} \label{BornRule}
\langle \pi(\bfk_1) \ldots \pi(\bfk_n)\rangle = \frac{\int \mathcal{D}  \pi \mathcal{D}  \sigma ~ \pi(\bfk_1) \ldots \pi(\bfk_n) \Psi[\pi, \sigma] \Psi^{*}[\pi, \sigma]}{\int \mathcal{D} \pi \mathcal{D}  \sigma \Psi[\pi, \sigma] \Psi^{*}[\pi, \sigma]} \,.
\end{align}
The wavefunction provides an amplitude for finding fields with certain spatial profiles at the end of inflation, then we compute the above average to extract correlation functions. For the above example, where we have a linear mixing between the inflaton and the spinning field, the bispectrum is given by
\begin{align} \label{BispectrumFromWavefunction}
B_3 & = \frac {\rho_3^{\pi\pi\pi}(\bfk_1,\bfk_2,\bfk_3) + \left (\dfrac {\rho_3^{\pi\pi\sigma}(\bfk_1,\bfk_2,\bfk_3) \rho_2^{\pi\sigma}(\bfk_3,-\bfk_3)}{\rho_2^{\sigma\sigma}(\bfk_3)} + \text{cyclic perms of }\{\bfk_1,\bfk_2,\bfk_3\}\right )}{\rho_2^{\pi\pi}(\bfk_1)\rho_2^{\pi\pi}(\bfk_2)\rho_2^{\pi\pi}(\bfk_3)}\,,
\end{align}
where $\rho_n(\{ \bfk \}) = \psi_n(\{ \bfk \}) + \psi^{*}_n(-\{ \bfk \})$, and again we have suppressed any indices. It is this combination that sources correlators since this is what appears in $\Psi \Psi^{*}$. To arrive at \eqref{BispectrumFromWavefunction}, we expand $\Psi \Psi^{*}$ around the Gaussian and then compute Gaussin integrals as dictated by \eqref{BornRule}. We can now use the wavefunction Feynman rules to deduce rules for computing the bispectrum. We have
\begin{align}
B_3 = 4\sum_{a,b=\pm} ab ~   \lambda_1 \lambda_2 \int d\eta d\eta' a^2(\eta) \partial_{\eta} G^{\pi}_{a}(\eta,k_1) k_2^i &G^{\pi}_a(\eta,k_2) G^{\sigma}_{ab,ij}(\eta,\eta',\bfk_3) a^2(\eta') k_3^j \partial_{\eta'} G^{\pi}_b(\eta',k_3) \nonumber \\
+& \text{perms of } \{ \bfk_1,\bfk_2,\bfk_3\}\,, \label{bispectrumCCMSK}
\end{align}
where the new internal propagators are
\begin{align}
G^{\sigma_{h,S}}_{++}(\eta,\eta',k)& = \sigma_{h,S}(\eta,k) \sigma^*_{h,S}(\eta',k) \theta (\eta - \eta') + \left (\eta \leftrightarrow \eta'\right ) \,, \\
G^{\sigma_{h,S}}_{+-} (\eta,\eta',k) & = \sigma^*_{h,S}(\eta,k) \sigma_{h,S}(\eta',k) \,,\\
G^{\sigma_{h,S}}_{-+}(\eta,\eta',k) & = \sigma_{h,S}(\eta,k) \sigma^*_{h,S} (\eta',k) \,, \\
G^{\sigma_{h,S}}_{--}(\eta,\eta',k) & = \sigma^*_{h,S}(\eta,k) \sigma_{h,S}(\eta',k) \theta (\eta - \eta') + \left (\eta \leftrightarrow \eta' \right ) \,, 
\end{align}
which combine into an indexed propagator:
\begin{align}
G^{\sigma}_{ab,i_1...i_Sj_1...j_S}(\eta,\eta',\bfk) & = \sum_{h=-S}^S G^{\sigma_{h,S}}_{ab}(\eta,\eta',k) e^{(h)}_{i_1...i_S}(\bfk) e^{(h)}_{j_1...j_S}(-\bfk) \,.
\end{align}
Note that $G^{\sigma_{h,S}}_{++}$ is the usual time-ordered two-point function i.e. the Feynman propagator. The propagators that connect to the boundary at $\eta_0$ are
\begin{align}
G^{\pi}_{+} (\eta, k) & = \pi^*(\eta,k) \pi(\eta_0,k) \,, \label{InflatonSK1} \\
G^{\pi}_{-}(\eta,k) & = \pi(\eta,k) \pi^*(\eta_0,k) \label{InflatonSK2} \,.
\end{align}
Instead of going via the wavefunction, we can compute correlators directly from the Schwinder-Keldysh path integral \cite{Chen:2017ryl}. Either way, the resulting Feynman rules for CCM exchanges are:
\begin{mdframed}
\underline{\bf General Feynman rules for inflaton correlators with CCM exchanges}
\begin{enumerate} [label=(\arabic*)]
\item Construct different diagrams with different colourings of vertices as follows:
\begin{enumerate} [label=(\roman*)]
    \item For an $n$-point correlator, draw $n$ external legs for $\pi$.
    \item Draw diagrams to connect the external legs, possibly with internal lines, where the kinds of vertices are dictated by the interaction Lagrangian $\mathcal{L}_{\text{int}}$, as with normal Feynman rules.
    \item Each vertex can be coloured black ($+$) 
    or white ($-$), so for a diagram with $V$ vertices, there are $2^V$ possible colourings.
\end{enumerate}
\item Compute the contributions of these diagrams as follows:
\begin{enumerate} [label=(\roman*)]
\item Associate each vertex with a time $\eta_1,...,\eta_V$.
\item The external legs are assigned a propagator $G^{\pi}_{\pm}(\eta_a,k)$, where $\pm$ is chosen based on whether it is connected to a black ($+$) or a white ($-$) vertex, and $\eta_a$ is the time associated with that vertex.
    \item Any line that connects two vertices is assigned a propagator $G^{\sigma}_{\pm \pm,i_1...i_Sj_1...j_S}(\eta_a,\eta_b,\bfk)$. There are four possible propagators ($++$, $+-$, $-+$, and $--$) and the correct one is dictated by the colouring of the two vertices it connects. $\bfk$ is the momentum that flows from vertex $\eta_a$ to vertex $\eta_b$.
    \item Assign appropriate factors for each vertex. This includes a factor of $\pm i$ for each ($\pm$) vertex, spatial derivatives (with spatial indices appropriately contracted with the propagators), and temporal derivatives that act on the propagators; as well as symmetry factors.
    \item Integrate over all times $\eta_1,...,\eta_V$, sum all the contributions from all $2^V$ colourings, and add up all the permutations of $\bfk_1,...,\bfk_n$.
\end{enumerate}
\end{enumerate}
\end{mdframed}
In the example we considered above, the four different diagrams we would need to compute are 
\begin{equation}
B_3=
\begin{gathered}
\begin{tikzpicture}
  \begin{feynman}
    \vertex (i1) at (0,0){\(\pi'\)};
    \vertex (i2) at (0,-2){\(\partial_i \pi\)};
    \vertex (i3) at (5/2,-1){\(\partial_i\pi'\)};
    \vertex[dot] (a) at (0.5,-1){};
    \vertex[dot] (b) at (3/2,-1){};
       \diagram* {
      (i1) -- [plain] (a) [dot] -- [plain] (i2),
      (i3) -- [plain] (b),
      (a) -- [gluon, edge label'=\(\sigma_i\)] (b),
    };
  \end{feynman}
\end{tikzpicture}
\end{gathered}+
\begin{gathered}
\begin{tikzpicture}
  \begin{feynman}
    \vertex (i1) at (0,0){\(\pi'\)};
    \vertex (i2) at (0,-2){\(\partial_i \pi\)};
    \vertex (i3) at (5/2,-1){\(\partial_i\pi'\)};
    \vertex[dot] (a) at (0.5,-1){};
    \vertex[empty dot] (b) at (3/2,-1){};
       \diagram* {
      (i1) -- [plain] (a) [dot] -- [plain] (i2),
      (i3) -- [plain] (b),
      (a) -- [gluon, edge label'=\(\sigma_i\)] (b),
    };
  \end{feynman}
\end{tikzpicture}
\end{gathered}+\begin{gathered}
\begin{tikzpicture}
  \begin{feynman}
    \vertex (i1) at (0,0){\(\pi'\)};
    \vertex (i2) at (0,-2){\(\partial_i \pi\)};
    \vertex (i3) at (5/2,-1){\(\partial_i\pi'\)};
    \vertex[empty dot] (a) at (0.5,-1){};
    \vertex[dot] (b) at (3/2,-1){};
       \diagram* {
      (i1) -- [plain] (a) [dot] -- [plain] (i2),
      (i3) -- [plain] (b),
      (a) -- [gluon, edge label'=\(\sigma_i\)] (b),
    };
  \end{feynman}
\end{tikzpicture}
\end{gathered}+\begin{gathered}
\begin{tikzpicture}
  \begin{feynman}
    \vertex (i1) at (0,0){\(\pi'\)};
    \vertex (i2) at (0,-2){\(\partial_i \pi\)};
    \vertex (i3) at (5/2,-1){\(\partial_i\pi'\)};
    \vertex[empty dot] (a) at (0.5,-1){};
    \vertex[empty dot] (b) at (3/2,-1){};
       \diagram* {
      (i1) -- [plain] (a) [dot] -- [plain] (i2),
      (i3) -- [plain] (b),
      (a) -- [gluon, edge label'=\(\sigma_i\)] (b),
    };
  \end{feynman}
\end{tikzpicture}
\end{gathered}
\,, \label{B3ExampleCCM}
\end{equation}
and they combine to give \eqref{bispectrumCCMSK}. Note that the above rules are valid for parity-even correlators in which case going from a black to a white vertex is equivalent to taking a complex conjugate. In parity-odd theories, going from black to white vertices requires a complex conjugation plus a sign flip of all momenta. This is manifest in the wavefunction approach since it is $\rho_n(\{ \bfk \}) = \psi_n(\{ \bfk \}) + \psi^{*}_n(-\{ \bfk \})$ that sources correlators. This ensures that parity-even correlators of $\pi$ are purely real functions, while parity-odd correlators are purely imaginary functions, as is expected on general grounds. See e.g. \cite{Stefanyszyn:2025yhq} for more details.    

In addition to the wavefunction and Schwinger-Keldysh approaches to arriving at $B_3$, and any other correlator, one could also use the in-in formalism \cite{Adshead:2009cb,Weinberg:2005vy}. A potential downfall of the latter is that we are required to compute the Hamiltonian whereas for the former two we need the Lagrangian. For this reason, in this paper we will focus on the wavefunction and Schwinger-Keldysh approaches. 

\section{Feynman rules for CC} \label{FeynmanRulesCC}
We now turn our attention to perhaps the more familiar description of massive spinning fields during inflation where the two-point function of the spinning field is de Sitter invariant. This set-up has been studied in the context of non-Gaussianities in \cite{Lee:2016vti}. We will refer to this set-up as {\it cosmological collider} (CC) physics. In this description, the free theories of the spinning fields are representations of the full de Sitter symmetries, described by a symmetric tensor $\Phi_{\mu_1...\mu_S}$, and these fields can interact with the inflaton and we will consider interactions that are consistent with the symmetries of the EFToI. As in the CCM case, there are still only $2S+1$ propagating degrees of freedom which means that some of the modes contained in $\Phi_{\mu_1...\mu_S}$ are non-dynamical and are solely required to maintain de Sitter invariance. The dynamical modes are actually contained in the traceless parts of the spatial components $\Phi_{i_1...i_S}$. In practice, when we compute e.g. the inflaton bispectrum or trispectrum, we will integrate out these non-dynamical modes, taking care of the interactions with the inflaton, and compute the Feynman rules in terms of the dynamical modes. However, our aim is to derive Feynman rules that automate this procedure and can therefore be read off from a Lagrangian that can include the non-dynamical modes. The general procedure that we will follow to compute wavefunction coefficients, which we will then use to extract correlators, is
\begin{mdframed}
\uline{{\bf General procedure for obtaining $\psi_n$}}
\begin{enumerate} [label=(\alph*)]
\item \label{CCstepA} Decompose the field into helicity modes: 
\begin{equation} \label{CCdecomposition}
    \Phi_{\eta...\eta i_1...i_n}(\eta, \bfk) = \sum_{h=-n}^n \Phi^h_{n,S} (\eta,k) e^{(h)}_{i_1...i_n}(\bfk) +  \text{traces},
\end{equation}
with the same polarisation tensors as in the CCM case. The $\Phi_{n,S}^h$ contain the dynamics of a mode with helicity $h$ that is contained within the component of $\Phi_{\mu_1 \ldots \mu_S}$ with $n$ spatial indices and $S-n$ temporal indices \cite{Lee:2016vti}. 
\item \label{CCstepB} Write down the action in terms of the helicity modes $S[\Phi^h_{n,S}, \text{traces},\pi]$.
\item \label{CCstepC} Using the equations of motion, write all $\Phi^h_{n,S}$ with $n < S$ and traces in terms of $\Phi^h_{S,S}$ and $\pi$. This is effectively writing all components of $\Phi_{\mu_1...\mu_S}$ in terms of the traceless part of the purely spatial-indexed $\Phi_{i_1...i_S}$ fields and the inflaton. It is important to include the inflaton corrections to the $\Phi^h_{n,S}$ solutions when working with exchange diagrams. 
\item \label{CCstepD} Express the action in terms of purely spatial helicity modes $S[\Phi^h_{S,S},\pi]$.
\item \label{CCstepE} Solve the equations of motion for $\Phi^h_{S,S}$ and $\pi$ with the introduction of bulk-boundary and bulk-bulk propagators.
\item \label{CCstepF} Plug the solutions back into $S[\Phi^h_{S,S},\pi]$ to obtain the wavefunction coefficients. These wavefunction coefficients will depend on the boundary data of the dynamical modes only.
\end{enumerate}
\end{mdframed}
The main difficulty in this CC description is that the non-dynamical components of $\Phi_{\mu_1...\mu_S}$ do not in general depend  algebraically on the dynamical modes. There are in general some differential operators acting on the spatial-indexed components, making the bulk-bulk propagators not as straightforward, as we will see explicitly below. Our main focus will be on the spin-$1$ and spin-$2$ cases. To the best of our knowledge, this procedure to accurately compute wavefunction coefficients and cosmological correlators in this CC set-up has not been presented before.  

As the reader can likely tell from the above procedure, the process will not maintain manifest de Sitter invariance. This will not be an issue for us as ultimately we are interested in interactions with the inflaton that break boosts and can therefore yield potentially observable non-Gaussianities. 
\subsection{Spin 1} \label{Spin1Section}
We begin our analysis by considering a spin-$1$ field in the CC scenario. We will first examine the free theory and then add interactions with the inflaton. 

\paragraph{Free theory} We follow the procedure outlined above.
\paragraph{\ref{CCstepA} Decompose the field into helicity modes.} We can decompose the spin-1 field $\Phi_{\mu}$ as (we use the notation of \cite{Lee:2016vti} but with slightly different conventions for the polarisation vectors)
\begin{align}
    \Phi_0 (\eta, \bfk) & = \Phi^0_{0,1}(\eta,\bfk) \label{phi0helicitydecomposition} \,, \\
    \Phi_i(\eta,\bfk) & = \Phi^0_{1,1}(\eta,\bfk) (i\hat{k}_i) + \sum_{h=\pm 1}\Phi^{h}_{1,1} (\eta,\bfk) e^{(h)}_i(\bfk) \,.\label{phiihelicitydecomposition}
\end{align}
\paragraph{\ref{CCstepB} Write the action in terms of helicity modes.} The free action for the spin-1 massive field is
\begin{equation}
    S_2 = \int d^4 x \sqrt{-g}\left (-\frac 14 F_{\mu\nu}F^{\mu\nu} - \frac 12 m^2 \Phi_{\mu}\Phi^{\mu}\right ) \,,
\end{equation}
where $F_{\mu\nu} = \partial_{\mu} \Phi_{\nu} - \partial_{\nu} \Phi_{\mu}$. Written in terms of the helicity modes, the action reads
\begin{align}
S_2 & = \frac 12 \int d\eta \int_{\bfk} \left [\partial_{\eta}\Phi^0_{1,1} \partial_{\eta} \Phi^0_{1,1} - 2k\Phi^0_{0,1} \partial_{\eta} \Phi^0_{1,1} + (k^2+m^2a^2) \left (\Phi^0_{0,1}\right )^2 - m^2a^2 \left (\Phi^0_{1,1}\right )^2\right ] \nonumber \\
& \qquad \qquad + \sum_{h=\pm 1} \frac 12 \int d\eta \int_{\bfk} \left [\partial_{\eta} \Phi^{h}_{1,1} \partial_{\eta} \Phi^{h}_{1,1} - (k^2 + m^2a^2)\left (\Phi^{h}_{1,1}\right )^2\right ] \,. \label{freetheoryCCspin1}
\end{align}
Note that the kinetic terms are the same as in flat-space i.e. with no scale factors, since the Maxwell kinetic term is conformally invariant. We see that the transverse modes decouple from each other and from the $h=0$ modes, as they should by the background $SO(3)$ symmetry, but that the two $h=0$ modes are coupled. As expected, the $\Phi^0_{0,1}$ mode does not have a kinetic term. 
\paragraph{\ref{CCstepC} Solve for $\Phi^0_{0,1}$ in terms of $\Phi^0_{1,1}$.} In this spin-1 case, there is only one mode of $\Phi^h_{n,S}$ where $n \neq S$, i.e. $\Phi^0_{0,1}$, so we should solve for this mode in terms of the others. The equations of motion for the free theory are
\begin{align}
0 = \frac {\delta S_2}{\delta \Phi^0_{0,1}} & = -k\partial_{\eta} \Phi^0_{1,1} + (k^2+m^2a^2)\Phi^0_{0,1} \,, \label{spin1freesigma001eom}\\
0 = \frac {\delta S_2}{\delta \Phi^0_{1,1}} & = -\partial_{\eta}^2 \Phi^0_{1,1} + k \partial_{\eta} \Phi^0_{0,1} - m^2a^2 \Phi^0_{1,1} \label{spin1freesigma011eom} \,, \\
0 = \frac {\delta S_2}{\delta \Phi^{\pm 1}_{1,1}} & = -\partial_{\eta}^2 \Phi^{\pm 1}_{1,1} - (k^2 + m^2a^2) \Phi^{\pm 1}_{1,1} =: \mathcal{O}^{\pm 1}_{1,1} \Phi^{\pm 1}_{1,1} \,.\label{spin1freesigma111eom}
\end{align}
The first equation of motion \eqref{spin1freesigma001eom} allows us to solve for $\Phi^0_{0,1}$ {\it algebraically}:
\begin{equation} \label{sigma001solutionfree}
    \Phi^0_{0,1} = \frac {k\partial_{\eta} \Phi^0_{1,1}}{k^2+m^2a^2} \,.
\end{equation}
We are therefore left with the expected three on-shell degrees of freedom.
\paragraph{\ref{CCstepD} Write the action in terms of $\Phi^h_{S,S}$.} The solution \eqref{sigma001solutionfree} can be plugged back into the free theory action \eqref{freetheoryCCspin1}, and we obtain
\begin{align}
    S_2 & = \frac 12 \int d\eta \int_{\bfk} \left [\frac {m^2a^2}{k^2+m^2a^2} \partial_{\eta} \Phi^0_{1,1} \partial_{\eta} \Phi^0_{1,1} -m^2a^2 \left (\Phi^0_{1,1}\right )^2\right ] \nonumber \\
    & \qquad \qquad + \sum_{h=\pm 1} \frac 12 \int d\eta \int_{\bfk} \left [\partial_{\eta} \Phi^{h}_{1,1} \partial_{\eta} \Phi^{h}_{1,1} - (k^2 + m^2a^2)\left (\Phi^{h}_{1,1}\right )^2\right ]\,. \label{freetheoryspin1pluggedin}
\end{align}
\paragraph{\ref{CCstepE} Solve the equations of motion of $\Phi^h_{S,S}$.} If we vary \eqref{freetheoryspin1pluggedin} with respect to $\Phi^0_{1,1}$, we have
\begin{equation} \label{sigma011eompluggedin}
    0 = \frac {\delta S_2}{\delta \Phi^0_{1,1}} = -\partial_{\eta} \left (\frac {m^2a^2}{k^2+m^2a^2}\partial_{\eta} \Phi^0_{1,1}\right ) - m^2a^2 \Phi^0_{1,1} =: \mathcal{O}^0_{1,1}\Phi^0_{1,1} \,,
\end{equation}
which is the same as plugging in the solution of $\Phi^0_{0,1}$ \eqref{sigma001solutionfree} into \eqref{spin1freesigma011eom}, as expected. The helicity $h=\pm 1$ sectors will give the same equations of motion as \eqref{spin1freesigma111eom}. Solving these equations of motion with Bunch-Davies initial conditions and the Wronskian conditions
\begin{align}
    \frac {m^2a^2}{k^2+m^2a^2} \left (\Phi^0_{1,1} \partial_{\eta} \Phi^{0^*}_{1,1} - \partial_{\eta} \Phi^0_{1,1} \Phi^{0^*}_{1,1} \right ) = i \,,\label{phi011Wronskian} \\
     \Phi^{\pm 1}_{1,1} \partial_{\eta} \Phi^{\pm 1^*}_{1,1} - \partial_{\eta} \Phi^{\pm 1}_{1,1} \Phi^{\pm 1^*}_{1,1} = i \,,\label{phi111Wronskian}
\end{align}
gives the mode functions \cite{Lee:2016vti}
\begin{align}
    \Phi^0_{1,1} & = \frac {\sqrt{\pi}H}{4m} e^{i\pi (\nu_1 + 1/2)/2} (-\eta)^{1/2} \left (k\eta [H^{(1)}_{\nu_1+1}(-k\eta) - H^{(1)}_{\nu_1-1}(-k\eta)]- H^{(1)}_{\nu_1}(-k\eta)\right ) \,, \label{phi011freesolution} \\
    \Phi^{\pm 1}_{1,1} & = \frac {\sqrt{\pi}}2 e^{i\pi(\nu_1+1/2)/2} (-\eta)^{1/2} H^{(1)}_{\nu_1}(-k\eta)\,, \label{phi111freesolution}
\end{align}
where
\begin{equation}
    \nu_1 = \sqrt{\frac 14 - \frac {m^2}{H^2}}\,.
\end{equation}
Having mode functions that are given by sums of Hankel functions with order parameters shifted by integers is a defining feature of the CC scenario and is something that never arises in the CCM scenario. This makes the question of if the CCM scenario can mimic the CC one non-trivial. 

Now that we have the action written in terms of the dynamical modes only, we can construct the bulk-boundary and bulk-bulk propagators for these modes in the same way we did for CCM. The bulk-boundary propagators are
\begin{equation}
    K^h_{1,1}(\eta,k) = \frac {\Phi^h_{1,1}(\eta,k)}{\Phi^h_{1,1}(\eta_0,k)} \,,
\end{equation}
for $h = -1,0,1$. They satisfy
\begin{equation}
    \mathcal{O}^h_{1,1} K^h_{1,1} = 0, \qquad K^h_{1,1}(\eta_0,k) = 1,\qquad \lim_{\eta \to -\infty(1-i\epsilon)}K^h_{1,1} (\eta,k) = 0 \,.
\end{equation}
The bulk-bulk propagators are
\begin{equation}
    G^h_{1,1}(\eta,\eta',k) = \Phi^h_{1,1}(\eta,k) \Phi^{h^*}_{1,1}(\eta',k) \theta (\eta - \eta') + (\eta \leftrightarrow \eta') - \frac {\Phi^h_{1,1}(\eta_0,k)}{\Phi^{h^*}_{1,1}(\eta_0,k)} \Phi^{h^*}_{1,1}(\eta,k) \Phi^{h^*}_{1,1}(\eta',k) \,,
\end{equation}
and they satisfy
\begin{equation} \label{sigma011Gcondition}
    \mathcal{O}^h_{1,1}(\eta,k) G^h_{1,1}(\eta,\eta',k) = i\delta (\eta - \eta'), \qquad \lim_{\eta,\eta' \to \eta_0} G^h_{1,1}(\eta,\eta',k) = 0, \qquad \lim_{\eta,\eta'\to -\infty(1-i\epsilon)}G^h_{1,1}(\eta,\eta',k) = 0 \,.
\end{equation}
In principle these are the propagators we need to compute wavefunction coefficients involving this massive spin-$1$ field. However, as we will see, it will sometimes be useful to \textit{rewrite} Feynman rules for certain exchange processes in terms of propagators constructed out of the mode function of $\Phi_{0,1}^{0}$. Using \eqref{sigma001solutionfree}, the mode function for $\Phi^0_{0,1}$ is
\begin{equation} \label{phi001freesolution}
    \Phi^0_{0,1} = \frac {\sqrt{\pi}}2  \frac {Hk}m e^{i\pi(\nu_1+1/2)/2} (-\eta)^{3/2} H^{(1)}_{\nu_1}(-k\eta) \,.
\end{equation}
We can then define the bulk-boundary and bulk-bulk propagators associated with this mode function:
\begin{equation}
    K^0_{0,1}(\eta,k) = \frac {\Phi^{0^*}_{0,1}(\eta,k)}{\Phi^{0^*}_{0,1}(\eta_0,k)} \,,
\end{equation}
and
\begin{equation} \label{G001spin1noneffective}
     G^0_{0,1}(\eta,\eta',k) = \Phi^0_{0,1}(\eta,k) \Phi^{0^*}_{0,1}(\eta',k) \theta (\eta - \eta') + (\eta \leftrightarrow \eta') - \frac {\Phi^0_{1,1}(\eta_0,k)}{\Phi^{0^*}_{1,1}(\eta_0,k)}\Phi^{0^*}_{0,1}(\eta,k) \Phi^{0^*}_{0,1}(\eta',k) \,.
\end{equation}
The bulk-boundary propagator $K^0_{0,1}(\eta,k)$ still satisfies the boundary conditions
\begin{align}
     K^0_{0,1}(\eta_0,k) = 1, \qquad \lim_{\eta \to -\infty (1-i\epsilon)} K^0_{0,1}(\eta,k) = 0 \,,
\end{align}
but while the bulk-bulk propagator $G^0_{0,1}$ vanishes at the infinite past, it only vanishes at the $\eta \to \eta_0$ boundary for light fields, i.e. $m^2/H^2 < 1/4$, when $\eta_0 \to 0$, since
\begin{align}
\lim_{\eta_0 \to 0} \frac {\Phi^0_{0,1}(\eta_0,k)}{\Phi^{0^*}_{0,1}(\eta_0,k)} = \lim_{\eta_0 \to 0} \frac {\Phi^0_{1,1}(\eta_0,k)}{\Phi^{0^*}_{1,1}(\eta_0,k)} \qquad \text{for }\frac {m^2}{H^2} < \frac 14 \,.
\end{align}
This is because we have defined $G^0_{0,1}(\eta,\eta',k)$ in a slightly different way to the usual definition of the wavefunction propagator. The reason for this will become clear when we include interactions and compute exchange diagrams. At this stage the reader should simply take it as a definition, and regard the $G_{1,1}^{h}$ as the true bulk-bulk propagators. 

Before turning our attention to interactions, we point out that we have not encountered the perhaps more familiar constraint equation $\nabla^{\mu} \Phi_{\mu} = 0$ that follows from taking the divergence of the covariant equation of motion $\nabla^{\mu}F_{\mu\nu} = m^2 \Phi_{\nu}$. Once we do a $3+1$ split and convert to helicity modes, we can solve this constraint equation to find
\begin{align} \label{LongitudinalModeConstraint}
\Phi_{1,1}^{0} = -\frac{1}{k}\left(\partial_{\eta} - \frac{2}{\eta} \right) \Phi_{0,1}^0 \,.
\end{align}
The remaining equations of motion then yield second-order differential equations for $\Phi_{0,1}^0$ and $\Phi_{1,1}^{\pm 1}$. From this perspective, the dynamical modes are $\Phi_{0,1}^0$ and $\Phi_{1,1}^{\pm 1}$ rather than $\Phi_{1,1}^0$ and $\Phi_{1,1}^{\pm 1}$ as we have above. This shows that what constitutes a dynamical mode is description-dependent. What really matters is the total number of dynamical modes. To arrive at \eqref{LongitudinalModeConstraint} from our above equations, we take the time derivative of \eqref{sigma001solutionfree} and plug it into \eqref{spin1freesigma011eom}. A dynamical equation for $\Phi_{0,1}^0$ can then be obtained by plugging \eqref{LongitudinalModeConstraint} into \eqref{spin1freesigma001eom}. However, in this work we will always take $\Phi_{0,1}^0$ to be the non-dynamical mode as it is the mode with no kinetic term and therefore no conjugate momentum.   
\paragraph{Example interactions} We now introduce interactions and go through a similar procedure to compute the quartic wavefunction coefficient of $\pi$ due to the exchange of a massive spin-$1$ theory. This example will be enough to provide us with insight into the Feynman rules. We then consider general interactions. We remind the reader that our aim is to derive Feynman rules that can be read off from the Lagrangian, with both dynamical and non-dynamical modes included, but to extract these rules we must first integrate out the non-dynamical modes following our general procedure. 

We choose to work with the interactions
\begin{align} \label{ExampleInteractionsSpinone}
S_{\text{int}} \supset \lambda_1 \int d^3 x d\eta \; a(\eta) \pi'^2 \Phi_0 + \lambda_2 \int d^3 x d\eta \; a(\eta) \pi' \partial_i \pi \Phi_i \,.
\end{align}
For illustrative purposes we keep $\lambda_1$ and $\lambda_2$ as independent parameters but in the EFToI construction they will be related. Indeed, these two interactions, in additional to others, can come from the building block:
\begin{align}
    \lambda \int d^3 x dt \sqrt{-g}\delta g^{00} \Phi^0 \,,
\end{align}
with $ -\lambda_1/3 =  -\lambda_2/2=\lambda$. Such a building block is allowed by the symmetries \cite{Lee:2016vti}. We will compute $\psi_4^{\pi\pi\pi\pi}$ at $\mathcal{O}(\lambda_1^2)$, $\mathcal{O}(\lambda_2^2)$ and $\mathcal{O}(\lambda_1 \lambda_2)$, going through steps \ref{CCstepA} to \ref{CCstepF}.
\paragraph{\ref{CCstepA} Decompose into helicity modes.} This decomposition has been done in \eqref{phi0helicitydecomposition} and \eqref{phiihelicitydecomposition}.
\paragraph{\ref{CCstepB} Write the action in terms of helicity modes.} 
The free theory in terms of helicity modes is as above, and the interactions are
\begin{align}
    S_{\text{int}} & \supset \lambda_1 \int d\eta \int_{\bfk_1,\bfk_2,\bfk} a(\eta) \pi'(\eta,\bfk_1) \pi'(\eta,\bfk_2) \Phi^0_{0,1}(\eta,\bfk) \delta^{(3)}(\bfk_1+\bfk_2+\bfk) \nonumber \\
    & - \lambda_2 \int d\eta \int_{\bfk_1,\bfk_2,\bfk} a(\eta) k_2^{i} \hat{k}^{i} \pi'(\eta, \bfk_1) \pi(\eta,\bfk_2) \Phi^0_{1,1}(\eta,\bfk) \delta^{(3)}(\bfk_1+\bfk_2+\bfk) \,.
\end{align}
We have only written out the helicity-$0$ modes of the full spin-$1$ field since only there is a there a subtlety with the Feynman rules. For the exchange of the transverse modes we have the usual rules which are discussed in e.g. \cite{Stefanyszyn:2023qov,Qin:2022fbv}. The full action should also include the $\pi$ quadratic action:
\begin{equation}
S_2[\pi] = \frac 12 \int d\eta \int_{\bfk} a^2(\eta) \left (\pi'^2 - c_{\pi}^2k^2 \pi^2\right ) \,.
\end{equation}
However, since we will not be exchanging the $\pi$ field, this sector does not contribute to $\psi_4^{\pi\pi\pi\pi}$. More generally, it is proved in Appendix \ref{nopisector} that $S_2[\pi]$ only contributes to the free-theory $\psi_2^{\pi\pi}$ and diagrams that contain $\pi$ internal lines. Therefore, we only focus on $S_2[\Phi^h_{n,1}]$ given in \eqref{freetheoryCCspin1}, and $S_{\text{int}}$. 
\paragraph{\ref{CCstepC} Solve for $\Phi^h_{n,S}$ in terms of $\Phi^h_{S,S}$.} The equation of motion coming from varying the action with respect to $\Phi_{0,1}^0$ is modified from \eqref{spin1freesigma001eom} to 
\begin{equation}
    0 = \frac {\delta S}{\delta \Phi^0_{0,1}} (-\bfk) = -k\partial_{\eta} \Phi^0_{1,1} (\bfk) + (k^2+ m^2a^2) \Phi^0_{0,1}(\bfk) + \lambda_1 a(\eta) \int_{\bfk_1,\bfk_2} \pi'(\bfk_1) \pi'(\bfk_2) \delta^{(3)}(\bfk_1+\bfk_2-\bfk),
\end{equation}
which means the solution \eqref{sigma001solutionfree} for the non-dynamical mode is modified to
\begin{equation} \label{sigma001solutionspecificinteraction}
    \Phi^0_{0,1}(\eta, \bfk)= \frac 1{k^2+m^2a^2(\eta)}\left [k\partial_{\eta} \Phi^0_{1,1}(\eta, \bfk) - \lambda_1 a(\eta)\int_{\bfk_1,\bfk_2} \pi'(\eta, \bfk_1)\pi'(\eta, \bfk_2) \delta^{(3)}(\bfk_1+\bfk_2-\bfk)\right ].
\end{equation}
We could phrase this solution in terms of Green's functions where the propagator would include an overall delta function in time. We will come back to this.
\paragraph{\ref{CCstepD} Write the action in terms of $\Phi^h_{S,S}$.} We now plug this full solution back into the action such that we have an action written in terms of the dynamical modes only. At zeroth-order in $\lambda_{1,2}$ we computed the action above and found \eqref{freetheoryspin1pluggedin}. The $\lambda$-dependent corrections are
\begin{align}
S & \supset \lambda_1 \int d\eta\int_{\bfk_1,\bfk_2,\bfk}  \frac{a(\eta)}{k^2 + m^2 a^2(\eta)} \pi'(\eta, \bfk_1) \pi'(\eta, \bfk_2) k \partial_{\eta} \Phi^0_{1,1}(\eta, \bfk) \delta^{(3)}(\bfk_1+\bfk_2+\bfk)  \nonumber \\
    & - \lambda_2 \int d\eta \int_{\bfk_1,\bfk_2,\bfk} a(\eta) k_2^i \hat{k}^i  \pi'(\eta, \bfk_1) \pi(\eta,\bfk_2)  \Phi^0_{1,1}(\eta,\bfk) \delta^{(3)}(\bfk_1+\bfk_2+\bfk) \nonumber \\
& -\frac {\lambda_1^2}2 \int d\eta \int_{\bfk_1,...,\bfk_4} \frac {a^2(\eta)}{s^2+m^2a^2(\eta)} \pi'(\eta, \bfk_1)...\pi'(\eta,\bfk_4) \delta^{(3)}(\bfk_1+\bfk_2+\bfk_3+\bfk_4)  \,,\label{actionextralambdasquared}
\end{align}
where $s = |\bfk_1 + \bfk_2|$ which arises from integrating over $\bfk$ in the third contribution. Note that there are no ``$t$" and ``$u$" channel contributions since the external momenta are integrated over in this expression and so there is no notion of channels. We also note that the $\mathcal{O}(\lambda_1^2)$ term has a contribution from the interacting part of the action but also from the free theory for the spin-$1$ field. This term corresponds to a quartic self-interaction for the inflaton that will contribute to $\psi_{4}^{\pi \pi \pi \pi}$ at $\mathcal{O}(\lambda_1^2)$ via a contact diagram. This interaction is non-local and is similar to the non-localities studied in \cite{Jazayeri:2022kjy,Jazayeri:2023xcj,Jazayeri:2023kji}. We will see, however, that this non-locality drops out in our final expressions. 

\paragraph{\ref{CCstepE} Solve the equations of motion of $\Phi^h_{S,S}$.} The equation of motion for $\Phi^0_{1,1}$ is now modified from \eqref{sigma011eompluggedin} to
\begin{align}
0 = \frac {\delta S}{\delta \Phi^0_{1,1}}  & = \mathcal{O}^0_{1,1}(\eta,k) \Phi^0_{1,1}(\eta,-\bfk) \nonumber \\ & \quad - \lambda_1 \partial_{\eta} \left (\frac { k a(\eta)}{k^2+m^2a^2(\eta)} \int_{\bfk_1,\bfk_2} \pi'(\eta, \bfk_1)\pi'(\eta, \bfk_2) \delta^{(3)}(\bfk_1+\bfk_2+\bfk)\right ) \nonumber \\
& \quad - \lambda_2 \int_{\bfk_1,\bfk_2} a(\eta) k_2^i \hat{k}^i  \pi'(\eta,\bfk_1) \pi(\eta,\bfk_2) \delta^{(3)}(\bfk_1+\bfk_2 + \bfk)\,,
\end{align}
where the operator $\mathcal{O}^0_{1,1}(\eta,k) $ is defined in \eqref{sigma011eompluggedin}. Note that this is equivalent to plugging the solution of $\Phi^0_{0,1}$ \eqref{sigma001solutionspecificinteraction} into the equation of motion for $\Phi^0_{1,1}$ \eqref{spin1freesigma011eom}, similar to the free theory case. We can now write down a formal solution to this equation by using the bulk-bulk propagator that satisfies \eqref{sigma011Gcondition}. We have 
\begin{align}
\Phi^0_{1,1}(\eta,\bfk) & = K^0_{1,1}(\eta,k)  \bar{\Phi}^0_{1,1} (\bfk) \nonumber \\
& + i \lambda_1\int d\eta' \int_{\bfk_1,\bfk_2} \frac {a(\eta')k}{k^2+m^2a^2(\eta')}  \pi'(\eta',\bfk_1)\pi'(\eta',\bfk_2) \delta^{(3)}(\bfk_1+\bfk_2-\bfk)\partial_{\eta'} G^0_{1,1}(\eta,\eta',k) \nonumber \\
&  + i\lambda_2 \int d\eta' \int_{\bfk_1,\bfk_2} a(\eta') k_2^i \hat{k}^i \pi'(\eta',\bfk_1)\pi(\eta',\bfk_2) \delta^{(3)}(\bfk_1+\bfk_2-\bfk) G^0_{1,1} (\eta,\eta',k) \,, \label{sigma011solutioninteracting}
\end{align}
where $\bar{\Phi}^0_{1,1}(\bfk)$ is the prescribed boundary value of $\Phi^0_{1,1}$ as $\eta \to \eta_0$, and we have integrated by parts on the second term to put the time derivative on the bulk-bulk propagator. A potential boundary term vanishes as we take $\eta_0 \rightarrow 0$.

The next step, step \ref{CCstepF}, is to plug the solution \eqref{sigma011solutioninteracting} back into the action, which is a sum of \eqref{freetheoryspin1pluggedin} and \eqref{actionextralambdasquared} so that we can read off the quartic wavefunction coefficients. We do this for the $\mathcal{O}(\lambda_1^2)$, $\mathcal{O}(\lambda_1\lambda_2)$, and $\mathcal{O}(\lambda_2^2)$ terms separately. Let's start with the $\mathcal{O}(\lambda_1^2)$ contributions. We can write the free part of the action as (again ignoring the transverse modes) 
\begin{align} \label{FreeTheoryWithBoundary}
S_2 = \frac{1}{2} \int d \eta \int_{\bfk} \Phi_{1,1}^{0}(\eta, - \bfk) \mathcal{O}_{1,1}^{0}(\eta, k ) \Phi_{1,1}^{0}(\eta,\bfk) + \frac{1}{2} \lim_{\eta \rightarrow \eta_0} \int_{\bfk} \frac{m^2 a^2}{k^2 + m^2 a^2} \Phi_{1,1}^{0}(\eta, - \bfk) \partial_{\eta} \Phi_{1,1}^{0}(\eta, \bfk) \,.
\end{align}
For all three possibilities that are quadratic in $\lambda_{1,2}$, the boundary term in \eqref{FreeTheoryWithBoundary} does not contribute. For the other part we have 
\begin{align}
S_{2} \supset \frac{- i \lambda_1^2}{2} &\int d \eta \int d \eta' \int_{\bfk_1, \ldots, \bfk_4} \delta^{(3)}\left(\sum_{a=1}^{4} \bfk_a \right) \nonumber \\
\times &\frac{a(\eta) s }{s^2+m^2 a^2(\eta)}\frac{a(\eta') s }{s^2+m^2 a^2(\eta')}\pi'(\eta, \bfk_1)\pi'(\eta, \bfk_2)\pi'(\eta', \bfk_3)\pi'(\eta', \bfk_4)\partial_{\eta} \partial_{\eta'}G_{1,1}^{0}(\eta, \eta', s)\,,
\end{align}
where again we have performed the integration over $\bfk$ which fixes $s = |\bfk_1 + \bfk_2|$ thereby leaving us with only a single momentum-conserving delta function. If we add this to the contribution from the interacting action, then the total contribution at $\mathcal{O}(\lambda_1^2)$ is  
\begin{align}
S \supset \frac{i \lambda_1^2}{2} &\int d \eta \int d \eta' \int_{\bfk_1, \ldots, \bfk_4} \delta^{(3)}\left(\sum_{a=1}^{4} \bfk_a \right) \nonumber \\
\times &\frac{a(\eta) s }{s^2+m^2 a^2(\eta)}\frac{a(\eta') s }{s^2+m^2 a^2(\eta')}\pi'(\eta, \bfk_1)\pi'(\eta, \bfk_2)\pi'(\eta', \bfk_3)\pi'(\eta', \bfk_4)\partial_{\eta} \partial_{\eta'}G_{1,1}^{0}(\eta, \eta', s) \nonumber \\
& -\frac {\lambda_1^2}2 \int d\eta \int_{\bfk_1,...,\bfk_4} \frac {a^2(\eta)}{s^2+m^2a^2(\eta)} \pi'(\eta, \bfk_1)...\pi'(\eta,\bfk_4) \delta^{(3)}\left(\sum_{a=1}^{4} \bfk_a \right) \,.
\end{align}
We can now plug in the solution for the inflaton i.e. $\pi(\eta, \bfk) = K_{\pi}(\eta, k) \bar{\pi}(\bfk)$ and read off the quartic wavefunction coefficient. We have 
\begin{align}
&\psi_{4}^{\pi \pi \pi \pi} \supset \nonumber \\
&- 4 \lambda_1^2 \int d \eta \int d \eta' \frac{a(\eta) s }{s^2+m^2 a^2(\eta)}\frac{a(\eta') s }{s^2+m^2 a^2(\eta')}K_{\pi}'(\eta, k_1)K_{\pi}'(\eta, k_2)K_{\pi}'(\eta', k_3)K_{\pi}'(\eta', k_4)\partial_{\eta} \partial_{\eta'}G_{1,1}^{0}(\eta, \eta', s) \nonumber \\
& ~~~~~~~~~~ + t,u\text{-channels} -4 i \lambda_1^2 \int d\eta  \frac {a^2(\eta)}{s^2+m^2a^2(\eta)} K_{\pi}'(\eta, k_1)...K_{\pi}'(\eta,k_4) + t,u\text{-channels}  \,,
\end{align}
where we have now introduced the notion of channels such that the wavefunction coefficient is permutation invariant. We therefore have an exchange contribution, as we might expect from the form of the interactions that we have, but also a contact diagram. Both of these contributions are ``non-local" in the sense that they contain inverse powers of the combination $s^2+m^2 a^2(\eta)$. However, as we alluded to when we discussed the free theory, we can instead write this wavefunction coefficient in terms of $G_{0,1}^{0}$ which is a propagator constructed out of the mode function of the non-dynamical mode. At zeroth-order in $\lambda$, the two propagators are related by 
\begin{equation} \label{Gspin1relation}
\begin{split} 
\left (\frac s{s^2+m^2a^2(\eta)} \partial_{\eta}\right )\left ( \frac s{s^2+m^2a^2(\eta')} \partial_{\eta'}\right ) & G^0_{1,1}(\eta,\eta',s) \\ & =  G^0_{0,1}(\eta,\eta',s)+ \frac {is^2}{m^2a^2(s^2+m^2a^2)} \delta (\eta -\eta')\,,
\end{split}
\end{equation}
where we have used \eqref{phi011Wronskian} to simplify the coefficient of the delta function. In terms of this propagator, the quartic wavefunction coefficient simplifies to  
\begin{align}
\psi_4^{\pi\pi\pi\pi}& \supset  -4\lambda_1^2 \int d\eta d\eta' a(\eta) K'_{\pi}(\eta, k_1)K'_{\pi}(\eta, k_2) G^0_{0,1}(\eta,\eta',s) a(\eta') K'_{\pi}(\eta', k_3) K'_{\pi}(\eta',k_4) \nonumber \\
& \quad + t,u\text{-channels} -\frac {12i\lambda_1^2}{m^2} \int d\eta K'_{\pi}(\eta, k_1)K'_{\pi}(\eta, k_2)K'_{\pi}(\eta, k_3)K'_{\pi}(\eta, k_4) \,.
\end{align}
Equivalently, we can write 
\begin{align} \label{QuarticWavefunctionLambda1}
\psi_4^{\pi\pi\pi\pi}& \supset -4\lambda_1^2 \int d\eta d\eta' a(\eta) K'_{\pi}(\eta,\bfk_1)K'_{\pi}(\eta,\bfk_2) \left (G^0_{0,1\to 0,1}\right )_{\text{eff}}(\eta,\eta',s) a(\eta') K'_{\pi}(\eta',\bfk_3) K'_{\pi}(\eta',\bfk_4)\nonumber \\
& \quad + t,u\text{-channels}, 
\end{align}
where we defined an effective propagator
\begin{keyeqn}
\begin{align}
    \left (G^0_{0,1\to 0,1}\right )_{\text{eff}}(\eta,\eta',k) = G^0_{0,1}(\eta,\eta',k) + \frac i{m^2a^2}\delta(\eta-\eta')\,. \label{Gspin1effective}
\end{align}
\end{keyeqn}
The delta function in this effective propagator relates to the fact that the mode we are exchanging here is not really a dynamical mode. For future notational consistency, we denote this delta function as
\begin{align}
\Delta G^0_{0,1\to 0,1}(\eta,\eta',k) = \frac{i}{m^2 a^2}\delta(\eta - \eta') \,. \label{extradeltaspin1}
\end{align}
Diagrammatically, we can represent this wavefunction coefficient as  
\begin{equation} \label{newFeynmanspin1trispectrum}
    \underbrace{\begin{gathered}
        \begin{tikzpicture}
  \begin{feynman}
    \vertex (a) at (0.5,0);
    \vertex (i1) at (1,0);
    \vertex (i2) at (2,0);
    \vertex (i3) at (3,0);
    \vertex (i4) at (4,0);
    \vertex (b) at (4.5,0);
    \vertex (c1) at (3/2,-2);
    \vertex (c2) at (7/2,-2);
        \vertex (i11) at (1,0.25) {\(\bfk_1\)};
    \vertex (i21) at (2,0.25) {\(\bfk_2\)};
    \vertex (i31) at (3,0.25) {\(\bfk_3\)};
    \vertex (i41) at (4,0.25) {\(\bfk_4\)};
       \diagram* {
      (a) -- [plain] (i1) -- [plain] (i2) -- [plain] (i3) -- [plain] (i4) -- [plain] (b),
      (c1) -- [plain] (i1),
      (c1) -- [plain] (i2),
      (c2) -- [plain] (i3),
      (c2) -- [plain] (i4),
      (c1) -- [double, edge label'=\(\Phi^0_{0,1}\)] (c2),
    };
  \end{feynman}
\end{tikzpicture}
\end{gathered}}_{\text{correct Feynman rules}}
= \underbrace{\begin{gathered}
\begin{tikzpicture}
  \begin{feynman}
   \vertex (a) at (0.5,0);
    \vertex (i1) at (1,0);
    \vertex (i2) at (2,0);
    \vertex (i3) at (3,0);
    \vertex (i4) at (4,0);
    \vertex (b) at (4.5,0);
    \vertex (c1) at (3/2,-2);
    \vertex (c2) at (7/2,-2);
        \vertex (i11) at (1,0.25) {\(\bfk_1\)};
    \vertex (i21) at (2,0.25) {\(\bfk_2\)};
    \vertex (i31) at (3,0.25) {\(\bfk_3\)};
    \vertex (i41) at (4,0.25) {\(\bfk_4\)};
       \diagram* {
      (a) -- [plain] (i1) -- [plain] (i2) -- [plain] (i3) -- [plain] (i4) -- [plain] (b),
      (c1) -- [plain] (i1),
      (c1) -- [plain] (i2),
      (c2) -- [plain] (i3),
      (c2) -- [plain] (i4),
      (c1) -- [ghost, edge label'=\(\Phi^0_{0,1}\)] (c2),
    };
  \end{feynman}
\end{tikzpicture}
    \end{gathered}}_{\text{naive Feynman rules}}
    + \; \underbrace{\begin{gathered}
\begin{tikzpicture}
  \begin{feynman}
    \vertex (a) at (0.5,0);
    \vertex (i1) at (1,0);
    \vertex (i2) at (2,0);
    \vertex (i3) at (3,0);
    \vertex (i4) at (4,0);
    \vertex (b) at (4.5,0);
    \vertex (c) at (5/2,-2);
    \vertex (i11) at (1,0.25) {\(\bfk_1\)};
    \vertex (i21) at (2,0.25) {\(\bfk_2\)};
    \vertex (i31) at (3,0.25) {\(\bfk_3\)};
    \vertex (i41) at (4,0.25) {\(\bfk_4\)};
    \vertex (d) at (5/2,-2.55) {\(\)};
       \diagram* {
      (a) -- [plain] (i1) -- [plain] (i2) -- [plain] (i3) -- [plain] (i4) -- [plain] (b),
      (c) -- [plain] (i1),
      (c) -- [plain] (i2),
      (c) -- [plain] (i3),
      (c) -- [plain] (i4),
    };
  \end{feynman}
\end{tikzpicture}
\end{gathered}}_{\text{new contact diagram}}\,.
\end{equation}
A few comments are in order:
\begin{enumerate}[label=(\roman*)]
    \item The first contribution in \eqref{newFeynmanspin1trispectrum} corresponds to the Feynman rules we would naively read off from the action \eqref{ExampleInteractionsSpinone} using the naive bulk-bulk propagator constructed from the mode functions. Recall that there is a subtlety in how we have defined $G_{0,1}^{0}$ for heavy fields since there the boundary term is not the usual one for a bulk-bulk propagator. This subtlety only arises for the wavefunction and will drop out once we convert to correlators in which case the analogue of this diagram is computed by the naive Schwinger-Keldysh rules for both light and heavy fields alike, as we will discuss below.
    \item The naive exchange diagram is not enough to compute the full wavefunction coefficient. The reason why reading off the rules from the action does not work in the usual way, is that the field $\Phi_{0,1}^{0}$ does not have a kinetic term and therefore no conjugate momentum. There is then no sense in which we can include it in exchange diagrams without going through the procedure we have done above. We then find the addition of the contact diagram. As we will discuss in Section \ref{DeltaSection}, such a contact diagram arises in the computation of flat-space amplitudes too, but is usually hidden.
\item The non-localities that were apparent when we wrote the wavefunction coefficient in terms of the $G_{1,1}^{0}$ propagator have cancelled once we have converted to the $G_{0,1}^{0}$ propagator. We note that the cancellation is non-trivial: 
    \begin{align}
        \underbrace{\frac {a^2}{k^2+m^2a^2}}_{\mathcal{O}(\lambda) \text{ soln. into }\mathcal{O}(\lambda) \text{ int.}} + \; \underbrace{\frac {k^2}{m^2(k^2+m^2a^2)}}_{\text{convert }G^0_{1,1} \text{ to }G^0_{0,1}} = \frac 1{m^2}. \label{nonlocalitiescancel}
    \end{align}
\end{enumerate}
We will now turn our attention to the other two contributions at $\mathcal{O}(\lambda^2)$ before considering general interactions thereby deriving general Feynman rules. 

Let's now consider the $\mathcal{O}(\lambda_1\lambda_2)$ contribution. We again get contributions from the free theory and the interacting theory and their contributions differ by a factor of $-1/2$. We find
\begin{align}
S  \supset - i\lambda_1\lambda_2 \int d\eta d\eta' & \int_{\bfk_1,...,\bfk_4} \delta^{(3)} \left(\sum_{a=1}^4 \bfk_a \right)a(\eta) a(\eta') \nonumber \\
 &\times k_4^i \hat{s}^i \pi'(\eta, \bfk_1)\pi'(\eta, \bfk_2) \pi'(\eta', \bfk_3) \pi(\eta', \bfk_4) \frac{s}{s^2 + m^2 a^2(\eta)} \partial_{\eta}G_{1,1}^0(\eta, \eta',s)\,,
\end{align}
where again $s^i = k_1^i + k_2^i$ and there are no channels since the external momenta are integrated over. Again, this way of writing the action is not particularly comforting, since it involves an apparent non-locality. As we did above, we can instead define a new propagator. Indeed the operator $\frac{s}{s^2 + m^2 a^2(\eta)} \partial_{\eta}$ acting on $G_{1,1}^0(\eta, \eta',s)$ is precisely the operator that converts between the mode functions of $\Phi_{1,1}^0$ and $\Phi_{0,1}^0$. This suggests that we can define a \textit{mixed propagator} that depends on the two different mode functions:  
\begin{keyeqn}
\begin{align} 
\left (G^0_{0,1\to 1,1}\right )_{\text{eff}}(\eta,\eta',k)& = \frac k{k^2+m^2a^2(\eta)} \partial_{\eta} G^0_{1,1}(\eta,\eta',k)\nonumber \\
&  = \Phi^0_{0,1}(\eta,k) \Phi^{0^*}_{1,1}(\eta',k) \theta(\eta-\eta') + \Phi^{0^*}_{0,1}(\eta,k)\Phi^0_{1,1}(\eta',k) \theta (\eta'-\eta) \nonumber \\
& \qquad \qquad \qquad \qquad \qquad \qquad \quad - \frac {\Phi^0_{1,1}(\eta_0,k)}{\Phi^{0^*}_{1,1}(\eta_0,k)} \Phi^{0^*}_{0,1}(\eta,k) \Phi^{0^*}_{1,1}(\eta',k). \label{Gspin1mixed}
\end{align}
\end{keyeqn}
A few comments are in order. Firstly, there is no delta function in this propagator since it is defined via a single time derivative acting on a bulk-bulk propagator. The delta functions therefore cancel, in contrast to what happened above where we defined $(G_{0,1 \rightarrow 0,1}^{0})_{\text{eff}}$. Secondly, while this mixed propagator vanishes in the far past and the $\eta'\to \eta_0$ boundary, it does not necessarily vanish at the $\eta \to \eta_0$ boundary. It vanishes at the boundary for light fields but not for heavy fields. Again, this subtlety will drop out once we convert to the correlator where we will also have this mixed propagator but without the final factorised part. We can now extract the quartic wavefunction coefficient and write it using this mixed propagator:
\begin{align}
\psi_4^{\pi\pi\pi\pi} & \supset -2 \lambda_1\lambda_2 \int d\eta d\eta' a(\eta) K_{\pi}'(\eta,k_1)K_{\pi}'(\eta,k_2) \left (G^0_{0,1\to 1,1}\right )_{\text{eff}}(\eta,\eta',s) \times \nonumber \\
& \qquad \qquad \qquad \left [a(\eta') k_4^i \hat{s}^i K_{\pi}'(\eta',k_3) K_{\pi}(\eta',k_4) + a(\eta') k_3^i \hat{s}^i  K_{\pi}' (\eta',k_4) K_{\pi}(\eta',k_3)\right ]\nonumber \\
& \quad +2\lambda_1\lambda_2 \int d\eta d\eta' a(\eta) K_{\pi}'(\eta,k_3)K_{\pi}'(\eta,k_4) \left (G^0_{0,1\to 1,1}\right )_{\text{eff}}(\eta,\eta',s) \times \nonumber \\
& \qquad \qquad \qquad \left [a(\eta')k_2^i \hat{s}^i  K_{\pi}'(\eta',k_1) K_{\pi}(\eta',k_2) + a(\eta') k_1^i \hat{s}^i  K_{\pi}' (\eta',k_2) K_{\pi}(\eta',k_1)\right ]+ t,u\text{-channels}\,.
\end{align}
For future consistency of notation, we also define
\begin{align}
    \left (G^0_{1,1\to 0,1}\right )_{\text{eff}} (\eta,\eta',k) = \left (G^0_{0,1\to 1,1}\right )_{\text{eff}} (\eta',\eta,k). \label{Gspin1mixedother}
    \end{align}
Writing the wavefunction in this way follows naturally from the action once we have introduced the notion of the mixed propagator i.e. we simply read off the vertices from the Lagrangian and use the mixed propagator in internal lines that connect $\Phi_{0,1}^0$ and $\Phi_{1,1}^0$. Let us also emphasise that in canonical quantisation we expect to see this mixed propagator.\footnote{We thank Hayden Lee for discussions on this.} Indeed, as explained in \cite{Higuchi:1986py}, when we quantise this massive spin-$1$ field, we do so with the same creation and annihilation operators for $\Phi_{0,1}^0$ and $\Phi_{1,1}^0$. This means that the time-ordered two-point function $\langle T \Phi^0_{0,1} (\eta,\bfk) \Phi^0_{1,1} (\eta',\bfk')\rangle$ is non-zero and gives rise to a mixed propagator. 
 This propagator has also been considered in \cite{Kumar:2017ecc}. Diagrammatically, we represent the mixed propagator as
\begin{equation}
    \begin{gathered}
        \begin{tikzpicture}
  \begin{feynman}
    \vertex (a) at (0.5,0);
    \vertex (i1) at (1,0);
    \vertex (i2) at (2,0);
    \vertex (i3) at (3,0);
    \vertex (i4) at (4,0);
    \vertex (b) at (4.5,0);
    \vertex (c1) at (3/2,-2);
    \vertex (c) at (5/2,-2);
    \vertex (c2) at (7/2,-2);
        \vertex (i11) at (1,0.25) {\(\bfk_1\)};
    \vertex (i21) at (2,0.25) {\(\bfk_2\)};
    \vertex (i31) at (3,0.25) {\(\bfk_3\)};
    \vertex (i41) at (4,0.25) {\(\bfk_4\)};
       \diagram* {
      (a) -- [plain] (i1) -- [plain] (i2) -- [plain] (i3) -- [plain] (i4) -- [plain] (b),
      (c1) -- [plain] (i1),
      (c1) -- [plain] (i2),
      (c2) -- [plain] (i3),
      (c2) -- [plain] (i4),
      (c1) -- [double, edge label'=\(\Phi^0_{0,1}\)] (c) -- [scalar, edge label'=\(\Phi^0_{1,1}\)] (c2)
    };
  \end{feynman}
\end{tikzpicture}
    \end{gathered}\,.
\end{equation}
Finally we have the $\mathcal{O}(\lambda_2^2)$ contribution. We plug the solution \eqref{sigma011solutioninteracting} into \eqref{freetheoryspin1pluggedin} and \eqref{actionextralambdasquared} and find
\begin{align}
S_{\text{int}} \supset - \frac {i\lambda_2^2}2 \int d\eta d\eta' \int_{\bfk_1,...,\bfk_4}& \delta^{(3)} \left(\sum_{a=1}^4 \bfk_a \right) \nonumber \\
&\times a(\eta) a(\eta') k_2^i \hat{s}^i k_4^j \hat{s}^j \pi'(\eta',\bfk_1)\pi(\eta',\bfk_2)  \pi'(\eta,\bfk_3)\pi(\eta,\bfk_4)   G^0_{1,1}(\eta,\eta',k) \,.
\end{align}
It follows that the $\mathcal{O}(\lambda_2^2)$ helicity-0 contribution to the wavefunction coefficient is
\begin{align}
\psi_4^{\pi\pi\pi\pi} & \supset \lambda_2^2 \int d\eta d\eta' a(\eta') \left [k_2^i \hat{s}^i K_{\pi}'(\eta',k_1) K_{\pi}(\eta',k_2) + k_1^i \hat{s}^i K_{\pi}'(\eta',k_2) K_{\pi}(\eta',k_1) \right ]\left (G^0_{1,1\to 1,1}\right )_{\text{eff}}(\eta,\eta',s)\times \nonumber \\& \qquad \qquad \qquad \quad a(\eta)  \left [k_4^j \hat{s}^j K_{\pi}'(\eta,k_3) K_{\pi}(\eta,k_4) + k_3^j \hat{s}^j \pi'(\eta,\bfk_4) \pi(\eta,\bfk_3)\right ] +t,u\text{-channels}\,,
\end{align}
where we have denoted
\begin{keyeqn}
    \begin{align}
        \left (G^0_{1,1\to 1,1}\right )_{\text{eff}} (\eta,\eta',k) = G^0_{1,1}(\eta,\eta',k) \,,\label{G011to11}
    \end{align}
\end{keyeqn}
for consistency of notation. This is exactly what we would expect using the standard Feynman rules read off the from interacting theory. Things work out in the normal way since here we are exchanging $\Phi_{1,1}^0$ which is the dynamical mode with its own kinetic term. Diagrammatically, we represent the propagator as
\begin{equation}
    \begin{gathered}
        \begin{tikzpicture}
  \begin{feynman}
    \vertex (a) at (0.5,0);
    \vertex (i1) at (1,0);
    \vertex (i2) at (2,0);
    \vertex (i3) at (3,0);
    \vertex (i4) at (4,0);
    \vertex (b) at (4.5,0);
    \vertex (c1) at (3/2,-2);
    \vertex (c2) at (7/2,-2);
        \vertex (i11) at (1,0.25) {\(\bfk_1\)};
    \vertex (i21) at (2,0.25) {\(\bfk_2\)};
    \vertex (i31) at (3,0.25) {\(\bfk_3\)};
    \vertex (i41) at (4,0.25) {\(\bfk_4\)};
       \diagram* {
      (a) -- [plain] (i1) -- [plain] (i2) -- [plain] (i3) -- [plain] (i4) -- [plain] (b),
      (c1) -- [plain] (i1),
      (c1) -- [plain] (i2),
      (c2) -- [plain] (i3),
      (c2) -- [plain] (i4),
      (c1) -- [scalar, edge label'=\(\Phi^0_{1,1}\)] (c2)
    };
  \end{feynman}
\end{tikzpicture}
    \end{gathered}\,.
\end{equation}
So far we have concentrated on the exchange of the helicity-$0$ modes since there we found some subtleties with the Feynman rules. For completeness let us at least state the quartic wavefunction coefficient due to helicity-$\pm 1$ mode exchange. We follow the same procedure as above and find the expected expression
\begin{align}
&\psi_4^{\pi\pi\pi\pi} \supset \nonumber \\
&\lambda_2^2 \sum_{h=\pm 1}\int d\eta d\eta' a(\eta')\left [k^i_2 {e}^{(h)}_i(-{\bf s}) K'_{\pi}(\eta',k_1) K_{\pi}(\eta',k_2) + k^i_1 {e}^{(h)}_i(-{\bf s}) K_{\pi}(\eta',k_1) K'_{\pi}(\eta',k_2)\right ] \nonumber \\
&\left (G^{h}_{1,1\to 1,1}\right )_{\text{eff}}(\eta,\eta',k) a(\eta) \left [k^i_4{e}^{(h)}_i({\bf s}) K'_{\pi}(\eta,k_3) K_{\pi}(\eta,k_4) + k^i_3 {e}^{(h)}_i({\bf s}) K_{\pi}(\eta,k_3) K'_{\pi}(\eta,k_4)\right ] \nonumber \\
& + t,u\text{-channels}, \label{Psi4TransverseExchange}
\end{align}
where we denoted
\begin{keyeqn}
    \begin{align}
        \left (G^{\pm 1}_{1,1\to 1,1}\right )_{\text{eff}} (\eta,\eta',k) = G^{\pm 1}_{1,1}(\eta,\eta',k) \,, \label{Gpm111to11}
    \end{align}
\end{keyeqn}
for consistency. This contribution to the  wavefunction coefficient can be diagrammatically represented as
\begin{equation}
    \begin{gathered}
        \begin{tikzpicture}
  \begin{feynman}
    \vertex (a) at (0.5,0);
    \vertex (i1) at (1,0);
    \vertex (i2) at (2,0);
    \vertex (i3) at (3,0);
    \vertex (i4) at (4,0);
    \vertex (b) at (4.5,0);
    \vertex (c1) at (3/2,-2);
    \vertex (c2) at (7/2,-2);
        \vertex (i11) at (1,0.25) {\(\bfk_1\)};
    \vertex (i21) at (2,0.25) {\(\bfk_2\)};
    \vertex (i31) at (3,0.25) {\(\bfk_3\)};
    \vertex (i41) at (4,0.25) {\(\bfk_4\)};
       \diagram* {
      (a) -- [plain] (i1) -- [plain] (i2) -- [plain] (i3) -- [plain] (i4) -- [plain] (b),
      (c1) -- [plain] (i1),
      (c1) -- [plain] (i2),
      (c2) -- [plain] (i3),
      (c2) -- [plain] (i4),
      (c1) -- [boson, edge label'=\(\Phi^{\pm 1}_{1,1}\)] (c2)
    };
  \end{feynman}
\end{tikzpicture}
    \end{gathered}\,.
\end{equation}
Let us briefly summarise the two main take-home messages from this example:
\begin{enumerate}[label=(\roman*)]
    \item When gluing together vertices that require a propagator for $\Phi_{0,1}^0$ we should use the effective propagator in \eqref{Gspin1effective} which contains a delta function. If we use this propagator, the Feynman rules are the usual ones with vertices read off from the Lagrangian. This delta function hints to us that this mode we are exchanging is really non-dynamical. In this example, the delta function yields a contact diagram in addition to the exchange but more generally it will introduce sums of new diagrams with fewer internal lines. Although we found non-localities during the calculation, they cancel once we write the final result using this effective propagator. 
    \item When gluing together vertices where one has a factor of $\Phi_{0,1}^0$ and the other has a factor of $\Phi_{1,1}^0$, we need to use the mixed propagator when computing wavefunction coefficients. If we use this propagator, the Feynman rules are the usual ones with vertices read off from the Lagrangian. 
\end{enumerate}
\paragraph{General interactions} To be sure that the above procedure does not depend on the exact form of the interactions, let's consider general interactions and derive the Feynman rules for wavefunction ceofficients. In our above examples we have seen that the general procedure we have used is too cumbersome. Indeed, there is no need to plug the solutions for the non-dynamical modes into the action and then derive new equations of motion for the dynamical modes. Instead, we can solve the equations for all modes at the same time and plug all solutions into the action. In Appendix \ref{eomonshell} we show that this more streamlined approach yields the same result. We therefore have the following updated procedure for obtaining wavefunction coefficients in this CC set-up:
\begin{mdframed}
\uline{{\bf (Updated) general procedure for obtaining Feynman rules for $\psi_n$}}
\begin{enumerate}[label=(\Alph*)]
\item \label{CCstep1} Decompose the field into helicity modes: 
\begin{equation}
    \Phi_{\eta...\eta i_1...i_n}(\eta, \bfk) = \sum_{h=-n}^n \Phi^h_{n,S} (\eta,k) e^{(h)}_{i_1...i_n}(\bfk) +  \text{traces},
\end{equation}
with the same polarisation tensors as in the CCM case. 
\item \label{CCstep2} Write down the action in terms of the helicity modes and the inflaton $S[\Phi^h_{n,S}, \text{traces},\pi]$.
\item \label{CCstep3} Solve the equations of motion for $\Phi^h_{n,S}$, the traces and the inflaton.
\item \label{CCstep4} Plug the solutions into the action to obtain the wavefunction coefficients. These will depend on the boundary data of the dynamical modes only. 
\end{enumerate}
\end{mdframed}
With the updated procedure, we can consider a general interaction.
\paragraph{\ref{CCstep1} Decompose into helicity modes.} This has been done in \eqref{phi0helicitydecomposition} and \eqref{phiihelicitydecomposition}.
\paragraph{\ref{CCstep2} Write the action in terms of helicity modes.} The total action is \eqref{freetheoryCCspin1} plus $S_{\text{int}}$ which we keep general.
\paragraph{\ref{CCstep3} Solve the equations of motion.} The equations of motion for the helicity-$0$ modes are
\begin{align}
    \frac {\delta S_{\text{int}}}{\delta \Phi^0_{0,1} (-\bfk)}  & = -\frac {\delta S_2}{\delta \Phi^0_{0,1}(-\bfk)} = k\partial_{\eta} \Phi^0_{1,1} (\bfk)-(k^2+m^2a^2)\Phi^0_{0,1}(\bfk) \,,\label{spin1interactingsigma001eom} \\
    \frac {\delta S_{\text{int}}}{\delta \Phi^0_{1,1} (-\bfk)} & = -\frac {\delta S_2}{\delta \Phi^0_{1,1}(-\bfk)}  = \partial_{\eta}^2 \Phi^0_{1,1}(\bfk) - k\partial_{\eta} \Phi^0_{0,1}(\bfk) + m^2a^2 \Phi^0_{1,1}(\bfk) \,,\label{spin1interactingsigma011eom}
\end{align}
where we have used \eqref{spin1freesigma001eom} and \eqref{spin1freesigma011eom} respectively. The first of these \eqref{spin1interactingsigma001eom} implies\footnote{If $S_{\text{int}}$ is quadratic or higher in $\Phi^0_{0,1}$ then there will still be $\Phi^0_{0,1}$ dependence on the right hand side, so it is not a ``solution" in that case. However, this statement is still true, and one can use it to perturbatively expand $\Phi^0_{0,1}$.}
\begin{equation} \label{spin1generalsigma001solution}
    \Phi^0_{0,1}(\bfk) = \frac 1{k^2+m^2a^2} \left [k\partial_{\eta} \Phi^0_{1,1} (\bfk) - \frac {\delta S_{\text{int}}}{\delta \Phi^0_{0,1}(-\bfk)} \right ]\,.
\end{equation}
We can now plug \eqref{spin1generalsigma001solution} into \eqref{spin1interactingsigma011eom} to obtain 
\begin{align}
\frac {\delta S_{\text{int}}}{\delta \Phi^0_{1,1}(-\bfk)}  & = - \mathcal{O}^0_{1,1}(\eta,k) \Phi^0_{1,1}(\eta,\bfk) + \partial_{\eta} \left (\frac k{k^2+m^2a^2}\frac {\delta S_{\text{int}}}{\delta \Phi^0_{0,1}(-\bfk)}\right )\,.
\end{align}
We can then provide a formal solution to this equation of
\begin{equation} \label{sigma011solutiongeneralinteraction}
    \begin{split}
    \Phi^0_{1,1} (\eta,\bfk) & = K^0_{1,1}(\eta,k) \bar{\Phi}^0_{1,1} (\bfk) + i\int d\eta' \frac {\delta S_{\text{int}}}{\delta \Phi^0_{0,1}}(\eta',-\bfk) \left (G^0_{1,1\to 0,1}\right )_{\text{eff}}(\eta,\eta',k) \\
& \qquad \qquad \qquad \qquad \quad + i\int d\eta' \frac {\delta S_{\text{int}}}{\delta \Phi^0_{1,1}}(\eta',-\bfk) \left (G^0_{1,1\to 1,1}\right )_{\text{eff}}(\eta,\eta',k) \,,    
    \end{split}
\end{equation}
where we have integrated by parts such that we have a differential operator acting on $G_{1,1}^{0}$ and then used the relation \eqref{Gspin1mixed} to introduce the mixed propagator. A potential boundary term vanishes. Note that $\left(G^0_{1,1\to 0,1}\right )_{\text{eff}}(\eta_0, \eta',k) \rightarrow 0$ for both light and heavy fields so this solution satisfies the appropriate boundary conditions. To find the solution for $\Phi^0_{0,1}$ we simply plug \eqref{sigma011solutiongeneralinteraction} into \eqref{spin1generalsigma001solution} and we find
\begin{equation} \label{sigma001solutiongeneralinteraction}
\begin{split}
    \Phi^0_{0,1}(\eta,\bfk) & = K^0_{0,1}(\eta,k) \left (\frac {\Phi^{0^*}_{0,1}(\eta_0,k)}{\Phi^{0^*}_{1,1}(\eta_0,k)}\bar{\Phi}^0_{1,1}(\bfk)\right ) + i\int d\eta' \frac {\delta S_{\text{int}}}{\delta \Phi^0_{1,1}}(\eta',-\bfk) \left (G^0_{0,1\to 1,1}\right )_{\text{eff}}(\eta,\eta',k) \\
    & \qquad \qquad \qquad \qquad \quad \qquad \qquad \qquad + i\int d\eta' \frac {\delta S_{\text{int}}}{\delta \Phi^0_{0,1}}(\eta',-\bfk) \left (G^0_{0,1\to 0,1}\right )_{\text{eff}}(\eta,\eta',k) \,,
    \end{split}
\end{equation}
where again we have traded any differential operators acting on the bulk-bulk propagators with other propagators as we saw for specific interactions above. The final term in \eqref{spin1generalsigma001solution} is now captured by the delta function of $\left(G^0_{0,1\to 0,1}\right )_{\text{eff}}$ c.f. \eqref{Gspin1effective}. In these two solutions we therefore find four different, but related, propagators: $\left (G^0_{n,1\to m,1}\right )_{\text{eff}}$ where $m,n=0,1$, and in all cases the boundary data is given in terms of the dynamical modes only. This ensures that the boundary wavefunction only depends on the data of the dynamical modes.
\paragraph{\ref{CCstep4} Plug the solutions back into action.} In the previous steps, we have kept the interactions completely general. For this step, however, we will focus on interactions that depend linearly on $\Phi_{\mu}$, and therefore contribute to $\psi_n$ from exchanging $\Phi_{\mu}$. The double-exchange case, arising when we have interactions that are quadratic in $\Phi_{\mu}$, is more complicated and we provide details in  Appendix \ref{factorexchange}. Since we are not exchanging $\pi$, the arguments in Appendix \ref{nopisector} imply that there are no contributions from $S_2[\pi]$. Since the free theory is quadratic in $\Phi^h_{n,S}$ and the interactions are assumed to be linear in these modes, the necessary parts of the action are
\begin{align}
    S^{(0)} & = \frac 12 \int d\eta \int_{\bfk} \Phi^0_{0,1}(\eta,\bfk) \frac {\delta S_2}{\delta \Phi^0_{0,1}}(\eta, \bfk) + \frac 12 \int d\eta \int_{\bfk} \Phi^0_{1,1}(\eta,\bfk) \frac {\delta S_2}{\delta \Phi^0_{1,1}}(\eta,\bfk) \nonumber \\
    & + \lim_{\eta \to \eta_0} \frac 12\int_{\bfk} \Phi^0_{1,1}(\eta,\bfk)\left (\partial_{\eta} \Phi^0_{1,1}(\eta,-\bfk) - k\Phi^0_{0,1}(\eta,-\bfk)\right ) \nonumber \\
    & +\int d\eta \int_{\bfk}\Phi^0_{0,1} (\eta,\bfk)\frac {\delta S_{\text{int}}}{\delta \Phi^0_{0,1}}(\eta,\bfk)+\int d\eta \int_{\bfk}\Phi^0_{1,1}(\eta,\bfk) \frac {\delta S_{\text{int}}}{\delta \Phi^0_{1,1}}(\eta,\bfk)\,,\label{generalspin1linearinteraction}
\end{align}
where we have integrated by parts. The boundary terms on the second line will be important for the wavefunction coefficients of the spinning field, but they are irrelevant for the inflaton wavefunction coefficients. Plugging the solutions \eqref{sigma011solutiongeneralinteraction} and \eqref{sigma001solutiongeneralinteraction} back into \eqref{generalspin1linearinteraction}, and extracting the parts of the action that can contribute to inflaton wavefunction coefficients, we have
\begin{align}
    iS^{(0)} & \supset -\frac 12 \sum_{m,n=0}^1\int d\eta d\eta' \int_{\bfk} \frac {\delta S_{\text{int}}}{\delta \Phi^0_{n,1}} (\eta,\bfk) \left (G^0_{n,1\to m,1}\right )_{\text{eff}}(\eta, \eta',k) \frac {\delta S_{\text{int}}}{\delta \Phi^0_{m,1}}(\eta',-\bfk) \,.\label{Feynmanspin1helicity0}
\end{align}
 For the $h=\pm 1$ modes, since they are both dynamical, running through the same arguments, we have
\begin{align} \label{sigma111solutiongeneralinteraction}
    \Phi^{\pm 1}_{1,1}(\eta,\bfk) = K^{\pm 1}_{1,1}(\eta,k) \bar{\Phi}^{\pm 1}_{1,1}(\bfk) + i\int d\eta \frac {\delta S_{\text{int}}}{\delta \Phi^{\pm 1}_{1,1}}(\eta',-\bfk) \left (G^{\pm 1}_{1,1\to 1,1}\right )_{\text{eff}}(\eta,\eta',k)\,,
\end{align}
and therefore
\begin{align}
    iS^{(\pm 1)} & \supset -\frac 12 \int d\eta d\eta' \int_{\bfk} \frac {\delta S_{\text{int}}}{\delta \Phi^{\pm 1}_{1,1}} (\eta, \bfk) \left (G^{\pm 1}_{1,1\to 1,1}\right )_{\text{eff}}(\eta,\eta',k) \frac {\delta S_{\text{int}}}{\delta \Phi^{\pm 1}_{1,1}}(\eta',-\bfk)\,, \label{Feynmanspin1helicity1}
\end{align}
and therefore the $\Phi^{\pm 1}_{1,1}$ propagator is as expected. The different helicities do not mix, so there are no mixed propagators for these other modes. The exact wavefunction coefficients can then be easily extracted from these expressions for the on-shell action. We now see that when we use the effective propagators, the wavefunction coefficients can be simply read off the interacting Lagrangian.

To summarise the Feynman rules for wavefunction coefficients, we first define the indexed propagator $G^{\Phi}_{\mu\nu}(\eta,\eta',\bfk)$ according to the decomposition of helicity modes in step \ref{CCstep1}:
\begin{align}
G^{\Phi}_{00}(\eta,\eta',\bfk) & = \left (G^0_{0,1\to 0,1}\right )_{\text{eff}}(\eta,\eta',k) \,, \label{Spin1Indexed1} \\
G^{\Phi}_{0i} (\eta,\eta',\bfk) & = \left (G^0_{0,1\to 1,1}\right )_{\text{eff}} (\eta,\eta',k) (-i\hat{k}_i)\,, \label{Spin1Indexed2} \\
G^{\Phi}_{i0}(\eta,\eta',\bfk) & = \left (G^0_{1,1\to 0,1}\right )_{\text{eff}} (\eta,\eta',k) (i\hat{k}_i) \,, \label{Spin1Indexed3} \\
G^{\Phi}_{ij}(\eta,\eta',\bfk) & = \sum_{h=-1}^1\left (G^h_{1,1\to 1,1}\right )_{\text{eff}}(\eta,\eta',k) e^{(h)}_i(\bfk) e^{(h)}_j (-\bfk)\,, \label{Spin1Indexed4}
\end{align}
where the effective propagators $\left (G^h_{n,1\to m,1}\right )_{\text{eff}}$ are given in the highlighted equations \eqref{Gspin1effective}, \eqref{Gspin1mixed}, \eqref{Gspin1mixedother}, \eqref{G011to11}, and \eqref{Gpm111to11}. The Feynman rules for wavefunction coefficients can then be summarised as follows:
\begin{mdframed}
\underline{\bf General Feynman rules for wavefunction coefficients with spin-1 CC exchanges}
\begin{enumerate} [label=(\arabic*)]
\item Construct Feynman-Witten diagrams as follows:
\begin{enumerate} [label=(\roman*)]
    \item For the wavefunction coefficient $\psi_n^{\pi...\pi\Phi...\Phi}$, where there are $m$ $\pi$'s and $(n-m)$ $\Phi$'s in the superscript, draw $m$ external legs for $\pi$ and $(n-m)$ external legs for $\Phi$. 
    \item Draw diagrams to connect the external legs with vertices dictated by the interaction Lagrangian $\mathcal{L}_{\text{int}}$, same as normal Feynman rules.
\end{enumerate}
\item Compute the contributions of these diagrams as follows:
\begin{enumerate} [label=(\roman*)]
\item Associate each vertex with a time $\eta_1,...,\eta_V$.
\item The external legs are assigned a propagator $K_{\pi}(\eta_a,k)$ or $K^{\Phi}_{\mu}(\eta_a,k)$ depending on the type of external leg, and $\eta_a$ is the time associated to that vertex.
    \item Any line that connects two vertices is assigned a propagator $G^{\Phi}_{\mu\nu}(\eta_a,\eta_b,\bfk)$, where $\bfk$ is the momentum that flows from vertex $\eta_a$ to vertex $\eta_b$. The components of this propagator correspond to the effective propagators we derived above. 
    \item Assign appropriate factors for each vertex. This includes a factor of $i$ for each vertex, spatial derivatives (with spatial indices appropriately contracted with the propagators) and temporal derivatives on $\pi$ and $\Phi$; as well as symmetry factors.
    \item Integrate over time $\eta_1,...,\eta_V$ and add up all the permutations of $\bfk_1,...,\bfk_n$.
\end{enumerate}
\end{enumerate}
\end{mdframed}
We will discuss how we convert to cosmological correlators in section \ref{FromWavefunctionToCorrelatorCC} and will provide further commentary on the delta function arising in the effective propagator for $\Phi_0$ in Section \ref{DeltaSection}.
\subsection{Spin 2} \label{Spin2CCSection}
We now turn our attention to a massive spin-$2$ field described by the CC scenario. As we did for spin-$1$, we will go through the updated general procedure \ref{CCstep1} to \ref{CCstep4} in order to extract the wavefunction propagators. The computations quickly become complicated so in this main body we only present the most important results with the details given in Appendix \ref{spin2details}.
\paragraph{\ref{CCstep1} Decompose the field into helicity modes.} We can decompose the spin-2 field $\Phi_{\mu\nu}$ as
\begin{align}
\Phi_{00}(\eta,\bfk) & = \Phi^0_{0,2}(\eta,\bfk) \,, \label{spin2decomposehelicity00} \\
\Phi_{0i}(\eta,\bfk) & = \Phi^0_{1,2}(\eta,\bfk) e^{(0)}_i(\bfk)+ \sum_{h=\pm 1}\Phi^{h}_{1,2} (\eta,\bfk) e^{(h)}_i(\bfk) \,, \label{spin2decomposehelicity0i} \\
\Phi_{ij}(\eta,\bfk) & = \Phi^0_{2,2}(\eta,\bfk) e^{(0)}_{ij}(\bfk)+ \sum_{h=\pm 1}\Phi^{h}_{2,2} (\eta,\bfk) e^{(h)}_{ij}(\bfk)+ \sum_{h=\pm 2} \Phi^h_{2,2}(\eta,\bfk) e^{(h)}_{ij}(\bfk)  + \frac 13 \delta_{ij} \Phi_{kk}(\eta, \bfk)\,,\label{spin2decomposehelicityij}
\end{align}
where again we have employed the notation of \cite{Lee:2016vti}. For the $\Phi_{ij}$ components we have further decomposed into pure trace and traceless parts such that we are working with irreps of $SO(3)$. The $e_i^{(h)}$ polarisation vectors are the same as those for the spin-$1$ field, and the $e_{ij}^{(h)}$ polarisation tensors are given in the Notation and Conventions section and are traceless. The reader may wonder why in this decomposition we have both $\Phi_{00}$ and $\Phi_{ii}$ as independent components. It is certainly true that on-shell $\Phi_{\mu\nu}$ is traceless are therefore $\Phi_{00}$ and $\Phi_{ii}$ are not independent (as we will see in Appendix \ref{spin2details}), however, we are doing this decomposition off-shell so at this level they are independent. For future convenience, we will write the trace as $\Phi_{kk} =: \Phi^0_{kk,2}$.
\paragraph{\ref{CCstep2} Write the action in terms of helicity modes.} The free action for the spin-2 massive field is \cite{Higuchi:1986py}
\begin{equation} \label{spin2covariant}
    \begin{split}
        S_2 & = \int d^4 x \sqrt{-g} \left (-\frac 12 \nabla^{\rho} \Phi^{\mu\nu} \nabla_{\rho} \Phi_{\mu\nu} + \nabla^{\mu}\Phi_{\mu\nu} \nabla_{\rho} \Phi^{\rho \nu} - \nabla^{\mu} \Phi_{\mu\nu} \nabla^{\nu} \Phi^{\rho}{}_{\rho} + \frac 12 \nabla^{\mu} \Phi^{\rho}{}_{\rho} \nabla_{\mu} \Phi^{\nu}{}_{\nu} \right . \\
        & \qquad \qquad \qquad \qquad  \left . -\frac 12 \left (m^2 + 2H^2\right )\left (\Phi^{\mu\nu}\Phi_{\mu\nu} - \Phi^{\mu}{}_{\mu} \Phi^{\nu}{}_{\nu} \right )-\frac 32 H^2 \Phi^{\mu}{}_{\mu} \Phi^{\nu}{}_{\nu}\right )\,,
    \end{split}
\end{equation}
where $m^2 \geq 2 H^2$ by the Higuchi bound \cite{Higuchi:1986py}. The case of $m^2 = 2H^2$ is the partially-massless limit. We want to now expand this action in terms of the helicity modes, and as expected the result is rather complicated. Since we are working with $SO(3)$ irreps the $h=0$, $h = \pm 1$ and $h = \pm 2$ modes decouple. The full expressions can be found in \eqref{fullspin2helicityaction} - \eqref{boundarytermsspin2action}.
\paragraph{\ref{CCstep3} Solve the equations of motion.} We include general interactions and vary the action with respect to each of the different helicity modes. In the free theory, we see that the modes $\Phi_{0,2}^0$, $\Phi_{1,2}^0$, $\Phi_{kk,2}^0$ and $\Phi_{1,2}^{\pm 1}$ can be solved for algebraically in terms of $\Phi_{2,2}^0$ and $\Phi_{2,2}^{\pm 1}$. The $\Phi_{2,2}^{\pm 2}$ modes are decoupled from the rest. From this perspective, the propagating modes are therefore those contained in the traceless part of $\Phi_{ij}$ which does indeed have the expected $5$ degrees of freedom. Similar to the spin-1 case, we can write the solution for each mode in terms of effective or mixed bulk-bulk propagators. In general, we can write the solutions in terms of the various mode functions as
\begin{align}
\Phi^h_{n,2}(\eta,\bfk) = K_{n,2}^h(\eta,k) &\left (\frac {\Phi^{h^*}_{n,2}(\eta_0,k)}{\Phi^{h^*}_{2,2}(\eta_0,k)} \bar{\Phi}^h_{2,2}(\bfk) \right ) \nonumber \\ & + \,i\int d\eta' \sum_{m} \frac {\delta S_{\text{int}}}{\delta \Phi^h_{m,2}}(\eta',-\bfk)\left (G^{h}_{n,2\to m,2}\right )_{\text{eff}}(\eta,\eta',k)\,,\label{spin2generalsolution}
\end{align}
where we have introduced the effective propagators $\left (G^{h}_{n,2\to m,2}\right )_{\text{eff}}(\eta,\eta',k)$ which conserve helicity as they should by the $SO(3)$ symmetry. The sum over $m$ is a sum over all modes, including any traces. As in the spin-$1$ case, the boundary data that appears in these solutions is always that of the dynamical modes $\bar{\Phi}_{2,2}^h$. These effective  propagators take the general form
\begin{align}
    \left (G^{h}_{n,2\to m,2}\right )_{\text{eff}} (\eta, \eta',k) & = G^{h}_{n,2\to m,2}(\eta, \eta',k) + \Delta G^h_{n,2\to m,2}(\eta,\eta',k)\,,
\end{align}
where
\begin{align}
G^{h}_{n,2\to m,2} (\eta, \eta',k)  = \Phi^h_{n,2}(\eta,k) \Phi^{h^*}_{m,2} (\eta',k) \theta (\eta - \eta') + & \Phi^{h^*}_{n,2}(\eta,k)  \Phi^h_{m,2}(\eta',k) \theta (\eta' - \eta) \nonumber \\
&  - \frac {\Phi^h_{2,2}(\eta_0,k)}{\Phi^{h^*}_{2,2}(\eta_0,k)} \Phi^{h^*}_{n,2}(\eta,k) \Phi^{h^*}_{m,2}(\eta',k)\,,
\end{align}
is the propagator we naively expected based on e.g. canonical quantisation but with a factorised term added. The mode functions are given in Appendix \ref{spin2details}. As with the spin-$1$ case, these propagators do not necessarily vanish at the boundary, but the boundary terms will drop out when we convert to correlators. The corrections, $\Delta G$, contain delta functions thereby suggesting that we are exchanging non-dynamical modes. The non-zero ones take the form
\begin{align}
\Delta G^0_{0,2\to 0,2}(\eta,\eta',k) & = - \frac {2iH^2\alpha_1\alpha_2}3 \partial_{\eta} \left (\eta^2\delta' (\eta - \eta')\right )- \frac {2i\alpha_1}3\left (1 - H^2k^2\eta^2\alpha_2\right )\delta (\eta -\eta') \,, \label{Deltaphi002phi002}\\
\Delta G^0_{0,2\to 1,2}(\eta,\eta',k) & = \frac {2iH^2k \alpha_1\alpha_2}3 \partial_{\eta} \left (\eta^2\delta (\eta - \eta')\right ) \,,  \\
\Delta G^0_{0,2\to kk,2}(\eta,\eta',k) & = 2iH^2\alpha_1\alpha_2\partial_{\eta} \left (\eta\delta(\eta - \eta')\right ) - \frac {i\alpha_1 (3 - 2H^2k^2\eta^2\alpha_2)}3 \delta (\eta - \eta') \,, \\
\Delta G^0_{0,2\to 2,2}(\eta,\eta',k) & = \frac {2iH^2k^2\eta^2 \alpha_1 \alpha_2}{3\sqrt 3} \delta (\eta - \eta') \,, \\
\Delta G^0_{1,2\to 1,2}(\eta,\eta',k)& = \frac {i\alpha_1(3+4H^2k^2\eta^2 \alpha_2)}6 \delta (\eta - \eta') \,,\\
\Delta G^0_{1,2\to kk,2} (\eta,\eta',k) & = 2iH^2k\eta \alpha_1\alpha_2 \delta (\eta - \eta') \,,  \\
\Delta G^0_{kk,2\to kk,2}(\eta,\eta',k) & = 6iH^2\alpha_1\alpha_2\delta (\eta - \eta') \,,  \\
\Delta G^{\pm 1}_{1,2\to 1,2}(\eta,\eta',k) & = \frac {i\alpha_1}2\delta (\eta - \eta')\,, \label{Deltaphi112phi112}
\end{align}
where we have defined
\begin{align}
    \alpha_1 = \frac 1{m^2}, \qquad \alpha_2 = \frac 1{m^2-2H^2}\,. \label{alphaHiguchi}
\end{align}
The other non-zero expressions can be obtained by the relation
\begin{align}
   \Delta G^h_{n,2\to m,2}(\eta,\eta',k) = \Delta G^h_{m,2\to n,2}(\eta',\eta,k) \,.
\end{align}
In the spin-$1$ case we saw that the corresponding non-zero $\Delta G$ contained a delta function and that arose due to the first-order differential operator that mapped between the $h=0$ modes. For this spin-$2$ case we see that derivatives of delta functions can appear. This is because the various modes can be related by differential operators that are second-order in time. We give the explicit expressions in Appendix \ref{spin2details} where we provide a full derivation of these effective propagators.
\paragraph{\ref{CCstep4} Plug the solutions back into action.} We now plug the general solutions \eqref{spin2generalsolution} into the action. If $S_{\text{int}}$ depends linearly on $\Phi_{\mu\nu}$, as will be the case for the leading contributions to the bispectra and trispectra of inflaton perturbations, we find
\begin{align}
iS & \supset -\frac 12 \sum_{h=-2}^2\sum_{m,n}\int d\eta d\eta' \int_{\bfk} \frac {\delta S_{\text{int}}}{\delta \Phi^h_{n,2}}(\eta,\bfk) \left (G^{h}_{n,2\to m,2}\right )_{\text{eff}}(\eta,\eta',k) \frac {\delta S_{\text{int}}}{\delta \Phi^h_{m,2}}(\eta',-\bfk)\,, \label{Feynmanspin2}
\end{align}
from which we can extract wavefunction coefficients. Here the sums over $m,n$ sum over all possible modes include any traces. We can again defined an indexed propagator as 
\begin{align}
G^{\Phi}_{\underbrace{\scriptstyle 0...0i_1...i_n}_{S \text{ indices}}\underbrace{\scriptstyle 0...0j_1...j_m}_{S \text{ indices}}}(\eta,\eta',k) = \sum_{h=-n}^{n} \left (G^h_{n,S\to m,S}\right )_{\text{eff}}(\eta,\eta',k) e^{(h)}_{i_1...i_n}(\bfk) e^{(h)}_{j_1...j_m}(-\bfk) + \text{traces}\,,
\end{align}
and then the Feynman rules for computing wavefunction coefficients due to exchanging spin-$2$ CC modes are the same as for exchanging spin-$1$: we simply read off the vertices for Feynman diagrams from the Lagrangian and glue vertices together using the effective propagator $G_{\mu\nu \rho \sigma}^{\Phi}$.

\subsection{Converting to cosmological correlators} \label{FromWavefunctionToCorrelatorCC}
Let's now convert wavefunction coefficients from CC exchanges to cosmological correlators. In Appendix \ref{WavefunctionToCorrelatorsDetails} we provide a detailed discussion for general interactions, here we will work with a concrete example in order to illustrate the most important points. We choose to work with the four-point function of inflaton perturbations due to spin-$1$ exchange. We allow for interactions such that all of the effective propagators we encountered in Section \ref{Spin1Section} are relevant. We parametrise the wavefunction as 
\begin{align}
\ln \Psi &= -\frac{1}{2} \int_{\bfk} \psi^{\pi \pi}_2(k) \bar{\pi}(-\bfk) \bar{\pi}(\bfk) -\frac{1}{2} \int_{\bfk} \psi^{\Phi \Phi }_{2,0}(k) \bar{\Phi }^{0}_{1,1}(-\bfk) \bar{\Phi }^{0}_{1,1}(\bfk)-\frac{1}{2} \int_{\bfk} \psi^{\Phi \Phi }_{2,\pm1}(k) \sum_{h=\pm1} \bar{\Phi}^{h}_{1,1}(-\bfk) \bar{\Phi}^{h}_{1,1}(\bfk) \nonumber \\
& + \frac{1}{2} \int_{\bfk_1 \ldots \bfk_3}\psi_{3,0}^{\pi \pi \Phi}(\{ \bfk \}) \hat{\delta}^{(3)}\left(\sum \bfk \right)\bar{\pi}(\bfk_1)\bar{\pi}(\bfk_2)\bar{\Phi}_{1,1}^{0}(\bfk_3) \nonumber \\
&+\frac{1}{2}\sum_{h=\pm 1} \int_{\bfk_1 \ldots \bfk_3}\psi_{3,h}^{\pi \pi \Phi}(\{ \bfk \}) \hat{\delta}^{(3)}\left(\sum \bfk \right)\bar{\pi}(\bfk_1)\bar{\pi}(\bfk_2)\bar{\Phi}_{1,1}^{h}(\bfk_3) \nonumber \\
& +\frac{1}{24}\int_{\bfk_1 \ldots \bfk_4}\psi_{4}^{\pi \pi \pi \pi}(\{ \bfk \}) \hat{\delta}^{(3)}\left(\sum \bfk \right)\bar{\pi}(\bfk_1)\ldots \bar{\pi}(\bfk_4)\,.
\end{align}
As we have expressed a number of times, this wavefunction only depends on the boundary data of the dynamical modes which in this case are $\bar{\pi}$, $\bar{\Phi}_{1,1}^0$ and $\bar{\Phi}_{1,1}^{\pm 1}$. Note that since the free theory for the spinning field is parity-even, we have $\psi^{\Phi \Phi}_{2,+1} = \psi^{\Phi \Phi}_{2,-1}$ which we have chosen to denote as $\psi^{\Phi \Phi}_{2, \pm 1}$. To compute cosmological correlators we need $|\Psi^2|$. The expression for $\ln |\Psi^2|$ is the same as above but with all $\psi$'s replaced by $\rho$'s where $\rho(\{ \bfk \}) = \psi(\{ \bfk \}) + \psi^{*}(- \{ \bfk \})$. Our four-point function of interest is then given by 
\begin{align}
\langle \pi(\bfk_1) \ldots \pi(\bfk_4) \rangle = \frac{\int  \mathcal{D} \bar{\pi}  \mathcal{D} \bar{\Phi} |\Psi^2| \bar{\pi}(\bfk_1) \ldots \bar{\pi}(\bfk_4)} {\int \mathcal{D} \bar{\pi} \mathcal{D} \bar{\Phi} | \Psi^2 |}\,,
\end{align}
where the integration over $\bar{\Phi}$ is shorthand for the integration over all dynamical modes of the spin-$1$ field. The tree-level expression is then
\begin{align}
B_4 = & \frac{\rho_4^{\pi \pi \pi \pi}(\bfk_1,\bfk_2,\bfk_3,\bfk_4)}{\Pi_{a=1}^4 \rho_2^{\pi \pi}(k_a)} \nonumber \\& +\frac{\rho_{3,0}^{\pi\pi\Phi}(\bfk_1, \bfk_2, -\mathbf{s})\rho_{3,0}^{\pi \pi \Phi}(\bfk_3, \bfk_4, \mathbf{s})}{\Pi_{a=1}^4 \rho_2^{\pi \pi}(k_a) \rho_{2,0}^{\Phi \Phi}(s)}+\sum_{h=\pm1}\frac{\rho_{3,h}^{\pi\pi\Phi}(\bfk_1, \bfk_2, -\mathbf{s})\rho^{\pi \pi \Phi}_{3,h}(\bfk_3, \bfk_4, \mathbf{s})}{\Pi_{a=1}^4 \rho_2^{\pi \pi}(k_a) \rho_{2,\pm 1}^{\Phi \Phi}(s)} +t+u \,.
\end{align}
We will now specialise to the example we considered in Section \ref{Spin1Section} with the interactions given by \eqref{ExampleInteractionsSpinone}. First consider the exchange of the $h = \pm 1$ with $\psi_4$ given by \eqref{Psi4TransverseExchange}. To go from this expression to $\rho_4^{\pi \pi \pi \pi}$ we simply need to take the real part and double the result since we are working with the parity-even sector: $\rho_4^{\pi \pi \pi \pi } = 2 \text{Re} \psi_4^{\pi \pi \pi \pi}$. To complete the computation we now need expressions for $\rho^{\pi \pi \Phi}_{3,h}$, $\rho^{\Phi \Phi}_{2, \pm 1}$ and $\rho^{\pi \pi}_2$.
We have
\begin{align}
\psi_{3,\pm 1}^{\pi \pi \Phi}(\bfk_1, \bfk_2, -\mathbf{s}) = -\lambda_2 \int d \eta a(\eta) K_{1,1}^{\pm 1}(\eta,s)[K'_{\pi}(\eta, k_1)&K_{\pi}(\eta, k_2) k^i_2 {e}^{(\pm 1)}_i(-{\bf s}) \nonumber \\
&+K'_{\pi}(\eta, k_2)K_{\pi}(\eta, k_1) k^i_1{e}^{(\pm 1)}_i(-{\bf s})]\,,
\end{align}
which we can read off from the action. In this case $\rho_{3,h}^{\pi \pi \Phi}$ is not simply given by the real part of the wavefunction coefficient but in the resulting expression the polarisation vector factorises out since ${\bf e}^{(\pm 1)}({\bf s})=[{\bf e}^{(\pm 1)}(-{\bf s})]^*$. The expressions for the $\rho_2$ are
\begin{align}
\rho_2^{\pi \pi}= [\pi(\eta_0, k) \pi^{*}(\eta_0,k)]^{-1} \,, \\
\rho_{2,\pm1}^{\Phi \Phi}= [\Phi_{1,1}^{\pm 1}(\eta_0, k) \Phi^{\pm 1*}_{1,1}(\eta_0,k)]^{-1} \,.
\end{align}
To arrive at these expressions we compute the on-shell action, take the real part, and use the Wronskians. Putting everything together, we find
\begin{align}
&B_4 \supset \nonumber \\
&\lambda_2^2 \sum_{h=\pm 1} \sum_{a,b=\pm}\int d\eta d\eta' a(\eta)\left [k^i_2{\bf e}^{(h)}_i(-{\bf s}) G'^{\pi}_{a}(\eta,k_1) G_{a}^{\pi}(\eta,k_2) + k^i_1{\bf e}^{(h)}_i(-{\bf s}) G_{a}^{\pi}(\eta,k_1) G'^{\pi}_{a}(\eta,k_2)\right ] \nonumber \\
&\left (G^{h}_{1,1\to 1,1}\right )_{ab,\text{eff}}(\eta,\eta',k) a(\eta') \left [k^i_4 {\bf e}^{(h)}_i({\bf s}) G'^{\pi}_{b}(\eta',k_3) G_{b}^{\pi}(\eta',k_4) + k^i_3{\bf e}^{(h)}_i({\bf s}) G_{b}^{\pi}(\eta',k_3) G'^{\pi}_{b}(\eta',k_4)\right ] \nonumber \\
& + t,u\text{-channels}\,,
\end{align}
where the propagators in this expression are the usual Schwinger-Keldysh ones for the transverse modes
\begin{align}
    \left (G^{\pm 1}_{1,1\to 1,1}\right )_{++,\text{eff}} (\eta,\eta',k) & = \Phi^{\pm 1}_{1,1}(\eta,k) \Phi^{\pm 1 ^*}_{1,1} (\eta',k) \theta (\eta - \eta') + \Phi^{\pm 1 ^*}_{1,1}(\eta,k) \Phi^{\pm 1}_{1,1}(\eta',k) \theta (\eta' - \eta)\,,  \\
    \left (G^{\pm 1}_{1,1 \to 1,1}\right )_{+-,\text{eff}} (\eta,\eta',k) & = \Phi^{\pm 1^*}_{1,1}(\eta,k) \Phi^{\pm 1}_{1,1}(\eta',k)\,, \\
    \left (G^{\pm 1}_{1,1 \to 1,1}\right )_{-+,\text{eff}} (\eta,\eta',k) & = \Phi^{\pm 1}_{1,1}(\eta,k) \Phi^{\pm 1^*}_{1,1}(\eta',k)\,, \\
    \left (G^{\pm 1}_{1,1\to 1,1}\right )_{--,\text{eff}} (\eta,\eta',k) & = \Phi^{\pm 1 ^*}_{1,1}(\eta,k) \Phi^{\pm 1 }_{1,1} (\eta',k) \theta (\eta - \eta') + \Phi^{\pm 1}_{1,1}(\eta,k) \Phi^{\pm 1 ^*}_{1,1}(\eta',k) \theta (\eta' - \eta)\,.
\end{align}
The inflaton propagators are given by \eqref{InflatonSK1} and \eqref{InflatonSK2}. 
Moving to the $h=0$ exchanges we have three different possibilities for each $\mathcal{O}(\lambda^2)$. For $\mathcal{O}(\lambda_1^2)$ the quartic wavefunction coefficient is given by \eqref{QuarticWavefunctionLambda1} and the relevant cubic wavefunction coefficient is given by
\begin{align}
\psi_{3,0}^{\pi\pi\Phi}(\bfk_1, \bfk_2, -\mathbf{s}) = 2 i \lambda_1 \frac{\Phi_{0,1}^{0*}(\eta_0,s)}{\Phi_{1,1}^{0*}(\eta_0,s)} \int d \eta a(\eta) K'_{\pi}(\eta, k_1) K'_{\pi}(\eta, k_1) K_{0,1}^{0}(\eta, s)\,,
\end{align}
where we have used the solution for $\Phi_{0,1}^0$ given by \eqref{sigma001solutiongeneralinteraction}. We also have 
\begin{align}
\rho_{2,0}^{\Phi \Phi} = (\Phi_{1,1}^{0}(\eta_0, k)\Phi_{1,1}^{0*}(\eta_0, k))^{-1} \,.
\end{align}
It follows that 
\begin{align} 
B_4& \supset -4\lambda_1^2 \sum_{a,b = \pm } ab \int d\eta d\eta' a(\eta) G'^{\pi}_a(\eta,\bfk_1)G'^{\pi}_a(\eta,\bfk_2) \left (G^0_{0,1\to 0,1}\right )_{ab,\text{eff}}(\eta,\eta',s) a(\eta') G'^{\pi}_b(\eta',\bfk_3) G'^{\pi}_b(\eta',\bfk_4)\nonumber \\
& \quad + t,u\text{-channels}\,,
\end{align}
where the propagators are 
\begin{align}
    \left (G^0_{0,1\to 0,1}\right )_{++,\text{eff}} (\eta,\eta',k) & = \Phi^0_{0,1}(\eta,k) \Phi^{0^*}_{0,1} (\eta',k) \theta (\eta - \eta') + \Phi^{0^*}_{0,1}(\eta,k) \Phi^0_{0,1}(\eta',k) \theta (\eta' - \eta) \nonumber \\
    & \quad + \Delta G^0_{0,1\to 0,1}(\eta,\eta',k)\,, \\
    \left (G^0_{0,1 \to 0,1}\right )_{+-,\text{eff}} (\eta,\eta',k) & = \Phi^{0^*}_{0,1}(\eta,k) \Phi^0_{0,1}(\eta',k)\,, \\
    \left (G^0_{0,1 \to 0,1}\right )_{-+,\text{eff}} (\eta,\eta',k) & = \Phi^{0}_{0,1}(\eta,k) \Phi^{0^*}_{0,1}(\eta',k)\,, \\
    \left (G^0_{0,1\to 0,1}\right )_{--,\text{eff}} (\eta,\eta',k) & = \Phi^{0^*}_{0,1}(\eta,k) \Phi^{0}_{0,1} (\eta',k) \theta (\eta - \eta') + \Phi^{0}_{0,1}(\eta,k) \Phi^{0^*}_{0,1}(\eta',k) \theta (\eta' - \eta) \nonumber \\
    & \quad - \Delta G^0_{0,1\to 0,1}(\eta,\eta',k)\,,
\end{align}
where $\Delta G^0_{0,1 \to 0,1}$ was introduced in \eqref{extradeltaspin1}.

The most important points here are that the factorised terms in the naive bulk-bulk propagators i.e. the terms required to satisfy the future boundary condition of the bulk-bulk propagator, have now cancelled with the factorised contributions coming from the Born rule. The $\Delta G$ term, however, that contains a delta function and therefore contributes to a contact diagram, remains. In this expression for $B_4$, all dependence on the boundary data of the spinning field has dropped out so there is no distinction between light and heavy fields. The structure is very similar for the other $\mathcal{O}(\lambda^2)$ terms and for more general interactions and spin-$2$ exchanges. We provide further details in Appendix \ref{WavefunctionToCorrelatorsDetails}. In general, the effective Schwinger-Keldysh propagators needed to compute cosmological correlators are:
\begin{align}
    \left (G^h_{n,S\to m,S}\right )_{++,\text{eff}} (\eta,\eta',k) & = \Phi^h_{n,S}(\eta,k) \Phi^{h^*}_{m,S} (\eta',k) \theta (\eta - \eta') + \Phi^{h^*}_{n,S}(\eta,k) \Phi^h_{m,S}(\eta',k) \theta (\eta' - \eta) \nonumber \\
    & \quad + \Delta G^h_{n,S\to m,S}(\eta,\eta',k) \,, \\
    \left (G^h_{n,S \to m,S}\right )_{+-,\text{eff}} (\eta,\eta',k) & = \Phi^{h^*}_{n,S}(\eta,k) \Phi^h_{m,S}(\eta',k)\,, \\
    \left (G^h_{n,S \to m,S}\right )_{-+,\text{eff}} (\eta,\eta',k) & = \Phi^{h}_{n,S}(\eta,k) \Phi^{h^*}_{m,S}(\eta',k) \,, \\
    \left (G^h_{n,S\to m,S}\right )_{--,\text{eff}} (\eta,\eta',k) & = \Phi^{h^*}_{n,S}(\eta,k) \Phi^{h}_{m,S} (\eta',k) \theta (\eta - \eta') + \Phi^{h}_{n,S}(\eta,k) \Phi^{h^*}_{m,S}(\eta',k) \theta (\eta' - \eta) \nonumber \\
    & \quad - \Delta G^h_{n,S\to m,S}(\eta,\eta',k)\,,
\end{align}
which combine according to the decomposition of helicity modes in step \ref{CCstep1} to form the indexed Schwinger-Keldysh propagator
\begin{align} \label{Spin2IndexedPropagator}
G^{\Phi}_{ab,\underbrace{\scriptstyle 0...0i_1...i_n}_{S \text{ indices}}\underbrace{\scriptstyle 0...0j_1...j_m}_{S \text{ indices}}}(\eta,\eta',k) = \sum_{h=-n}^{n} \left (G^h_{n,S\to m,S}\right )_{ab,\text{eff}}(\eta,\eta',k) e^{(h)}_{i_1...i_n}(\bfk) e^{(h)}_{j_1...j_m}(-\bfk) + \text{traces}\,,
\end{align}
where $a,b\in \{+ ,-\}$. Using the indexed propagator, the Feynman rules for correlators arising from CC exchanges are summarised as follows:
\begin{mdframed}
\underline{\bf General Feynman rules for correlators with CC exchanges}
\begin{enumerate} [label=(\arabic*)]
\item Construct different diagrams with different colourings of vertices as follows:
\begin{enumerate} [label=(\roman*)]
    \item For an $n$-point correlator, draw $n$ external legs for $\pi$.
    \item Draw diagrams to connect the external legs with vertices dictated by the interaction Lagrangian $\mathcal{L}_{\text{int}}$, same as normal Feynman rules.
    \item Each vertex can be coloured black 
    or white, so for a diagram with $V$ vertices, there are $2^V$ possible colourings.
\end{enumerate}
\item Compute the contributions of these diagrams as follows:
\begin{enumerate} [label=(\roman*)]
\item Associate each vertex with a time $\eta_1,...,\eta_V$.
\item The external legs are assigned a propagator $G^{\pi}_{\pm}(\eta_m,k)$, where $\pm$ is chosen based on whether it is connected to a black ($+$) or a white ($-$) vertex, and $\eta_m$ is the time associated to that vertex.
    \item Any line that connects two vertices is assigned a propagator $G^{\Phi}_{\pm \pm,\mu_1...\mu_S\nu_1...\nu_S}(\eta_m,\eta_n,\bfk)$. There are 4  propagators per line ($++$, $+-$, $-+$, and $--$) depending on which kinds of vertices it is connected to. $\bfk$ is the momentum that flows from vertex $\eta_m$ to vertex $\eta_n$. The components of this propagator are the effective Schwinger-Keldysh propagators we derived above.
    \item Assign appropriate factors for each vertex. This includes a factor of $\pm i $ for each $(\pm)$ vertex, spatial derivatives (with spatial indices appropriately contracted with the propagators) and temporal derivatives on $\pi$ and $\Phi$; as well as symmetry factors.
    \item Integrate over times $\eta_1,...,\eta_V$, sum all the contributions from all $2^V$ colourings, and add up all the permutations of $\bfk_1,...,\bfk_n$.
\end{enumerate}
\end{enumerate}
\end{mdframed}
These rules are the same as those derived in \cite{Chen:2017ryl} but now with effective propagators that can potentially include delta functions and derivatives thereof, and can connect vertices with different modes (but with the same helicity). 

\subsection{Properties of Wick-rotated propagators}
The main focus of \cite{Stefanyszyn:2023qov} was to determine when wavefunction coefficients are purely real functions of the momenta. In general, they are complex functions but the real and imaginary parts ultimately give rise to distinct cosmological correlators. Indeed, for massless scalars and gravitons, it is the real part that contributes to parity-even correlators, and the imaginary part that contributes to parity-odd correlators. It was shown in \cite{Stefanyszyn:2023qov} that under some mild assumptions, wavefunction coefficients with external massless scalars or gravitons, possibly exchanging additional degrees of freedom, are purely real thereby making the existence of non-zero parity-odd correlators somewhat more non-trivial than their parity-even counterparts \cite{Liu:2019fag,Cabass:2022rhr,Stefanyszyn:2025yhq,Stefanyszyn:2024msm,Creque-Sarbinowski:2023wmb,Orlando:2025fec,Lee:2023jby,Thavanesan:2025kyc,Goodhew:2024eup}. The technique employed in \cite{Stefanyszyn:2023qov} was to make use of Wick rotations of the integrated time variables to prove reality. This relied on various properties of the Wick-rotated bulk-bulk propagators. For the CCM scenario, it was shown that for light fields the \textit{indexed} bulk-bulk propagator i.e. the propagator with the polarisation tensors included and helicities summed, is purely real after we rotate both time variables by $90^{\circ}$ in the complex plane. The situation is slightly more complicated for heavy fields but one can always make the time-ordered parts of the propagator purely real after a rotation with any imaginary components factorised i.e. without any $\theta$-functions. A similar story applies to the CC scenario, the indexed bulk-bulk propagator is purely real for light fields, while for heavy fields only the time-ordered parts are purely real. However, when considering the CC case, \cite{Stefanyszyn:2023qov} did not consider the mixed propagators that we have seen in this work and did not consider the corrections $\Delta G$ that contain delta functions. In this subsection we will therefore complete the analysis by considering the more general CC propagators. Ultimately, all of the reality properties of Wick-rotated bulk-bulk propagators found in \cite{Stefanyszyn:2023qov} hold for the more general propagators we have found in this work. This is to be expected since all of these modes form a single multiplet and it ensures that the wavefunction reality theorems derived in \cite{Stefanyszyn:2023qov} hold more generally.  

To be concrete, let us first work with the spin-$1$ CC set-up. The bulk-bulk propagator of the transverse modes is the same as that considered in \cite{Stefanyszyn:2023qov} i.e. it does not have a delta function correction, so we will focus on the $h=0$ modes with the indexed propagators given in \eqref{Spin1Indexed1} - \eqref{Spin1Indexed4}. The propagator for the $\Phi_{1,1}^{0}$ mode, contained as the $h=0$ part of \eqref{Spin1Indexed4}, is also as in \cite{Stefanyszyn:2023qov} so we won't consider it. Any reality properties of \eqref{Spin1Indexed3} will follow from \eqref{Spin1Indexed2}, so we focus on \eqref{Spin1Indexed1} and \eqref{Spin1Indexed2} only. Let's start with the propagator of $\Phi_{0,1}^0$ i.e. \eqref{Spin1Indexed1} whose explicit form is given by \eqref{Gspin1effective}. The only difference between this propagator and the one considered in \cite{Stefanyszyn:2023qov} is the delta function correction $i \delta(\eta-\eta')/(m^2 a^2)$. After a Wick rotation of $\eta$ and $\eta'$ by $90^{\circ}$ this correction becomes a purely real function of the rotated variables and therefore the reality properties of $G_{0,1}^{0}$ extend to the full \eqref{Spin1Indexed1}. The more non-trivial case is the mixed propagator \eqref{Spin1Indexed2}. If we strip-off the factor of $i \hat{k}$ i.e the polarisation factor, the helical propagator is simply a differential operator acting on the helical propagator $G_{1,1}^0$ c.f. \eqref{Gspin1mixed}. We see that this differential operator is first-order in time and therefore becomes imaginary after we rotate $\eta$ by $90^{\circ}$. The reality properties of $G_{1,1}^0$ therefore do not extend to the helical part of \eqref{Spin1Indexed2}. However, this is where the polarisation factor $i \hat{k}$ comes to the rescue since it is purely imaginary. So, the indexed propagator \eqref{Spin1Indexed2} does indeed inherit all of the reality properties of the $h=0$ part of \eqref{Spin1Indexed4}. For the spin-$1$ case, the reality properties discussed in \cite{Stefanyszyn:2023qov} therefore apply to our more general propagators. For the same reason, our more general Schwinger-Keldysh propagators, i.e. including any $\Delta G$ parts and the mixed terms, also satisfy all of the reality properties of the Schwinger-Keldysh propagators derived in \cite{Stefanyszyn:2023qov}.   

The spin-$2$ case works similarly and we will spare the reader the full details. The upshot is that the delta function terms in $\Delta G$ are either purely real or purely imaginary after the rotation. When they are purely real, the polarisation factor also is, and when they are purely imaginary, the polarisation factor also is. So these parts of our more general propagators satisfy all of the reality properties derived in \cite{Stefanyszyn:2023qov}. The rest of the propagators work in precisely the same way. The helical parts correspond to differential operators acting on propagators that do satisfy the reality properties e.g $G_{2,2}^h$ for $h=0, \pm 1, \pm 2$. When this operator is real after the rotation, the polarisation factor is too and when this operator is imaginary, the polarisation factor is too. We note that to see these properties of the polarisation factors we need to sum over helicities. We refer the reader to \cite{Stefanyszyn:2023qov} for details.       

\section{Correlator comparison} \label{CorrelatorComparison}
Although the free theories of the massive spinning fields in the two models adhere to different linearly-realised symmetries (the two-point function in the CC case is invariant under the full de Sitter symmetries, while for the CCM case it is only invariant under rotations), their interactions with the Goldstone boson $\pi$ ultimately have identical symmetries since boosts are spontaneously broken by the inflaton's vev. We are therefore naturally led to the question: if the production and decay of such massive spinning fields gives rise to correlation functions of $\pi$, do they yield distinct signatures? i.e. can we in principle use observations of the CMB and LSS to distinguish between these two different set-ups? This is a timely question to ask given that the signatures of new degrees of freedom are currently being looked for in the data \cite{Cabass:2024wob,Sohn:2024xzd}. 

An obvious obstacle to the possibility of the two set-ups yielding identical correlators is that the CC modes propagate with unit sound speed, whereas the CCM modes propagate according to a free parameter $c_{h,S}$ and different helicities can propagate with different speeds. Furthermore, the mass of the CC modes is restricted by the Higuchi bound \cite{Higuchi:1986py} whereas the CCM modes are not \cite{Bordin:2018pca}. The question we are therefore really asking is if the correlators produced in the CC set-up can be mimicked by those produced in the CCM scenario when we fix $c_{h,S} = 1$ (i.e. $\delta c=0$, $c=1$, c.f. \eqref{CCMspeeds}) and restrict the CCM masses. We answer this question in this section and find the following:
\begin{itemize}
    \item At the level of $\pi$ correlators, the CC correlators can be mimicked by sums of local CCM ones, up to local contact contributions from $\pi$ self-interactions, if \textit{only a single helicity contributes}. This is the case for the bispectrum where in both the CC and CCM cases \textit{only} the exchange of $h=0$ modes are relevant at tree level \cite{Pimentel:2022fsc}. It can also be the case for higher-point correlators, but unlike for the bispectrum it would depend on the type of diagram. This statement holds for generic $SO(3)$ invariant theories, and continues to hold when we impose the EFToI symmetries (as we show in Section \ref{CCvCCMEFToI}). 
    \item If a correlator is produced by the exchange of \textit{multiple helicities}, CC and CCM exchanges yield distinct signatures in $\pi$ correlators that cannot be made to be degenerate by local interactions. This is the case for e.g. the trispectrum and indeed any other higher-point correlator.
\end{itemize}
Processes that are subject to single-helicity exchanges only are those for which the propagator of the spinning field is connected to at least one \textit{linear-mixing vertex} i.e. the spinning propagator must appear in the form:
 \begin{align} 
 \begin{gathered}
  \begin{tikzpicture}
  \begin{feynman}
    \vertex[blob] (a) at (0,0){\(A\)};
    \vertex[dot] (b) at (2,0){};
    \vertex[blob] (c) at (4,0){\(B\)};
       \diagram* {
      (a) -- [boson, edge label = \(\Phi\)] (b) -- [plain, edge label = \(\pi\)] (c),
    };
  \end{feynman}
\end{tikzpicture}
\end{gathered} \,. \label{singlehelicitybuildingblock}
\end{align}
There are no restrictions on the vertices $A$ and $B$ and no restriction on how the propagators of $\pi$ enter the diagram. The linear mixing isolates the $h=0$ mode since rotations forbids different helicities from coupling at quadratic order.  

\subsection{Exchanges of spin-$1$ fields} \label{only1helicity}
We start with spin-$1$ exchanges and first relate the mode functions in the two cases. Recalling that $\Phi$ labels the CC field and $\sigma$ labels the CCM one, we use the CC mode functions \eqref{phi011freesolution}, \eqref{phi111freesolution}, and \eqref{phi001freesolution} and the parity-even CCM mode function \eqref{sigmahsCCMfreesolution} with $c_{h,S} = 1$ and $m_{\text{CCM}} = \sqrt{m_{\text{CC}}^2 +2H^2}$ to write\footnote{With $c_{h,S}=1$ we have $\sigma_{0,1} = \sigma_{1,1}$ so we can write all relations in terms of $\sigma_{0,1}$ only.}
\begin{align}
\Phi^0_{0,1} (\eta,k)& = \frac k{m_{\text{CC}}} \sigma_{0,1} (\eta,k)\,, \label{phi001intermsofCCM} \\
\Phi^0_{1,1}(\eta,k)  & = -\frac 1{m_{\text{CC}}} \left (\partial_{\eta} - \frac 2{\eta} \right ) \sigma_{0,1}(\eta,k)\,, \\
\Phi^{\pm 1}_{1,1} (\eta,k) & = a (\eta) \sigma_{0,1} (\eta,k)\,.
\end{align}
Note that to arrive at this correspondence we needed to shift the mass of the CCM field and in doing so we have removed some of the complementary series mass values of the CCM field. As we explained above, we are asking if the CC scenario can be mimicked by the CCM one so it is consistent to restrict the masses on the CCM side if necessary. From these expressions we can find relations between the various propagators required to compute cosmological correlators. We find  
\begin{align}
\left (G^0_{0,1\to 0,1}\right )_{\pm \pm, \text{eff}}(\eta,\eta',k) &= \frac{k^2}{m_{\text{CC}}^2} G^{\sigma_{0,1}}_{\pm \pm}(\eta, \eta', k) \pm  \frac i{m_{\text{CC}}^2a^2} \delta (\eta - \eta')\,, \label{PhiToSigma1} \\
\left (G^0_{0,1\to 0,1}\right )_{\pm \mp, \text{eff}}(\eta,\eta',k)  &= \frac{k^2}{m_{\text{CC}}^2} G^{\sigma_{0,1}}_{\pm \mp}(\eta, \eta', k)\,. \label{PhiToSigma2}
\end{align}
\begin{align}
\left (G^0_{0,1\to 1,1}\right )_{\pm \pm, \text{eff}}(\eta,\eta',k) &= -\frac{k}{m_{\text{CC}}^2}\left(\partial_{\eta'} - \frac{2}{\eta'} \right)  G^{\sigma_{0,1}}_{\pm \pm}(\eta, \eta', k)\,, \label{PhiToSigma3} \\
\left (G^0_{0,1\to 1,1}\right )_{\pm \mp, \text{eff}}(\eta,\eta',k)  &= -\frac{k}{m_{\text{CC}}^2}\left(\partial_{\eta'} - \frac{2}{\eta'} \right) G^{\sigma_{0,1}}_{\pm \mp}(\eta, \eta', k)\,. \label{PhiToSigma4}
\end{align}
\begin{align}
\left (G^0_{1,1\to 1,1}\right )_{\pm \pm, \text{eff}}(\eta,\eta',k) &= \frac 1{m_{\text{CC}}^2} \left (\partial_{\eta} - \frac 2{\eta} \right )\left (\partial_{\eta'}-\frac 2{\eta'}\right ) G^{\sigma_{0,1}}_{\pm \pm}(\eta,\eta',k) \mp \frac i{m_{\text{CC}}^2a^2} \delta (\eta - \eta') \label{PhiToSigma5} \,,\\
\left (G^0_{1,1\to 1,1}\right )_{\pm \mp, \text{eff}}(\eta,\eta',k)  &= \frac 1{m_{\text{CC}}^2} \left (\partial_{\eta} - \frac 2{\eta} \right )\left (\partial_{\eta'}-\frac 2{\eta'}\right ) G^{\sigma_{0,1}}_{\pm \mp}(\eta,\eta',k) \,. \label{PhiToSigma6}
\end{align}
\begin{align}
\left (G^{\pm 1}_{1,1\to 1,1}\right )_{\pm \pm, \text{eff}}(\eta,\eta',k) &= a(\eta) a(\eta') G^{\sigma_{0,1}}_{\pm \pm}(\eta,\eta',k)\,, \label{PhiToSigma7} \\
\left (G^{\pm 1}_{1,1\to 1,1}\right )_{\pm \mp, \text{eff}}(\eta,\eta',k)  &= a(\eta) a(\eta') G^{\sigma_{0,1}}_{\pm \mp}(\eta,\eta',k) \label{PhiToSigma8} \,.
\end{align}
In arriving at the delta function in \eqref{PhiToSigma5}, we used the Wronskian condition \eqref{CCMWronskian}. In the following we will consider single-helicity and multi-helicity exchanges separately. 
\paragraph{Single-helicity exchange}
We start with processes that only have $h=0$ exchange i.e. where the spinning propagator enters only via \eqref{singlehelicitybuildingblock}. To illustrate the main points, we initially concentrate on the leading contribution to the bispectrum and will discuss other diagrams and processes afterwards. We start by considering the CC set-up with interactions
\begin{align}
S_{\text{int}} = \int d^3 x d \eta \left[ (f^{(1)}[\pi]+f^{(2)}[\pi])\Phi_{0} + g^{(1)}[\pi]\partial_{i} \Phi_{i} + g^{(2)}_{i}[\pi]\Phi_i \right]\,,
\end{align}
where the superscripts on $f$ and $g$ indicate the power of $\pi$ in the corresponding operator. For $g^{(1)}$ we are able to integrate by parts to always move the necessary spatial derivative to act on the spinning field, whereas for $g^{(2)}$ this is not generically possible so we have a spatial index which must ultimately correspond to a spatial derivative acting on one of the two factors of $\pi$. We derived the Feynman rules for calculating the bispectrum in such a theory in Section \ref{FromWavefunctionToCorrelatorCC} and diagrammatically we have 
 \begin{align} 
 B_3 = \sum_{\text{S/K diagrams}}   \begin{gathered}
  \begin{tikzpicture}
  \begin{feynman}
    \vertex[blob] (a) at (0,0){\(f^{(2)}\)};
    \vertex[blob] (b) at (2,0){\(f^{(1)}\)};
    \vertex (i1) at (-0.5,1) {};
    \vertex (i2) at (-0.5,-1) {};
    \vertex (i3) at (3,0) {};
       \diagram* {
      (i1) -- (a)  -- (i2),
      (b) --  (i3),
      (a) -- [double] (b),
    };
  \end{feynman}
\end{tikzpicture}
\end{gathered} +  \sum_{\text{S/K diagrams}}    \begin{gathered}
     \begin{tikzpicture}
  \begin{feynman}
    \vertex[blob] (a) at (0,0){\(g^{(2)}\)};
    \vertex[blob] (b) at (2,0){\(g^{(1)}\)};
    \vertex (i1) at (-0.5,1) {};
    \vertex (i2) at (-0.5,-1) {};
    \vertex (i3) at (3,0) {};
       \diagram* {
      (i1) -- (a)  -- (i2),
      (b) --  (i3),
      (a) -- [scalar] (b),
    };
  \end{feynman}
\end{tikzpicture}
\end{gathered} \nonumber \\ +  \sum_{\text{S/K diagrams}}    \begin{gathered}
   \begin{tikzpicture}
  \begin{feynman}
    \vertex[blob] (a) at (0,0){\(f^{(2)}\)};
    \vertex[blob] (b) at (2,0){\(g^{(1)}\)};
    \vertex (i1) at (-0.5,1) {};
    \vertex (i2) at (-0.5,-1) {};
    \vertex (i3) at (3,0) {};
    \vertex (d) at (1,0);
       \diagram* {
      (i1) -- (a)  -- (i2),
      (b) --  (i3),
      (a) -- [double] (d) -- [scalar] (b),
    };
  \end{feynman}
\end{tikzpicture}
\end{gathered}+ \sum_{\text{S/K diagrams}}    \begin{gathered}
   \begin{tikzpicture}
  \begin{feynman}
    \vertex[blob] (a) at (0,0){\(g^{(2)}\)};
    \vertex[blob] (b) at (2,0){\(f^{(1)}\)};
    \vertex (i1) at (-0.5,1) {};
    \vertex (i2) at (-0.5,-1) {};
    \vertex (i3) at (3,0) {};
    \vertex (d) at (1,0);
       \diagram* {
      (i1) -- (a)  -- (i2),
      (b) --  (i3),
      (a) -- [scalar] (d) -- [double] (b),
    };
  \end{feynman}
\end{tikzpicture}
\end{gathered} \,, \label{B3Example}
\end{align}
where $\sum_{\text{S/K diagrams}}$ indicates that we sum over $\pm$ for each vertex so each sum contains four diagrams, as in \eqref{B3ExampleCCM}. Initially, consider the first diagram in \eqref{B3Example} corresponding to the $\Phi_0$ exchange. Given the relations \eqref{PhiToSigma1} and \eqref{PhiToSigma2}, and the fact that the only difference between the Feynman rules for CC and CCM is in the propagators, we see that this contribution is equivalent to that generated by 
\begin{align} \label{CCtoCCM1}
S_{\text{int}} \supset \int d^3 x d \eta \left[ (f^{(1)}[\pi] + f^{(2)}[\pi]) \frac{(-\partial_i \sigma_i)}{m_{\text{CC}}} \right] - \frac{f^{(1)}[\pi] f^{(2)}[\pi]}{m^2_\text{CC} a^2} \,,
\end{align}
in the CCM scenario. We see that the delta function in \eqref{PhiToSigma1} necessitates a self-interaction for the inflaton for the correspondence to work. As long as the original vertices in the CC theory are local, these self-interactions will also be local. In momentum space, $(-\partial_i \sigma_i)$ generates a factor of $k$ where $k_i$ is the momentum of the internal line. The CCM exchange diagram coming from \eqref{CCtoCCM1} therefore yields a factor of $k^2$ which accounts for those factors in \eqref{PhiToSigma1} and \eqref{PhiToSigma2}. For the second diagram in \eqref{B3Example} we use the relations \eqref{PhiToSigma5} and \eqref{PhiToSigma6} and see that this contribution is equivalent to that generated by
\begin{align} \label{CCtoCCM2}
S_{\text{int}} \supset \int d^3 x d \eta \left[ g^{(1)}[\pi] (\partial_{\eta} + 2 H a) \frac{(-\partial_i \sigma_i)}{m_{\text{CC}}} + g_i^{(2)}[\pi] (\partial_{\eta} + 2 H a) \frac{(-\sigma_i) }{m_{\text{CC}}} \right] + \frac{\partial_i g^{(1)}[\pi]g_{i}^{(2)}[\pi]}{m^2_{\text{CC}} a^2}  \,,
\end{align}
in the CCM scenario. Again, a cubic self-interaction is required to make the match and the original locality is inherited. 

At this stage we have matched the $\mathcal{O}(f^2)$ and $\mathcal{O}(g^2)$ contributions to the bispectrum and we are now left with the $\mathcal{O}(fg)$ contributions corresponding to the final two diagrams in \eqref{B3Example}. It is straightforward to see that the interactions we have been required to introduce in \eqref{CCtoCCM1} and \eqref{CCtoCCM2} are exactly the interactions required to match the $\mathcal{O}(fg)$ contributions to the bispectrum. We therefore conclude that the bispectrum generated in the CC scenario can be mimicked by the CCM scenario with the addition of some contact contributions. If the interactions on the CC side are local, those on the CCM side will be too. In summary, we have 
\begin{keyeqn}
\begin{align} \label{CCtoCCMGeneralB3}
\begin{gathered}
        \begin{tikzpicture}
  \begin{feynman}
    \vertex[dot] (a) at (0,0){};
    \vertex[dot] (b) at (2,0){};
    \vertex (i1) at (-0.5,1) {};
    \vertex (i2) at (-0.5,-1) {};
    \vertex (i3) at (3,0) {};
    \vertex (c) at (1,-0.3) {\(\text{CC}\)};   
       \diagram* {
      (i1) -- (a) [blob] -- (i2),
      (b) --  (i3),
      (a) -- [boson] (b),
    };
  \end{feynman}
\end{tikzpicture}
\end{gathered}
= \begin{gathered}
\begin{tikzpicture}
  \begin{feynman}
    \vertex[dot] (a) at (0,0){};
    \vertex[dot] (b) at (2,0){};
    \vertex (i1) at (-0.5,1) {};
    \vertex (i2) at (-0.5,-1) {};
    \vertex (i3) at (3,0) {};
    \vertex (c) at (1,-0.3){\(\text{CCM}\)};
       \diagram* {
      (i1) -- (a) [blob] -- (i2),
      (b) --  (i3),
      (a) -- [boson] (b),
    };
  \end{feynman}
\end{tikzpicture}
    \end{gathered}
    +  \begin{gathered}
    \begin{tikzpicture}
  \begin{feynman}
    \vertex[dot] (a) at (0,0){};;
    \vertex (i1) at (-0.5,1) {};
    \vertex (i2) at (-0.5,-1) {};
    \vertex (i3) at (1,0) {};
       \diagram* {
      (i1) -- (a) -- (i3),
      (a) --  (i2),
    };
  \end{feynman}
\end{tikzpicture}
\end{gathered}\,.
\end{align}
\end{keyeqn}
It is perhaps surprising that the vertices required on the RHS of \eqref{CCtoCCMGeneralB3} are local given that the polarisation vector for $\Phi_{1,1}^0$ is $i \hat{k}_i$. Indeed, such a momentum-dependence would come from $\frac{\partial_i}{\sqrt{- \partial^2}}$ when acting on a scalar. For the $\Phi_0$ exchanges we need to account for the factor of $k^2$ in \eqref{PhiToSigma1} and \eqref{PhiToSigma2} which can be achieved by a factor of $\partial_i \sigma_i$ in each vertex connected by the spinning field. The contact diagram arising from the delta function in \eqref{PhiToSigma1} is guaranteed to be local if the interactions on the LHS of \eqref{CCtoCCMGeneralB3} are local since there is no extra $k$-dependence to worry about. For the mixed exchanges there are no delta functions so no contact digrams. For the exchange we have a single factor of $k$ in both of \eqref{PhiToSigma3} and \eqref{PhiToSigma4} which can be accounted for by replacing the $\Phi_0$ factor with $\partial_i \sigma_i$ and it doesn't matter if $\Phi_0$ enters via the linear-mixing or the cubic vertex since by momentum conservation it will always carry the same momentum. Finally, for the $\Phi_{1,1}^0$ exchange there are no additional factors of $k$ to worry about in the exchange part. For the contact diagram that arises from the delta function in \eqref{PhiToSigma5} there are no extra factors of $k$ but the contact diagram would still inherit the polarisation vectors. In this case the linear-mixing must come with a factor of $\partial_i \Phi_i$ by rotations. If we denote the momentum of $\Phi_i$ as $k_i$, this contraction will yield a factor of $k$ (ignoring any minus signs). The polarisation vector from the cubic vertex is also $\hat{k}_i$ by momentum conservation, so we will always have a factor of $k \hat{k}_i = k_i$ in the bispectrum. The inverse powers of $k$ therefore always cancel in the final form of the diagram and the self-interactions therefore require $\partial_i$ rather than $\frac{\partial_i}{\sqrt{- \partial^2}}$. In arriving at this conclusion it was imperative that one of the vertices was forced to contain a factor of $\partial_i \Phi_i$. This would not be the case for the trispectrum and other higher-point correlators so in that sense the bispectrum is special. 

As long as the spinning propagator enters as in \eqref{singlehelicitybuildingblock}, this argument extends to any other tree-level diagram: the CC correlators can be mimicked by sums of \textit{local} CCM ones. As above, the delta functions will necessitate different topologies for the diagrams on the CCM side with fewer internal lines. For example, consider the bispectrum generated when there is a cubic self-interaction for the spinning field in addition to the linear-mixing. In this case the correspondence reads
 \begin{align} 
 \underbrace{
\begin{gathered}
  \begin{tikzpicture}
  \begin{feynman}
    \vertex[dot] (a) at (0,0){};
    \vertex[dot] (b) at (-1,0){};
    \vertex (i1) at (1,0){};
    \vertex [dot] (c) at (-1.5,0.866025403784){};
    \vertex (i2) at (-2,1.73205080757){};
    \vertex [dot](d) at (-1.5,-0.866025403784){};
     \vertex (i3) at (-2,-1.73205080757){};
       \diagram* {
      (i1) -- [plain] (a) -- [boson] (b) -- [boson] (c) -- [plain] (i2),
      (b) -- [boson] (d) -- [plain] (i3),
    };
  \end{feynman}
\end{tikzpicture}
\end{gathered}}_{\text{CC}}=  \underbrace{ \begin{gathered}
     \begin{tikzpicture}
  \begin{feynman}
     \vertex[dot] (a) at (0,0){};
    \vertex[dot] (b) at (-1,0){};
    \vertex (i1) at (1,0){};
    \vertex [dot] (c) at (-1.5,0.866025403784){};
    \vertex (i2) at (-2,1.73205080757){};
    \vertex [dot](d) at (-1.5,-0.866025403784){};
     \vertex (i3) at (-2,-1.73205080757){};
       \diagram* {
      (i1) -- [plain] (a) -- [boson] (b) -- [boson] (c) -- [plain] (i2),
      (b) -- [boson] (d) -- [plain] (i3),
    };
  \end{feynman}
\end{tikzpicture}
\end{gathered} +  3\begin{gathered}
   \begin{tikzpicture}
  \begin{feynman}
     \vertex (a) at (0,0){};
    \vertex[dot] (b) at (-1,0){};
    \vertex [dot] (c) at (-1.5,0.866025403784){};
    \vertex (i2) at (-2,1.73205080757){};
    \vertex [dot](d) at (-1.5,-0.866025403784){};
     \vertex (i3) at (-2,-1.73205080757){};
       \diagram* {
     (a) -- [plain] (b) -- [boson] (c) -- [plain] (i2),
      (b) -- [boson] (d) -- [plain] (i3),
    };
  \end{feynman}
\end{tikzpicture}
\end{gathered} +  3 \begin{gathered}
   \begin{tikzpicture}
  \begin{feynman}
    \vertex[dot] (a) at (0,0){};
    \vertex[dot] (b) at (-1,0){};
    \vertex (i1) at (1,0){};
    \vertex (c) at (-1.5,0.866025403784){};
    \vertex(d) at (-1.5,-0.866025403784){};
       \diagram* {
      (i1) -- [plain] (a) -- [boson] (b) -- [plain] (c),
      (b) -- [plain] (d),
    };
  \end{feynman}
\end{tikzpicture}
\end{gathered}+   \begin{gathered}
   \begin{tikzpicture}
  \begin{feynman}
    \vertex (a) at (0,0){};
    \vertex[dot] (b) at (-1,0){};
    \vertex (c) at (-1.5,0.866025403784){};
    \vertex(d) at (-1.5,-0.866025403784){};
       \diagram* {
      (a) -- [plain] (b) -- [plain] (c),
      (b) -- [plain] (d),
    };
  \end{feynman}
\end{tikzpicture}
\end{gathered}}_{\text{CCM}}\,,
\end{align}
with all vertices on the RHS being local ones. Note that the delta functions effectively collapse the spinning propagators into a vertex, so with 3 spinning propagators, there are $\binom 31 = 3$ and $\binom 32 = 3$ ways to collapse those into double-exchange and single-exchange diagrams respectively. We also note that the diagrams on the CCM side that are required by the delta functions in the CC propagators, include interactions that did not contribute on the CC side. 

For the inflaton bispectrum, the exchange of spinning fields can only occur thanks to a linear-mixing vertex $\pi \Phi$ and therefore such diagrams are \textit{always} restricted to single-helicity exchange. Given our above discussion, if we are interested in the signatures of massive spinning fields in the bispectrum, we can compute such signatures exclusively in the CCM scenario and automatically capture all the signatures of the CC scenario. This is a positive result since the CCM computations are practically more straightforward since we do not need to deal with the effective and mixed propagators we derived in Section \ref{FeynmanRulesCC}. The CCM mode functions are those of a canonical massive scalar and various techniques have now been developed to deal with the corresponding computations. We expect a similar conclusion to hold for correlators that involve the graviton and we leave this for future work.

\paragraph{Multi-helicity exchange} Processes that come under the banner of multi-helicity are those whose Feynman diagrams have at least one spinning propagator that does not enter as in \eqref{singlehelicitybuildingblock}. For illustrative purposes, we focus on the trispectrum and restrict ourselves to CC interactions of the form
\begin{align}
S_{\text{int}} = \int d^3 x d \eta ~ h^{(2)}_i[\pi] \Phi_i \,.
\end{align}
The trispectrum arises from both $\Phi_{1,1}^{\pm 1}$ and $\Phi_{1,1}^0$ exchange:
\begin{equation} 
 B_4 =    \sum_{\text{S/K diagrams}}   \begin{gathered}
  \begin{tikzpicture}
  \begin{feynman}
    \vertex[blob] (a) at (0,0){\(h^{(2)}\)};
    \vertex[blob] (b) at (2,0){\(h^{(2)}\)};
    \vertex (i1) at (-0.5,1) {};
    \vertex (i2) at (-0.5,-1) {};
    \vertex (i3) at (2.5,1) {};
    \vertex (i4) at (2.5,-1) {};
       \diagram* {
      (i1) -- (a)  -- (i2),
     (i4) -- (b) --  (i3),
      (a) -- [boson] (b),
    };
  \end{feynman}
\end{tikzpicture}
\end{gathered} +  \sum_{\text{S/K diagrams}}   \begin{gathered}
  \begin{tikzpicture}
  \begin{feynman}
    \vertex[blob] (a) at (0,0){\(h^{(2)}\)};
    \vertex[blob] (b) at (2,0){\(h^{(2)}\)};
    \vertex (i1) at (-0.5,1) {};
    \vertex (i2) at (-0.5,-1) {};
    \vertex (i3) at (2.5,1) {};
    \vertex (i4) at (2.5,-1) {};
       \diagram* {
      (i1) -- (a)  -- (i2),
     (i4) -- (b) --  (i3),
      (a) -- [scalar] (b),
    };
  \end{feynman}
\end{tikzpicture}
\end{gathered} \,.
\end{equation}
We now ask if such a trispectrum can be generated in the CCM set-up. The exchange of transverse modes is simple given the relations \eqref{PhiToSigma7} and \eqref{PhiToSigma8}, we simply need to use  
\begin{align}
S_{\text{int}} = \int d^3 x d \eta ~ h^{(2)}_i[\pi] a(\eta) \sigma_i \,,
\end{align}
in the CCM scenario. For the $h=0$ mode exchange, however, to match that contribution we would need to use 
\begin{align}
S_{\text{int}} = \int d^3 x d \eta ~ h^{(2)}_i[\pi]  (\partial_{\eta} + 2 H a) \frac{(-\sigma_i) }{m_{\text{CC}}} + \text{self-interactions} \,.
\end{align}
We are therefore required to use different interactions to match the different exchanges but this doesn't seem to be possible since all of the $h = \pm 1, 0$ modes are packaged into a single field $\sigma_i$. The only way we could mimic the CC trispectrum using CCM interactions is to therefore decouple the different helicity modes and build the necessary interactions for each separately. However, this can not be done in a local manner. Indeed, in order to project $\sigma_i$ onto the transverse modes only, we need to use the combination
\begin{align}
\sigma_i - \frac{\partial_i \partial_j}{\nabla^2} \sigma_j\,,
\end{align}
which is not a local operator. We therefore conclude, in contrast to the single-helicity exchange case, that we cannot mimic CC exchanges with local CCM ones. We could in principle decide to go ahead and only consider CCM fields and use non-local interactions. However, it is not clear how one should go about building such theories. What are the allowed operators? What is the correct way to organise the EFT? Such non-localities are ubiquitous in cosmology arising from solving for the non-dynamical parts of the metric \cite{Maldacena:2002vr,Seery:2006vu,Bonifacio:2022vwa}, but there we always build the theories with the non-dynamical modes and solve the constraints after. With this logic in mind, it therefore seems more sensible to treat CC and CCM as distinct possibilities for describing massive spinning fields during inflation when our interests are in the trispectrum and beyond. 

\subsection{Exchanges of spin-$2$ fields}
We now consider the exchange of massive spin-$2$ fields. Again, we first relate the CC and CCM mode functions. The CC mode functions are given in \eqref{phi022modefunction}, \eqref{phi002modefunction}, \eqref{phi012modefunction}, \eqref{phi122modefunction}, \eqref{phi112modefunction}, and \eqref{phi222modefunction}; and the parity-even CCM mode function is given by \eqref{sigmahsCCMfreesolution} and as above we will fix $c_{h,S} = 1$ and $m_{\text{CCM}} = m_{\text{CC}}\equiv m$ (c.f. \eqref{Spin2Order} and \eqref{Spin2OrderCCM}) so that we have a hope of matching the correlators. The relations between the mode functions are then\footnote{Again we write all relations in terms of $\sigma_{0,2}$ since we impose unit speed of sound $c_{h,S}=1$ for all modes in CCM.}
\begin{align}
\Phi^0_{kk,2} (\eta,k) = \Phi^0_{0,2}(\eta,k) & = \sqrt{\frac {2\alpha_1\alpha_2}3}k^2 \sigma_{0,2}(\eta,k) \,,\label{phi002CCM}\\
\Phi^0_{1,2}(\eta,k) & = -\sqrt{\frac {2\alpha_1\alpha_2}3} k  \left (\partial_{\eta} - \frac 2{\eta} \right )  \sigma_{0,2}(\eta,k) \,, \\
\Phi^0_{2,2}(\eta,k) & = \sqrt {\frac {\alpha_1 \alpha_2}2} \left [\frac 2{\eta} \partial_{\eta} - \frac 6{\eta^2}+\frac {2k^2}3+m^2a^2(\eta)\right ] \sigma_{0,2}(\eta,k)\,,\\
\Phi^{\pm 1}_{1,2} (\eta,k) & = \frac {\sqrt {2\alpha_1} k}{2}a(\eta) \sigma_{0,2}(\eta,k) \,,\\
\Phi^{\pm 1}_{2,2}(\eta,k) & = -\sqrt{\frac {\alpha_1}2} a(\eta) \left (\partial_{\eta} - \frac 3{\eta}\right ) \sigma_{0,2}(\eta,k)\,, \label{phi222CCMA} \\
\Phi^{\pm 2}_{2,2}(\eta,k) & = \frac{1}{\sqrt{2}}a^2(\eta) \sigma_{0,2}(\eta,k)\,, \label{phi222CCM}
\end{align}
where $\alpha_1$, $\alpha_2$ are defined in \eqref{alphaHiguchi}. For notational convenience, we denote the operators that convert the CCM mode functions to CC ones as $\mathcal{D}^h_{m,2}$ so that \eqref{phi002CCM}-\eqref{phi222CCM} are written as
\begin{align}
\Phi^h_{m,2}(\eta) = \mathcal{D}^h_{m,2}(\eta,k) \sigma_{0,2}(\eta,k)\,. \label{Dhm2notation}
\end{align}
For example,
\begin{align}
    \mathcal{D}^0_{2,2} (\eta) & = \sqrt{\frac {\alpha_1\alpha_2}2} \left [\frac 2{\eta} \partial_{\eta} - \frac 6{\eta^2}+\frac {2k^2}3+m^2a^2(\eta)\right ]\,.
\end{align}
To compare the $\pi$ correlators generated by coupling the inflaton to the CCM and CC fields we need to compare the Schwinger-Keldysh propagators in the two set-ups including any additional delta function terms we found in Section \ref{FeynmanRulesCC}. We have
\begin{align}
\left (G^h_{n,2\to m,2}\right )_{\pm\pm,\text{eff}}(\eta,\eta',k) & = \mathcal{D}^h_{n,2} (\eta) \mathcal{D}^h_{m,2}(\eta')G^{\sigma_{0,2}}_{\pm\pm}(\eta,\eta',k) \pm \Delta G^h_{n,2\to m,2}(\eta,\eta',k) \pm C^h_{n,2\to m,2}(\eta,\eta',k) \,,\label{spin2CCtoCCMSK++} \\
\left (G^h_{n,2\to m,2}\right )_{\pm\mp,\text{eff}}(\eta,\eta',k)& = \mathcal{D}^h_{n,2} (\eta) \mathcal{D}^h_{m,2}(\eta')G^{\sigma_{0,2}}_{\pm\mp}(\eta,\eta',k)\,.
\end{align}
Note that $\Delta G$ are the delta functions in the effective spin-2 propagators given by \eqref{Deltaphi002phi002}-\eqref{Deltaphi112phi112}, while the new contributions $C^h_{n,2\to m,2}$ are additional delta function terms that arise when we match the CC and CCM propagators. Using the Wronskian condition \eqref{CCMWronskian}, the non-zero ones are given by
\begin{align}
    C^0_{1,2\to 1,2}(\eta,\eta',k) &= -\frac {2iH^2k^2\eta^2\alpha_1\alpha_2}{3} \delta (\eta - \eta')\,, \label{C01212}\\
    C^0_{1,2\to 2,2}(\eta,\eta',k) & = \frac {2i H^2k\eta\alpha_1\alpha_2 }{\sqrt 3} \delta (\eta - \eta')\,, \\
    C^0_{2,2 \to 2,2}(\eta,\eta',k) & = -2iH^2\alpha_1\alpha_2 \delta (\eta-\eta')\,,\\
    C^{\pm 1}_{2,2\to 2,2} (\eta,\eta',k) & = -\frac {i\alpha_1}2 \delta (\eta -\eta')\,. \label{C12222}
\end{align}
All other non-zero combinations of $C$ are obtained by 
\begin{align}
    C^h_{m,2\to n,2}(\eta,\eta',k) = C^h_{n,2\to m,2}(\eta',\eta,k)\,.
\end{align}
Inspired by the spin-1 case, we expect that for spin-2 we can mimic CC correlators with {\it local} CCM ones when there is a single-helicity exchange, but when there are multiple-helicities being exchanged, we cannot establish the correspondence with local interactions. We will now show that this is indeed the case.

\paragraph{Single-helicity exchange} To establish the correspondence for single-helicity exchange, we can rewrite \eqref{phi002CCM}-\eqref{phi222CCM} by combining the helicity-0 modes with their polarisation tensors to relate the indexed CC field ($\Phi_{\mu\nu}$) to the CCM field ($\sigma_{ij})$ as follows:
\begin{align}
\Phi_{kk}(\eta,k)=\Phi_{00}(\eta,k) & = -\sqrt{\frac {\alpha_1\alpha_2}2} \partial_i\partial_j\sigma_{ij} (\eta,k)\,, \\
\Phi_{0i} (\eta,k) & \supset -\sqrt{\frac {\alpha_1\alpha_2}2} \left (\partial_{\eta} - \frac 2{\eta}\right ) \partial_j \sigma_{ij} (\eta,k)\,, \\
\widehat{\Phi}_{ij}(\eta,k) & \supset \sqrt {\frac {\alpha_1 \alpha_2}2} \left [\frac 2{\eta} \partial_{\eta} - \frac 6{\eta^2}+m^2a^2(\eta) - \frac 23 \partial_l\partial_l\right ] \sigma_{ij} (\eta,k)\,,
\end{align}
where $\widehat{\Phi}_{ij}$ denotes the traceless part of $\Phi_{ij}$ i.e. $\widehat{\Phi}_{ij} = \Phi_{ij}-\frac{1}{3} \delta_{ij} \Phi_{kk}$. For simplicity we first consider a single-exchange contribution with interactions
\begin{align} \label{CCInteractionsCorrespondence}
S_{\text{int}} & = \int d^3 x d\eta \left [f[\pi] \Phi_{00} + g_i[\pi]\Phi_{0i} + h_{ij}[\pi] \widehat{\Phi}_{ij} + t[\pi] \Phi_{kk}\right ]\,,
\end{align}
with $f,g,h$ and $t$ being non-linear in $\pi$. To match the resulting CC correlators, the delta functions in the relations between the CC and CCM propagators given above require both exchange and contact diagrams on the CCM side. To extract the necessary exchange diagrams we simply replace $\Phi_{\mu\nu}$ by their expressions in terms of $\sigma_{ij}$. This will give us the correct correlators up to contact diagrams that arise from self-interactions of the inflaton. On the CCM side we therefore have
\begin{align}
S_{\text{int}} & =  \sqrt{\frac {\alpha_1\alpha_2}2}\int d^3 x d\eta \left [-(f[\pi] + t[\pi])\partial_i\partial_j \sigma_{ij} - g_i[\pi]\left (\partial_{\eta}-\frac 2{\eta}\right ) \partial_j\sigma_{ij} \right . \nonumber \\
&\qquad \qquad \qquad \qquad \qquad \left . +\; h_{ij}[\pi] \left (\frac 2{\eta}\partial_{\eta} - \frac 6{\eta^2} + m^2a^2 - \frac 23 \partial_l\partial_l\right )\sigma_{ij}\right ] +\text{self interactions of $\pi$}\,. \label{CCMspin2correspondnoexplicitcontact}
\end{align}
Note that these interactions on the CCM side are local whenever the original ones on the CC side are. We then need to check the locality of the required contact terms, which come from the sum of $\Delta G^h_{n,2 \to m,2}$ (\eqref{Deltaphi002phi002}-\eqref{Deltaphi112phi112}) and $C^h_{n,2\to m,2}$ (\eqref{C01212}-\eqref{C12222}) multiplied by the appropriate polarisation tensors.  Explicitly, the required self-interactions are given by the following Lagrangian:
\begin{align}
\mathcal{L}_{\text{contact}} & = \frac {H^2\eta^2\alpha_1\alpha_2}{3} \left (\partial_{\eta}f[\pi] \right )^2 - \frac {\alpha_1}3 f^2[\pi] +\frac {H^2\eta^2\alpha_1\alpha_2}3 \left (\partial_if[\pi]\right )^2 -  \frac {2H^2\eta^2\alpha_1\alpha_2}3 \partial_{\eta} f[\pi]\partial_ig_i[\pi] \nonumber \\
& \quad + \frac {2H^2\eta^2\alpha_1\alpha_2}3 f[\pi]\left (\partial_i\partial_j h_{ij}[\pi] - \frac 13 \nabla^2 h_{ii}[\pi]\right ) + 2H^2\eta \alpha_1\alpha_2 \partial_{\eta}f[\pi]t[\pi] + \alpha_1 f[\pi]t[\pi] \nonumber \\
& \quad - \frac {2H^2\eta^2\alpha_1\alpha_2}3 \partial_if[\pi]\partial_it[\pi] + \frac {\alpha_1}4 \left (\frac {\partial_i}{\sqrt{-\nabla^2}}g_i[\pi]\right )^2- 2H^2 \eta \alpha_1\alpha_2 \partial_ig_i[\pi] \left (\frac {\partial_j\partial_k}{\nabla^2} h_{jk}[\pi]- \frac 13 h_{jj} [\pi]\right ) \nonumber \\
& \quad - 2H^2 \eta \alpha_1\alpha_2 \partial_i g_i[\pi] t[\pi] - 3H^2 \alpha_1 \alpha_2 \left (\frac {\partial_i\partial_j}{\nabla^2} h_{ij}[\pi] - \frac 13 h_{ii}[\pi] \right )^2 + 3H^2 \alpha_1\alpha_2 t^2[\pi]\,.
\end{align}
We therefore see that some non-local terms arise. However, they are precisely those that only involve $g_i[\pi]$ and/or $h_{ij}[\pi]$. The way we have written the interactions in \eqref{CCInteractionsCorrespondence} does not restrict us to the bispectrum only. To zoom in on the bispectrum we write $g_i[\pi] = \partial_i g[\pi] + \widetilde{g}_i[\pi]$ and $h_{ij}[\pi] = \partial_i\partial_j h[\pi] + \widetilde{h}_{ij}[\pi]$ for some $g$, $\widetilde{g}_i$, $h$, and $\widetilde{h}_{ij}$. Diagrams that contribute to the bispectrum are those those involving at least one of $\partial_i g[\pi]$ and $\partial_i\partial_j h[\pi]$ since they are required for us to have at least one linear-mixing vertex. In such cases we have
\begin{align}
\left (\frac {\partial_i}{\sqrt{-\nabla^2}}g_i[\pi]\right )^2 & \supset \left (\partial_i g[\pi]\right )^2 - 2 g[\pi] \partial_i \widetilde{g}_i[\pi] \,, \\
\partial_ig_i[\pi] \frac {\partial_j\partial_k}{\nabla^2} h_{jk}[\pi] & \supset \partial_ig_i[\pi] \nabla^2 h[\pi]\,, \\
h_{ii}[\pi] \frac {\partial_j\partial_k}{\nabla^2} h_{jk}[\pi] & \supset h_{ii}[\pi] \nabla^2 h[\pi] \,,\\
\left (\frac {\partial_i\partial_j}{\nabla^2} h_{ij}[\pi]\right )^2 & \supset \left (\nabla^2 h[\pi]\right )^2 + 2h[\pi] \partial_i\partial_j \widetilde{h}_{ij}[\pi]\,.
\end{align}
We therefore see that the self-interactions become local when we restrict ourselves to the bispectrum of $\pi$, in comparison to the spin-$1$ case.

\paragraph{Multi-helicity exchange} For multiple helicity exchanges let's first consider the relations in \eqref{phi222CCMA} and \eqref{phi222CCM} which both come from $\widehat{\Phi}_{ij}$. We see that these two different modes are mapped to $\sigma_{0,2}$ in different ways and therefore in the correspondence we would need to treat the $h = \pm 2$ modes of $\widehat{\Phi}_{ij}$ in a different way to the $h = \pm 1$ modes. As in the spin-$1$ case, we cannot disentangle these different helicity modes in a local way and therefore any CC to CCM correspondence when processes are sensitive to multiple helicities will necessarily be non-local.
\subsection{Correlator comparison in the EFT of inflation} \label{CCvCCMEFToI}
So far we have shown that the bispectrum arising from a {\it general} CC interaction with $\pi$ can be mimicked by some CCM interactions with the inflaton and self-interactions of the inflaton. However, ultimately we are interested in computing correlators from interactions that are constrained by the symmetries of the EFToI \cite{Cheung:2007st}. Here we restrict ourselves to CC exchanges consistent with the EFToI, and see if the corresponding interactions on the CCM side are also permitted by the EFToI. We would expect this to indeed be the case given that both cases will satisfy the consistency relations, however, it could be that the consistency relations on the CCM side require cancellations between exchange and contact diagrams rather than each contribution satisfying the consistency relations on its own. If a cancellation was required, then it would be very non-trivial to write down the appropriate operators. In the following we check a number of examples and we show that the couplings between the inflaton and the CCM fields, and the self-interactions of the inflaton, can be separately constructed using the building blocks of the EFToI.

Let's start by consider the bispectrum generated by the EFToI interaction:\footnote{There is an abuse of notation: on the left hand side of \eqref{CCtoCCMspin1EFTexample}, the upper 0 index refers to $t$; while on the right hand side, the upper 0 index in $\delta g^{00}$ still refers to $t$, just written as a function of $\eta$, but the lower 0 index in $\Phi_0$ refers to $\eta$.}
\begin{align}
S_{\text{int}} = \lambda \int d^3x dt \sqrt{-g} \delta g^{00} \Phi^0 = \lambda \int d^3 x d\eta \left (- a^2 \delta g^{00} (a+\pi') \Phi_0 + a^2 \delta g^{00} \partial_i\pi \Phi_i \right )\,. \label{CCtoCCMspin1EFTexample}
\end{align}
Using the above rules, the corresponding bispectrum can be mimicked by
\begin{align}
S_{\text{int}} & = \frac {\lambda}{m_{\text{CC}}} \int  d^3 x d\eta \left (-a^2 \partial_i \left (\delta g^{00} (a+\pi')\right ) \sigma_i + \left (\partial_{\eta} + \frac 2{\eta}\right )\left (a^2 \delta g^{00} \partial_i \pi \right )\sigma_i\right ) \nonumber \\
& \quad +\frac {\lambda^2}{2m_{\text{CC}}^2} \int d^3 x d\eta a^2 \left (\delta g^{00}(a+\pi')\right )^2 \\
& = \frac {\lambda}{m_{\text{CC}}} \int d^3x dt \sqrt{-g} \left (-\partial_i \delta g^{00} \Sigma^i + \dot{\pi} \partial_i \delta g^{00} \Sigma^i + \partial_t \delta g^{00} \partial_i \pi \Sigma^i\right ) \nonumber \\
& \quad +\frac {\lambda^2}{2m_{\text{CC}}^2} \int d^3 x dt \sqrt{-g} \left (\delta g^{00}\right )^2 \left (1+ 2\dot{\pi} + \dot{\pi}^2 \right ). \label{beforerepackageCCEFTcorrespondence}
\end{align}
Using $\Sigma^0 = -\partial_i\pi \Sigma^i + \mathcal{O}(\pi^2) $ and that under the Stuckelberg trick, $\delta g^{00} \mapsto -2\dot{\pi} + \mathcal{O}(\pi^2)$, we can repackage \eqref{beforerepackageCCEFTcorrespondence} into
\begin{align}
S_{\text{int}} & = -\frac {\lambda}{m_{\text{CC}}} \int d^3 x dt \sqrt{-g} \left (\nabla_{\mu} \delta g^{00} \Sigma^{\mu} - \frac 14 \nabla_{\mu} \left (\delta g^{00}\right )^2 \Sigma^{\mu} + \mathcal{O}(\pi^3 \Sigma)\right ) \nonumber \\
& \quad + \frac {\lambda^2}{2m_{\text{CC}}^2} \int d^3 x dt \sqrt{-g} \left [\left (\delta g^{00}\right )^2 - \left (\delta g^{00}\right )^3 + \mathcal{O}(\pi^4)\right ],
\end{align}
which are all EFToI building blocks (up to higher-order terms that don't contribute to the bispectrum). In the table below we present further examples for both spin-$1$ and spin-$2$, where we denoted $g_{\alpha} = \delta g^{00} - \alpha \left (\delta g^{00} \right )^2$ for simplicity and $\delta K_{\mu\nu}$ is the extrinsic curvature. We see that in each case the correspondence is consistent with the EFToI, although the correspondence becomes rather complicated. 
\SetTblrInner{rowsep=4pt}
\begin{table}[H]
\centering
\begin{tblr}{ c | c }
CC EFToI block & CCM EFToI blocks  \\ \hline
$\lambda_1 \delta g^{00} \Phi^0$ & $\displaystyle -\frac {\lambda_1}{m_{\text{CC}}} \nabla_{\mu} g_{\frac 14} \Sigma^{\mu} + \frac {\lambda_1^2}{2m_{\text{CC}}^2}g_0g_1$ \\ \hline
$\displaystyle \lambda_2 \left (\delta g^{00}\right )^2 \Phi^0$ & $\displaystyle -\frac {\lambda_2}{m_{\text{CC}}} \nabla_{\mu}g_0^2 \Sigma^{\mu}$ \\ \hline 
\SetCell[r=2]{c}{$\displaystyle \lambda_3 \delta g^{00} \Phi^{00}$} & $\displaystyle \sqrt{\frac {\alpha_1\alpha_2}2} \lambda_3 \left [-\nabla_{\mu} \nabla_{\nu}g_{\frac 12}+2(n^{\lambda}\nabla_{\lambda}+H) g_0 \delta K_{\mu\nu} \right ]\Sigma^{\mu\nu} -\frac {\alpha_1\lambda_3^2}3 g_1^2$ \\ 
& $\displaystyle  +\frac {\alpha_1\alpha_2\lambda_3^2}3 \left [\nabla_{\mu} g_1 \nabla^{\mu} g_1 + 2 (n^{\lambda}\nabla_{\lambda}g_1)(n^{\sigma}\nabla_{\sigma} g_1) - 4 g_1 (n^{\lambda}\nabla_{\lambda}+H) g_1 \delta K \right ]$\\ \hline
$\displaystyle \lambda_4 \left (\delta g^{00}\right )^2 \Phi^{00}$ & $\displaystyle -\sqrt{\frac {\alpha_1\alpha_2}2} \lambda_4 \nabla_{\mu} \nabla_{\nu}g_0^2 \Sigma^{\mu\nu}$
\end{tblr}
\caption{The correspondence between the CC EFToI blocks with the CCM EFToI blocks for the bispectrum.}
\label{CCtoCCMtable}
\end{table}
\noindent

\subsection{Delta functions in propagators}\label{DeltaSection}
In both the spin-$1$ and spin-$2$ cases, we have seen that some of the bulk-bulk propagators, and Schwinger-Keldysh propagators, contain delta functions corresponding to instantaneous propagation. There are two questions one might naturally ask at this point:
\begin{enumerate}
    \item Why would \textit{any} propagator have a delta function term?
    \item Why do such delta functions not appear in the flat-space theory and break Lorentz invariance?
\end{enumerate}
We answer these in turn. 
\begin{enumerate}
    \item Delta functions corresponding to instantaneous propagation are ubiquitous in cosmology and General Relativity where we often encounter non-dynamical modes. Indeed, in the ADM formalism, the lapse and shift are non-dynamical modes that are solved for in perturbation theory with their solutions plugged back into the action to yield an action for the dynamical modes only. This procedure was worked through for the $\zeta$ bispectrum in \cite{Maldacena:2002vr,Chen:2006nt}, the parity-even $\zeta$ trispectrum in \cite{Seery:2006vu}, the parity-odd $\zeta$ trispectrum in \cite{Stefanyszyn:2025yhq,Orlando:2025fec} and the graviton trispectrum in \cite{Bonifacio:2022vwa}. As a simple toy model for what happens in these cases we refer the reader to Appendix B of \cite{Pajer:2020wxk}. The upshot is that the non-dynamical modes can be integrated out as explicitly done in the above works, or retained in the action and used in Feynman diagrams but with propagators that are given by delta functions. In the above examples, the lapse and shift can be solved for in terms of the external states only:
    \begin{align}
    \text{non-dynamical modes} = f(\text{external states})\,.
    \end{align}
    This is the primary reason why their propagators are given purely by delta functions and not with additional dynamical pieces. This is in contrast to what we have seen here. For example, when we integrate out the temporal mode of the massive spin-$1$ field, its solution depends on the external states of interest, in our case the inflaton $\pi$, but also depends on the longitudinal part of the spatial components of the spin-$1$ field which is a dynamical mode that appears in internal lines c.f. \eqref{spin1generalsigma001solution}. The same is true for spin-$2$. We therefore have 
    \begin{align}
    \text{non-dynamical modes} = f(\text{internal states, external states})\,,
    \end{align}
    and therefore we encounter more non-trivial propagators for the non-dynamical modes as in e.g. \eqref{Gspin1effective}.
    \item  The most relevant source to answer this question for a massive spin-$1$ field is provided by Weinberg \cite{Weinberg:1995mt}. In canonical quantisation, one writes the Hamiltonian in terms of $A_i$ and the conjugate momentum $\Pi_i$ with the temporal component $A_0$ solved for in terms of the dynamical variables (it does not have its own conjugate momentum). In the presence of a source $\mathcal{L} \supset J_{\mu}A^{\mu}$, the $A_0$ equation of motion is corrected by $J_0$ and so is the solution, as has been the case for us in this work. Upon plugging this solution into the Hamiltonian, a term of the form $J_0 J^0$ arises which introduces additional interactions that seem to violate Lorentz invariance. However, upon deriving the flat-space Feynman propagator $\mathcal{G}$, by computing the time-ordered two-point function from the free Hamiltonian, one sees that it is not Lorentz covariant (even after combining all modes). It is given by \cite{Weinberg:1995mt}
    \begin{align} 
    \mathcal{G}_{\mu\nu}(x, x') = i  \int \frac{d^3 k}{(2 \pi)^{3}}\frac{\eta_{\mu\nu} + \frac{k_\mu k_\nu}{m^2}}{2 k_0}\left[\theta(t - t')e^{i k^{\mu}(x_\mu - x'_{\mu})}+\theta(t' - t)e^{i k^{\mu}(x'_\mu - x_{\mu})}  \right] \,,
    \end{align}
    where $k_0$ takes its on-shell value i.e. $k_0 = \sqrt{k^2 + m^2}$ since this time-ordered two-point function comes from the on-shell Wightman functions. Using $\eta_{00}  + k_{0}k_{0}/m^2 = \frac{k^2}{m^2}$ we can write the $00$ component in terms of the scalar propagator as
    \begin{align} \label{FlatSpaceProca}
    \mathcal{G}_{00}(t, t', k) = \frac{k^2}{m^2} \mathcal{G}_{\text{scalar}}(t, t', k) \,.
    \end{align}
   This expression might not be familiar because we choose to work in the mixed domain where we only Fourier transform in the spatial directions so we can easily compare to what we have in cosmology. This formula is completely analogous to the first term on the RHS of \eqref{PhiToSigma1}. If we are performing computations using the canonical formalism we therefore have a non-covariant propagator and non-invariant interactions that ultimately combine to yield Lorentz-invariant results for e.g. amplitudes \cite{Weinberg:1995mt}. Rather than keeping the $J_0 J^0$ interactions in the Hamiltonian, we can account for them by adding a delta function to \eqref{FlatSpaceProca}. Such a delta function is then completely analogous to what we have in \eqref{Gspin1effective} and \eqref{PhiToSigma1}. 

    In the path integral approach we don't encounter these Lorentz-breaking intermediate steps. Indeed, there we read of interactions from the original Lorentz-invariant Lagrangian i.e. from $A_\mu J^\mu$ and use the Lorentz-covariant position-space propagator \cite{Weinberg:1995mt}
    \begin{align} \label{LIprocaprop}
    \widetilde{\mathcal{G}}_{\mu\nu}(x,x') = \int \frac{ d^4 k}{(2 \pi)^4} \frac{\left(\eta_{\mu\nu} + \frac{k_\mu k_\nu}{m^2}\right) e^{i k^{\mu}(x_{\mu} - x'_{\mu})}}{k^{\mu}k_{\mu} + m^2 - i \epsilon} \,.
    \end{align}
    The claim is that this propagator contains \eqref{FlatSpaceProca} in its $00$ component but also a delta function term that captures the $J_0 J^0$ contributions in the canonical approach. The $00$ component can be written as 
    \begin{align} 
    \widetilde{\mathcal{G}}_{00}(x,x') &=  \int \frac{ d^4 k}{(2 \pi)^4} \frac{\left(-1 + \frac{k_0 k_0}{m^2}\right) e^{i k^{\mu}(x_{\mu} - x'_{\mu})}}{k^{\mu}k_{\mu} + m^2 - i \epsilon} \nonumber \\
    &   = -\left(1 - \frac {\partial_{t}\partial_{t'}}{m^2}  \right) \int \frac{d^4 k}{(2 \pi)^4} \frac{e^{i k^{\mu}(x_{\mu} - x'_{\mu})}}{k^{\mu}k_{\mu} + m^2 - i \epsilon} \,.
    \end{align}
    In this expression $k_0$ is not on-shell rather it is an auxiliary variable that will take on its on-shell value once we perform the $k_0$ integral using Cauchy's theorem. Performing that integral using the standard approach of closing the contour in the upper-half complex plane for $t' > t$ and in the lower-half for $t'<t$ yields
     \begin{align} 
    \widetilde{\mathcal{G}}_{00}(x,x') =& - \left(1 - \frac {\partial_{t}\partial_{t'}}{m^2}  \right) \int \frac{d^3 k}{(2 \pi)^3}\frac{i}{2 k_0} \left[\theta(t-t')e^{i k^{\mu}(x_\mu - x'_\mu)}+\theta(t'-t)e^{-i k^{\mu}(x_\mu - x'_\mu)} \right]\,, \\ \nonumber
    =& -  \int \frac{d^3 k}{(2 \pi)^3}\frac{i}{2 k_0} \left(1 - \frac {k_0^2}{m^2}  \right) \left[\theta(t-t')e^{i k^{\mu}(x_\mu - x'_\mu)}+\theta(t'-t)e^{-i k^{\mu}(x_\mu - x'_\mu)} \right] \\ \nonumber 
     &-\frac{1}{m^2}\int \frac{d^3 k}{(2 \pi)^3} e^{i k^{i}(x_{i}-x'_{i})} \delta(t - t') \,,
    \end{align}
    and we therefore see the emergence of the expected delta function coming from hitting the $\theta$-functions with time derivatives. Since we now have $k_{0} = \sqrt{k^2 + m^2}$, we can write 
     \begin{align} 
    \widetilde{\mathcal{G}}_{00}(x,x') =\;& i \int \frac{d^3 k}{(2 \pi)^3} \frac{k^2}{m^2}  \frac{1}{2 k_0} \left[\theta(t-t')e^{i k^{\mu}(x_\mu - x'_\mu)}+\theta(t'-t)e^{-i k^{\mu}(x_\mu - x'_\mu)} \right] \nonumber \\ 
     &- \frac{1}{m^2}\int \frac{d^3 k}{(2 \pi)^3} e^{i k^{i}(x_{i}-x'_{i})} \delta(t - t') \,.
    \end{align}
    Once we convert to the mixed domain and introduce the scalar propagator, this propagator is therefore given by \eqref{FlatSpaceProca} plus the addition of a delta function term that accounts for the $J_0 J^0$ interactions in the canonical approach:
    \begin{align}
    \widetilde{\mathcal{G}}_{00}(t, t', k) = \frac{k^2}{m^2} \mathcal{G}_{\text{scalar}}(t, t', k) - \frac{1}{m^2} \delta(t - t')\,.
    \end{align}
    This confirms that while $\widetilde{\mathcal{G}}_{\mu\nu}$ is indeed covariant, $\mathcal{G}_{\mu\nu}$ is not. It is therefore comforting to see that the same delta function we encountered in this work in the $A_0$ propagator also appears in flat-space (although it is usually hidden), and it further emphasises that $\eqref{Gspin1effective}$ and \eqref{PhiToSigma1} are the correct expressions.

    The other delta function we have in the spin-$1$ case is in \eqref{PhiToSigma5} which actually has no flat-space counterpart. Indeed, in flat space the mode functions for the massive spinning field are simply those of a massive scalar field and the subtitles and delta functions ultimately arise from the polarisation factor, as we have just seen. However, in de Sitter space the longitudinal mode of $A_i$ truly propagates in a different fashion to $A_0$ and in a different fashion to a massive scalar. The delta function in \eqref{PhiToSigma5} is therefore a consequence of the expanding background.
\end{enumerate}

\section{Conclusion and outlook} \label{Conclusion}
In this paper we reviewed the Feynman rules for computing inflaton wavefunction coefficients and cosmological correlators for the exchange of massive spinning fields in the CCM description, and derived them for the CC description. With the rules at hand, we then compared the correlators computed in the two descriptions.

We found that the Feynman rules for CC differ from the what we may have expected in two ways: (1) there are {\it mixed propagators}, which allows modes with the same helicity, but coming from different components of the spinning fields, to propagate into each other, and (2) propagators that involve the non-dynamical modes can additionally carry a delta function term that signifies instantaneous propagation. The delta function must be included to arrive at the correct perturbative results. With regards to the correlator comparison, we found that a CC exchange contribution with local interactions can be recast as a sum of CCM exchange contributions and contact diagrams if, and only if, only a single helicity contributes to the exchange process. This is the case for the bispectrum of inflaton perturbations but not for the trispectrum and other higher-point correlators. We confirmed that this correspondence continues to hold when we work within the EFToI and impose the corresponding non-linearly realised symmetries. Our results suggest that the cosmological collider signals being searched for in the data \cite{Cabass:2024wob,Sohn:2024xzd} are fully captured by the CCM scenario which is actually the simpler set-up for performing computations. 

There are a few avenues for future research:
\begin{itemize}
\item {\bf Graviton correlators}: we focused on inflaton correlators, but we expect very similar conclusions to apply to correlators involving gravitons too, i.e. we expect that the CC exchanges can be mimicked by CCM ones if we restrict to the bispectrum. However, the situation is a little more involved than for inflaton correlators. Indeed, consider the mixed correlator $\langle \pi \pi \gamma_{ij}\rangle $ which has contributions from the following two diagrams:
\begin{align}
\langle \pi \pi \gamma_{ij}\rangle \supset
    \begin{gathered}
   \begin{tikzpicture}
  \begin{feynman}
    \vertex[dot] (a) at (0,0){};
    \vertex[dot] (b) at (-1,0){};
    \vertex (i1) at (1,0){\(\pi\)};
    \vertex (c) at (-1.5,0.866025403784){\(\gamma_{ij}\)};
    \vertex(d) at (-1.5,-0.866025403784){\(\pi\)};
       \diagram* {
      (i1) -- [plain] (a) -- [boson] (b) -- [scalar] (c),
      (b) -- [plain] (d),
    };
  \end{feynman}
\end{tikzpicture}
\end{gathered}
+
 \begin{gathered}
   \begin{tikzpicture}
  \begin{feynman}
    \vertex[dot] (a) at (0,0){};
    \vertex[dot] (b) at (-1,0){};
    \vertex (i1) at (1,0){\(\gamma_{ij}\)};
    \vertex (c) at (-1.5,0.866025403784){\(\pi\)};
    \vertex(d) at (-1.5,-0.866025403784){\(\pi\)};
       \diagram* {
      (i1) -- [scalar] (a) -- [boson] (b) -- [plain] (c),
      (b) -- [plain] (d),
    };
  \end{feynman}
\end{tikzpicture}
\end{gathered}\,.
\end{align}
In these two diagrams {\it different} helicities are being exchanged due to the different linear-mixing vertex for each diagram: the left-hand diagram only exchanges the $h=0$ mode of the CC field, while the right-hand diagram only exchanges the $h=2$ mode of the CC field. Even though we expect both diagrams to individually map to a sum of CCM diagrams, one needs to verify that the correspondence respects locality of the vertices. If the correspondence holds, the CCM computation would be the simpler route to take in order to calculate such bispectra and would complement the results of \cite{Cabass:2022jda}. 
    \item {\bf Higher spins}: in this work we focused on the exchange of massive spin-1 and spin-2 CC fields, however it would be interesting to consider higher spins too with the aim of deriving the CC rules and performing the correlator comparison.
    \item {\bf Loops}: so far we only considered tree-level diagrams when deriving the CC Feynman rules and comparing CC and CCM bispectra. It would be interesting to perform the correlator comparison at loop level too. Loops have been considered in e.g. \cite{Lee:2023jby,Jain:2025maa,Qin:2024gtr,Gorbenko:2019rza,Cespedes:2023aal,Cohen:2020php,Premkumar:2022bkm,Senatore:2009cf,Huenupi:2024ksc,Palma:2025oux,Zhang:2025nzd,Weinberg:2005vy,Melville:2021lst}. We expect the bispectrum correspondence to break down at loop level. Indeed, consider the contribution
    \begin{align}
    B_3 \supset
    \begin{gathered}
   \begin{tikzpicture}
  \begin{feynman}
    \vertex[dot] (a) at (0,0){};
    \vertex[dot] (b) at (-1,0){};
    \vertex (i1) at (1,0){};
    \vertex (c) at (-1.5,0.866025403784){};
    \vertex(d) at (-1.5,-0.866025403784){};
       \diagram* {
      (i1) -- [plain] (a) -- [boson, half left] (b) -- [plain] (c),
       (a) -- [boson, half right] (b) -- [plain] (d),
    };
  \end{feynman}
\end{tikzpicture}
\end{gathered}
\,.
\end{align}
In contrast to the tree-level contributions, this diagram does not depend on the linear mixing vertex and therefore it represents the exchange of multiple helicities and our experience at tree level leads us to expect that such contributions cannot be mimicked by CCM loops with local vertices. 

\item {\bf More general EFTs}: it is possible that other descriptions of massive spinning fields during inflation are possible. It would be interesting to search for other possibilities and to check if any new descriptions yield distinct signatures in inflationary bispectra.
\end{itemize}

\paragraph*{Acknowledgements} We thank Zongzhe Du, Sadra Jazayeri, Austin Joyce, Hayden Lee, Enrico Pajer, Zhehan Qin, Paul Saffin, Santiago Ag\"{u}\'{i} Salcedo, Xi Tong, Dong-Gang Wang, Yi Wang, and Yuhang Zhu for helpful discussions. T.C. is supported by an STFC studentship [grant number ST/Y509437/1] and a scholarship from the University of Nottingham. D.S. is supported by a UKRI Stephen Hawking Fellowship [grant number EP/W005441/1] and a Nottingham Research Fellowship from the University of Nottingham. For the purpose of open access, the authors have applied a CC BY public copyright licence to any Author Accepted Manuscript version arising.

\appendix
\appendixpage
\addappheadtotoc
\section{Further details on computing wavefunction coefficients}
In the main text, we have asserted without proof that:
\begin{itemize}
    \item For diagrams with no $\pi$ internal lines, the $\pi$ free theory $S_2[\pi]$ does not contribute to the $\pi$ wavefunction coefficients other than $\psi_2^{\pi\pi}$. The free theory of the spinning field does contribute, however.
    \item The equations of motion for the dynamical modes $\Phi^h_{S,S}$ in step \ref{CCstepE} {\it after} rewriting the action is equivalent to directly plugging in the solution of the non-dynamical modes $\Phi^h_{n,S}$ in step \ref{CCstepC} into the equations of motion for $\Phi^h_{S,S}$ {\it before} rewriting the action.
\end{itemize}
In this appendix we provide proofs of these statements. 

\subsection{No contribution from $S_2[\pi]$ to $\pi$ wavefunction coefficients} \label{nopisector}
We can integrate by parts to write the free theory as
\begin{align}
S_2[\pi] & = \frac 12 \int d\eta \int_{\bfk} \pi(\eta, \bfk) \mathcal{O}_{\pi}(\eta, k) \pi(\eta, -\bfk) + \frac 12 \lim_{\eta \to \eta_0} \int_{\bfk} a^2(\eta)\pi (\eta,\bfk) \pi'(\eta,-\bfk)\,. \label{pifreetheoryibp}
\end{align}
In a general theory where the free theory is augmented by interactions, the general solution for the inflaton is 
\begin{align}
    \pi(\eta,\bfk) = K_{\pi}(\eta,k) \bar{\pi}(\bfk) + i\int d\eta' G_{\pi}(\eta,\eta',k) \frac {\delta S_{\text{int}}}{\delta \pi}(\eta',-\bfk)\,,\label{generalsolutionofpi}
\end{align}
where the bulk-bulk propagator satisfies 
\begin{align}
\lim_{\eta \rightarrow \eta_0}G_{\pi}(\eta, \eta', k) \sim \eta_0^3 \,,
\end{align}
and
\begin{align}
\lim_{\eta \to \eta_0} \partial_{\eta} G_{\pi} (\eta,\eta',k) & = \lim_{\eta\to \eta_0} \left (\partial_{\eta} \pi(\eta,k)\pi^*(\eta',k) - \frac {\pi(\eta_0,k)}{\pi^*(\eta_0,k)} \partial_{\eta} \pi^*(\eta,k) \pi^*(\eta',k) \right ) \\
& = K(\eta',k) \lim_{\eta \to \eta_0} \left (\pi^*(\eta_0,k) \partial_{\eta} \pi (\eta,k) - \pi(\eta_0,k) \partial_{\eta} \pi^*(\eta,k)\right ) \\
& = -ia^{-2}(\eta_0) K_{\pi}(\eta',k)\,,
\end{align}
and 
\begin{align}
\mathcal{O}_{\pi} (\eta,k)G_{\pi}(\eta,\eta',k) = i \delta (\eta -\eta') \,.
\end{align}
Terms with no bulk-bulk propagators only come from the boundary term in \eqref{pifreetheoryibp} and ultimately lead to the power spectrum of $\pi$. Terms with two copies of the bulk-bulk propagator will ultimately contribute to exchange diagrams with $\pi$ internal lines. So the case of interest is with one factor of the bulk-boundary propagator and one copy of the bulk-bulk propagator. These contributions are
\begin{align}
S_2[\pi]&  \supset \frac 12 \int d\eta \int_{\bfk} K_{\pi}(\eta,k) \bar{\pi}(\bfk)\left (-\frac {\delta S_{\text{int}}}{\delta \pi}(\eta,\bfk)\right )  + \frac 12 \int_{\bfk} a^2(\eta_0) \bar{\pi}(\bfk) \left ( a^{-2}(\eta_0)\int d\eta' K_{\pi}(\eta',k) \frac {\delta S_{\text{int}}}{\delta \pi} (\eta',\bfk)\right )\,,
\end{align}
and we see that these two terms cancel. Therefore, unless we are considering an exchange of $\pi$, or the power spectrum, we do not need to consider $S_2[\pi]$ when computing wavefunction coefficients. 

\subsection{Equations of motion for $\Phi^h_{S,S}$}  \label{eomonshell}
In the main body we took two approaches to deriving wavefunction coefficients. In the first approach we integrated out the non-dynamical modes to give us an action for the dynamical modes only. We then derived and solved the resulting equations of motion to compute the on-shell action. In examples we saw that it wasn't necessary to write down an action for the dynamical modes only, rather we can solve the full set of equations of motion, including those of the non-dynamical modes, and then compute the on-shell action. Here we show that these two methods yield the same result. 

We consider the action $S[\Phi^h_{n,S},\text{traces},\pi]$, and in step \ref{CCstepC}, we solve the equations of motion for the spinning field:
\begin{align}
&\frac {\delta S[\Phi^h_{n,S},\text{traces},\Phi^h_{S,S},\pi]}{\delta \Phi^h_{n,S}(\eta,\bfk)} = 0 \,, \label{verygeneraleom1} \\
&\frac {\delta S[\Phi^h_{n,S},\text{traces},\Phi^h_{S,S},\pi]}{\delta (\text{traces})(\eta,\bfk)} = 0\,, \label{verygeneraleom2}
\end{align}
but only for the non-dynamical modes i.e. for $0\leq n<S$, so that 
\begin{align}
\Phi^h_{n,S} = \Phi^h_{n,S} (\Phi^h_{S,S},\pi),\qquad \text{traces}=\text{traces}(\Phi^h_{S,S},\pi)\,.
\end{align}
Note that these solutions are differential operators acting on $\Phi^h_{S,S}$ and $\pi$.\footnote{This might not be easily written down explicitly if the interactions are not linear in $\Phi$, but we can in theory solve the differential equation given $\Phi^h_{S,S}$ and $\pi$, and these are best regarded as formal solutions.} If we now plug these solutions for the non-dynamical modes into the action, it becomes
\begin{equation}
    S = S[\Phi^h_{n,S}(\Phi^h_{S,S},\pi),\text{traces}(\Phi^h_{S,S},\pi),\Phi^h_{S,S},\pi]\,.
\end{equation}
Now we compute the variational derivative with respect to $\Phi^h_{S,S}$ using the equivalent of the ``multivariate chain rule" in functional calculus:
\begin{align}
& \frac {\delta S[\Phi^h_{n,S}(\Phi^h_{S,S},\pi),\text{traces}(\Phi^h_{S,S},\pi),\Phi^h_{S,S},\pi]}{\delta \Phi^h_{S,S}(\eta,\bfk)} \nonumber \\
=\; & \int d\eta' \int_{\bfk'} \left (\sum_{n=0}^{S-1}\cancelto{0}{\frac {\delta S[\Phi^h_{n,S},\text{traces},\Phi^h_{S,S},\pi]}{\delta \Phi^h_{n,S}(\eta',\bfk')}} \frac {\delta \Phi^h_{n,S}(\Phi^h_{S,S},\pi)(\eta',\bfk')}{\delta \Phi^h_{S,S}(\eta,\bfk)} \right . \nonumber \\
& \qquad \qquad \left .+ \cancelto{0}{\frac {\delta S[\Phi^h_{n,S},\text{traces},\Phi^h_{S,S},\pi]}{\delta (\text{traces})(\eta',\bfk')}} \frac {\delta \text{traces}(\Phi^h_{S,S},\pi)(\eta',\bfk')}{\delta \Phi^h_{S,S}(\eta,\bfk)}\right ) + \frac {\delta S[\Phi^h_{n,S},\text{traces},\Phi^h_{S,S},\pi]}{\delta \Phi^h_{S,S}(\eta,\bfk)}\,,
\end{align}
and hence it is the same as the variational derivative {\it before} we have used the equations of motion \eqref{verygeneraleom1} and \eqref{verygeneraleom2} to eliminate $\Phi^h_{n,S}$ and the traces. Exactly the same argument goes for the variational derivative with respect to $\pi$, and therefore the equations of motion before and after plugging in the solutions of $\Phi^h_{n,S}$ and the traces are identical.

\section{Double-exchange diagrams} \label{factorexchange}
In the main text we focused on single exchanges of a massive spinning field. In this appendix we consider a double-exchange diagram focusing on the type that can contribute to the cubic wavefunction coefficient and the bispectrum e.g.
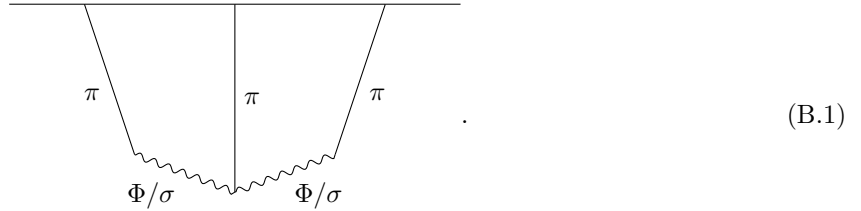
\begin{equation} \label{DoubleExchangeDiagram}
    \begin{gathered}
        \begin{tikzpicture}
  \begin{feynman}
    \vertex (a) at (1,0);
    \vertex (i1) at (2,0);
    \vertex (i2) at (4,0);
    \vertex (i3) at (6,0);
    \vertex (b) at (7,0);
    \vertex (c1) at (8/3,-2);
    \vertex (c2) at (4,-5/2);
    \vertex (c3) at (16/3,-2);
       \diagram* {
      (a) -- [plain] (i1) -- [plain] (i2) -- [plain] (i3) -- [plain] (b),
      (c1) -- [plain, edge label = \(\pi\)] (i1),
      (c2) -- [plain, edge label' = \(\pi\)] (i2),
      (c3) -- [plain, edge label' = \(\pi\)] (i3),
      (c1) -- [boson, edge label'=\(\Phi / \sigma\)] (c2) -- [boson, edge label'=\(\Phi / \sigma\)] (c3),
    };
  \end{feynman}
\end{tikzpicture}
    \end{gathered}
    .
\end{equation}
The procedure is somewhat similar to the single-exchange case. In general, the solutions to the equations of motion of the exchanged fields are of the form
\begin{align}
\varphi_i(\eta,\bfk) = K_{\varphi_i}(\eta,k)\bar{\varphi_i}(\bfk) + i\sum_{j}\int d\eta' \widetilde{G}_{\varphi_i\varphi_j}(\eta,\eta',k) \frac {\delta S_{\text{int}}}{\delta \varphi_j} (\eta',-\bfk)\,,\label{generalsolutionofCCandCCMfields}
\end{align}
where $\varphi_i$ covers both CCM and CC fields, and the $\widetilde{G}_{\varphi_i\varphi_j}$ are the {\it mixed effective propagators} we derived in the main text. In the CCM case, these fields are $\sigma_{h,S}$, and the labels $i$ refer to the helicities $h$. Since different helicities decouple, and there is only one mode for each helicity, the mixed effective propagators for CCM fields are just the usual bulk-bulk propagators:
\begin{equation}
\widetilde{G}_{\sigma_{h,S}\sigma_{h',S}}(\eta,\eta',k) = G_{\sigma_{h,S}} (\eta,\eta',k) \delta_{hh'}\,,
\end{equation}
where the $G_{\sigma_{h,S}}$ are the CCM propagators defined in \eqref{CCMpropagators}. For CC fields, the labels $i$ refer to both helicities $h$ and ``spatial spin" $n$, as well as the traces of $\Phi$. Different helicities still decouple, but the CC fields can mix between different $n$'s. For example, when $S = 1$,
\begin{align}
\widetilde{G}_{\Phi^0_{0,1}\Phi^0_{0,1}}(\eta,\eta',k) & = \left (G^0_{0,1 \to 0,1}\right )_{\text{eff}} (\eta,\eta',k)\,, \\
\widetilde{G}_{\Phi^0_{0,1}\Phi^0_{1,1}} (\eta,\eta',k) & = \left (G^0_{0,1 \to 1,1}\right )_{\text{eff}}(\eta,\eta',k)\,,
\end{align}
where the effective and mixed propagators are defined in \eqref{Gspin1effective} and \eqref{Gspin1mixed} respectively. Similar to \eqref{generalsolutionofCCandCCMfields}, we also have the solution for $\pi$ which is given in \eqref{generalsolutionofpi}. Now we assume that the interaction Lagrangian is comprised of
\begin{align}
    S_{\text{int}} & = S^{(\text{linear})}_{\text{int}} +  S^{(\text{quadratic})}_{\text{int}},
    \end{align}
where $S^{(\text{linear})}_{\text{int}}$ denotes the part of the interaction that depends linearly on the massive fields $\varphi_i$, and $S^{(\text{quadratic})}_{\text{int}}$ for the part that depends quadratically. These are the only interactions that will contribute to a diagram of the form \eqref{DoubleExchangeDiagram}. We can then write the action, only including terms with the spinning field, as
\begin{align}
S & = \underbrace{\frac 12 \sum_{i} \int d\eta \int_{\bfk} \varphi_i (\bfk) \frac {\delta S_2}{\delta \varphi_i} (\bfk)}_{\text{free theory}} + \underbrace{\sum_{i} \int d\eta \int_{\bfk} \varphi_i (\bfk) \frac {\delta S^{(\text{linear})}_{\text{int}}}{\delta \varphi_i}(\bfk) + \frac 12 \sum_{i} \int d\eta \int_{\bfk} \varphi_i (\bfk) \frac {\delta S^{(\text{quadratic})}_{\text{int}}}{\delta \varphi_i}(\bfk)}_{\text{interactions}} \nonumber \\
& \quad + \text{boundary terms}\\
& = \frac 12 \sum_{i} \int d\eta \int_{\bfk} \varphi_i (\bfk) \underbrace{\frac {\delta S^{(\text{linear})}_{\text{int}}}{\delta \varphi_i}(\bfk)}_{\text{contains } \pi \text{ only}} + \;\text{boundary terms}\,, \label{Sintlineargeneraldoubleexchange}
\end{align}
where we have used the equations of motion to arrive at the final line. Substituting the solutions \eqref{generalsolutionofCCandCCMfields} into \eqref{Sintlineargeneraldoubleexchange} gives
\begin{align}
S & \supset \frac 12 \sum_i \int d\eta \int_{\bfk} \frac {\delta S^{(\text{linear})}_{\text{int}}}{\delta \varphi_i}(\eta, \bfk) \left [i\sum_j \int d\eta' \widetilde{G}_{\varphi_i \varphi_j} (\eta,\eta',k) \underbrace{\frac {\delta S^{(\text{quadratic})}_{\text{int}}}{\delta \varphi_j} (\eta',-\bfk)}_{\text{linear in }\varphi_k}\right ]\,, \label{intermediatestepgeneraldoubleexchange}
\end{align}
where we have kept only the part relevant for a double-exchange diagram i.e. we have dropped the contribution with two copies of $S_{\text{int}}^{(\text{linear})}$ since that term yields the single exchange contribution and have dropped the term that depends on the boundary value of the spinning fields. We now denote
\begin{align}
\frac {\delta S^{(\text{quadratic})}_{\text{int}}}{\delta \varphi_j} (\eta',-\bfk) & = \sum_k \int_{\bfk_1,\bfk_2} \mathcal{O}^{\pi}_{jk} (\eta') \pi (\eta',\bfk_1) \mathcal{O}^{\varphi}_{jk} (\eta') \varphi_k (\eta',\bfk_2) \delta^{(3)}(\bfk_1+\bfk_2+\bfk)\,,
\end{align}
for some differential operators $\mathcal{O}^{\pi}_{jk}$ and $\mathcal{O}^{\varphi}_{jk}$. We have taken this interaction to be linear in $\pi$ since we are concentrating on \eqref{DoubleExchangeDiagram}. Then \eqref{intermediatestepgeneraldoubleexchange} contains a double-exchange term of the form
\begin{align}
S& \supset -\frac 12 \sum_{i,j,k} \int d\eta d\eta' d\eta'' \int_{\bfk,\bfk_1,\bfk_2} \frac {\delta S^{(\text{linear})}_{\text{int}}}{\delta \varphi_i}(\eta,\bfk) \widetilde{G}_{\varphi_i \varphi_j} (\eta,\eta',k) \mathcal{O}^{\pi}_{jk} (\eta') \pi (\eta',\bfk_1) \times \nonumber \\
& \qquad \qquad \qquad \qquad \qquad \qquad \qquad \mathcal{O}^{\varphi}_{jk}(\eta') \widetilde{G}_{\varphi_j\varphi_k} (\eta',\eta'',k_2) \frac {\delta S^{(\text{linear})}_{\text{int}}}{\delta \varphi_k}(\eta'',\bfk_2) \delta^{(3)}(\bfk_1+\bfk_2 +\bfk)\,.
\end{align}
One can additionally use integration by parts to remove the differential operator $\mathcal{O}^{\varphi}_{jk}$ from $\widetilde{G}$ and transfer them onto $\pi$, keeping track of any boundary terms. The main difference compared with the single-exchange case is that these operators $\mathcal{O}^{\varphi}_{jk}$ come from variational derivative of the {\it quadratic} part of $S_{\text{int}}$, and come with a factor of 2. Therefore, while the symmetry factor for a single-exchange diagram is 1/2 (see \eqref{Feynmanspin1helicity0}, \eqref{Feynmanspin1helicity1}, and \eqref{Feynmanspin2}), the symmetry factor for a double-exchange diagram is 1. Such diagrams have been computed in \cite{Aoki:2024uyi}.  

As an example, and to relate this with our discussion of CC Feynman rules in Section \ref{Spin1Section}, consider a CC spin-1 field interacting with $\pi$ via
\begin{align}
S_{\text{int}} = \int d^3x d\eta \left (\lambda_1  a^2 \pi'\Phi_0 + \lambda_2 a \pi' \Phi_0^2 \right ) \,,
\end{align}
then 
\begin{align}
    \frac {\delta S^{(\text{quadratic})}_{\text{int}}}{\delta \Phi^0_{0,1}}(\eta',-\bfk) & = 2 \lambda_2 \int_{\bfk_1,\bfk_2} a(\eta') \pi'(\eta',\bfk_1) \Phi^0_{0,1}(\eta',\bfk_2)\delta^{(3)}(\bfk_1+\bfk_2 + \bfk) \,,
\end{align}
and the double-exchange diagram is then given by
\begin{align}
\psi_3^{\pi\pi\pi}(\bfk,\bfk_1,\bfk_2) & \supset -i\lambda_1^2\lambda_2 \int d\eta d\eta' d\eta'' \left (a^2(\eta) K'_{\pi} (\eta,k)\right ) \left (G^0_{0,1\to 0,1}\right )_{\text{eff}}(\eta,\eta',k) \left (a(\eta') K'_{\pi}(\eta',k_1)\right ) \times \nonumber \\
& \qquad \qquad \qquad \left (G^0_{0,1\to 0,1}\right )_{\text{eff}}(\eta',\eta'',k_2)\left (a^2(\eta'') K'_{\pi}(\eta'',k_2)\right ) + \text{perms of }\{\bfk,\bfk_1,\bfk_2\}\,.
\end{align}
If we now write the effective propagators using $G^0_{0,1}$ using \eqref{Gspin1effective}, then we have
\begin{align}
\psi_3^{\pi\pi\pi}(\bfk,\bfk_1,\bfk_2) & \supset -i\lambda_1^2\lambda_2 \int d\eta d\eta' d\eta'' \left (a^2(\eta) K'_{\pi} (\eta,k)\right ) G^0_{0,1}(\eta,\eta',k) \left (a(\eta') K'_{\pi}(\eta',k_1)\right ) \times \nonumber \\
& \qquad \qquad \qquad \qquad \qquad G^0_{0,1\to 0,1} (\eta',\eta'',k_2)\left (a^2(\eta'') K'_{\pi}(\eta'',k_2)\right ) \nonumber \\
& \quad + \frac {2 \lambda_1^2 \lambda_2}{m^2} \int d\eta d\eta'' a(\eta) K'_{\pi}(\eta,k) K'_{\pi}(\eta,k_1) G^0_{0,1}(\eta',\eta'',k_2) \left (a^2(\eta'') K'_{\pi}(\eta'',k_2)\right ) \nonumber \\
& \quad + \frac {i\lambda_1^2\lambda_2}{m^4} \int d\eta a(\eta) K'_{\pi}(\eta,k) K'_{\pi}(\eta',k_1) K'_{\pi}(\eta'',k_2) + \text{perms of }\{\bfk,\bfk_1,\bfk_2\}\,,
\end{align}
so diagrammatically, using similar notation as \eqref{newFeynmanspin1trispectrum}, we have
\begin{equation}
    \underbrace{\begin{gathered}
        \begin{tikzpicture}
  \begin{feynman}
    \vertex (a) at (1.25,0);
    \vertex (i1) at (1.75,0);
    \vertex (i2) at (3,0);
    \vertex (i3) at (4.25,0);
    \vertex (b) at (4.75,0);
    \vertex (c1) at (2,-3/2);
    \vertex (c2) at (3,-15/8);
    \vertex (c3) at (4,-3/2);
       \diagram* {
      (a) -- [plain] (i1) -- [plain] (i2) -- [plain] (i3) -- [plain] (b),
      (c1) -- [plain] (i1),
      (c2) -- [plain] (i2),
      (c3) -- [plain] (i3),
      (c1) -- [double] (c2) -- [double] (c3),
    };
  \end{feynman}
\end{tikzpicture}
    \end{gathered}}_{\text{correct Feynman rules}}
    = \underbrace{\begin{gathered}
        \begin{tikzpicture}
  \begin{feynman}
    \vertex (a) at (1.25,0);
    \vertex (i1) at (1.75,0);
    \vertex (i2) at (3,0);
    \vertex (i3) at (4.25,0);
    \vertex (b) at (4.75,0);
    \vertex (c1) at (2,-3/2);
    \vertex (c2) at (3,-15/8);
    \vertex (c3) at (4,-3/2);
       \diagram* {
      (a) -- [plain] (i1) -- [plain] (i2) -- [plain] (i3) -- [plain] (b),
      (c1) -- [plain] (i1),
      (c2) -- [plain] (i2),
      (c3) -- [plain] (i3),
      (c1) -- [ghost] (c2) -- [ghost] (c3),
    };
  \end{feynman}
\end{tikzpicture}
    \end{gathered}}_{\text{naive Feynman rules}}+ \underbrace{\begin{gathered}
\begin{tikzpicture}
  \begin{feynman}
   \vertex (a) at (1.25,0);
    \vertex (i1) at (1.75,0);
    \vertex (i2) at (2.75,0);
    \vertex (i4) at (4.25,0);
    \vertex (b) at (4.75,0);
    \vertex (c1) at (2.25,-15/8);
    \vertex (c2) at (3.75,-15/8);
       \diagram* {
      (a) -- [plain] (i1) -- [plain] (i2) -- [plain] (i4) -- [plain] (b),
      (c1) -- [plain] (i1),
      (c1) -- [plain] (i2),
      (c2) -- [plain] (i4),
      (c1) -- [ghost] (c2),
    };
  \end{feynman}
\end{tikzpicture}
    \end{gathered}}_{\text{naive Feynman rules}}+ \underbrace{\begin{gathered}
\begin{tikzpicture}
  \begin{feynman}
   \vertex (a) at (1.25,0);
    \vertex (i1) at (1.75,0);
    \vertex (i2) at (3,0);
    \vertex (i4) at (4.25,0);
    \vertex (b) at (4.75,0);
    \vertex (c1) at (3,-15/8);
       \diagram* {
      (a) -- [plain] (i1) -- [plain] (i2) -- [plain] (i4) -- [plain] (b),
      (c1) -- [plain] (i1),
      (c1) -- [plain] (i2),
      (c1) -- [plain] (i4),
    };
  \end{feynman}
\end{tikzpicture}
    \end{gathered}}_{\text{contact diagram}}.
\end{equation}
One can generalise even further to triple or higher number of exchanges, with the complications being the symmetry factors. Barring time derivatives hitting the massive field, all statements in the main text generalise (relatively) easily to such more complicated exchanges.

\section{Details of derivation of spin-$2$ CC rules} \label{spin2details}
In this appendix, we present the full details of the spin-$2$ CC Feynman rules.
\paragraph{\ref{CCstep1} Decompose the field into helicity modes.} This has been done in \eqref{spin2decomposehelicity00} -- \eqref{spin2decomposehelicityij}.
\paragraph{\ref{CCstep2} Write the action in terms of helicity modes.} From the covariant action \eqref{spin2covariant}, we can  define $m_2^2=m^2+2H^2$ and expand in helicity modes. We denote the free theory action as
\begin{equation} \label{fullspin2helicityaction}
    S_2 = \sum_{h=-2}^2 \int d\eta \int_{\bfk} L^{(h)}_2 + S_{\text{boundary}}\,, 
\end{equation}
where 
\begin{align}
L^{(0)}_2 & = \frac {1}{a^2}\partial_{\eta}\Phi^0_{2,2}(-\bfk) \partial_{\eta}\Phi^0_{2,2}(\bfk) - \left ((m_2^2-4H^2)-\frac {k^2}{3a^2}\right ) \Phi^0_{2,2}(-\bfk)\Phi^0_{2,2}(\bfk)  - \frac 1{3a^2}\partial_{\eta}\Phi_{kk}(-\bfk)\partial_{\eta}\Phi_{kk}(\bfk) \nonumber \\ &   + \left (\frac 13(m_2^2-4H^2)+\frac {k^2}{9a^2}\right )\Phi_{kk}(-\bfk)\Phi_{kk}(\bfk) +(m_2^2-2H^2) \Phi^0_{1,2}(-\bfk) \Phi^0_{1,2}(\bfk)  \nonumber \\
& -3H^2 \Phi^0_{0,2}(-\bfk)\Phi^0_{0,2}(\bfk) - \frac {2k^2}{3\sqrt{3}a^2}\Phi^0_{2,2}(-\bfk)\Phi_{kk}(\bfk) - \frac {4k}{\sqrt{3}a^2}\partial_{\eta}\Phi^0_{1,2}(-\bfk)\Phi^0_{2,2}(\bfk)   \nonumber  \\
&  + \frac {2k^2}{\sqrt{3}a^2}\Phi^0_{0,2}(-\bfk)\Phi^0_{2,2}(\bfk) + \frac {4k}{3a^2}\partial_{\eta}\Phi^0_{1,2}(-\bfk)\Phi_{kk}(\bfk)  +\frac{2H}{a} \partial_{\eta}\Phi^0_{0,2}(-\bfk)\Phi_{kk}(\bfk) \nonumber  \\
&- \left (m_2^2 -4H^2 + \frac {2k^2}{3a^2}\right )\Phi^0_{0,2}(-\bfk)\Phi_{kk}(\bfk)- \frac {4 H k}{a} \Phi^0_{0,2}(-\bfk)\Phi^0_{1,2}(\bfk) \,, \\
L^{(\pm 1)}_2 & =\frac 1{a^2}\partial_{\eta} \Phi^{\pm 1}_{2,2}(-\bfk) \partial_{\eta} \Phi^{\pm 1}_{2,2}(\bfk) - (m_2^2 - 4H^2)\Phi^{\pm 1}_{2,2}(-\bfk)\Phi^{\pm 1}_{2,2}(\bfk) \nonumber \\
&+\left (m_2^2 - 2H^2 + \frac {k^2}{a^2}\right ) \Phi^{\pm 1}_{1,2}(-\bfk)\Phi^{\pm 1}_{1,2}(\bfk) + \frac {2k}{a^2} \Phi^{\pm 1}_{2,2}(-\bfk) \partial_{\eta} \Phi^{\pm 1}_{1,2}(\bfk) \,, \\
L^{(\pm 2)}_2 & = \frac 1{a^2} \partial_{\eta}\Phi^{\pm 2}_{2,2}(-\bfk) \partial_{\eta}\Phi^{\pm 2}_{2,2}(\bfk) -\left (m_2^2-4H^2 + \frac {k^2}{a^2}\right )\Phi^{\pm 2}_{2,2}(-\bfk)\Phi^{\pm 2}_{2,2}(\bfk) \,,\\ 
 S_{\text{boundary}}  &= \lim_{\eta \to \eta_0}\int_{\bfk} \left [\sum_{h=-2}^2 -\frac {2H}{a} \Phi^h_{2,2}(-\bfk)\Phi^h_{2,2}(\bfk) + \sum_{h=-1}^1 \frac {4H}{a} \Phi^h_{1,2}(-\bfk)\Phi^h_{1,2}(\bfk) - \frac{3H}{2 a}\Phi^0_{0,2}(-\bfk)\Phi^0_{0,2}(\bfk) \right . \nonumber \\
&  \left . + \frac H{6a} \Phi_{kk}(-\bfk)\Phi_{kk}(\bfk) 
- \frac k{a^2} \Phi^0_{1,2}(-\bfk)\Phi_{kk}(\bfk)   - \frac {3H}{a}\Phi^0_{0,2}(-\bfk) \Phi_{kk}(\bfk) - \frac k{a^2}\Phi^0_{0,2}(-\bfk)\Phi^0_{1,2}(\bfk)  \right ]\,. \label{boundarytermsspin2action} 
\end{align}
In addition to this action for the free theory, we will also include interactions between this spin-$2$ field and the inflaton. We keep those interactions general. We see from the above expressions that the modes $\Phi_{S,S}^h$ have standard kinetic terms which is expected since these will ultimately correspond to the five dynamical modes. We also see, however, that $\Phi_{kk,2}^0$ has a kinetic term with the wrong sign. As we will see below, this mode is indeed non-dynamical in the sense that it can be solved for algebraically and therefore it does not correspond to a ghostly degree of freedom despite appearances.  
\paragraph{\ref{CCstep3} Solve for $\Phi^h_{n,S}$ and traces in terms of $\Phi^h_{S,S}$.} If we vary the full action, free theory plus interactions, we obtain the following equations of motion (with the arguments suppressed):
\begin{align}
-\frac {\delta S_{\text{int}}}{\delta \Phi^0_{0,2}} = \frac {\delta S_2}{\delta \Phi^0_{0,2}}&= -6H^2 \Phi^0_{0,2} - \frac {4 H k}{a}\Phi^0_{1,2} + \left (-m_2^2+6H^2 - \frac {2k^2}{3a^2} - \frac{2 H }{a} \partial_{\eta} \right )\Phi_{kk} + \frac {2k^2}{\sqrt{3}a^2} \Phi^0_{2,2}\,, \label{sigma002eom}\\
-\frac {\delta S_{\text{int}}}{\delta \Phi^0_{1,2}} = \frac {\delta S_2}{\delta \Phi^0_{1,2}} &= -\frac {4 H k}{a} \Phi^0_{0,2} + 2(m_2^2-2H^2) \Phi^0_{1,2} +\left (\frac {8 H k}{3a} - \frac {4k}{3a^2} \partial_{\eta}\right )\Phi_{kk} - \left (\frac {8 H k}{\sqrt{3} a} - \frac {4k}{\sqrt 3a^2} \partial_{\eta}\right )\Phi^0_{2,2}\,, \label{sigma012eom} \\-\frac {\delta S_{\text{int}}}{\delta \Phi_{kk}} = \frac {\delta S_2}{\delta \Phi_{kk}} & = \left (-m_2^2+4H^2-\frac {2k^2}{3a^2} + \frac{2 H }{a}\partial_{\eta}\right )\Phi^0_{0,2} + \frac {4k}{3a^2}\partial_{\eta}\Phi^0_{1,2} \nonumber \\
& \quad + \left (\frac 2{3a^2} \partial_{\eta}^2 - \frac {4 H }{3a} \partial_{\eta} + \frac 23 (m_2^2-4H^2) + \frac {2k^2}{9a^2}\right )\Phi_{kk} - \frac {2k^2}{3\sqrt{3}a^2} \Phi^0_{2,2} \,,\label{sigmakkeom} \\ 
-\frac {\delta S_{\text{int}}}{\delta \Phi^0_{2,2}} = \frac {\delta S_2}{\delta \Phi^0_{2,2}} & = \frac {2k^2}{\sqrt 3 a^2}\Phi^0_{0,2} - \frac {4k}{\sqrt 3 a^2}\partial_{\eta} \Phi^0_{1,2} - \frac {2 k^2}{3 \sqrt 3 a^2} \Phi_{kk} \nonumber \\
& \quad -\left (\frac 2{a^2} \partial_{\eta}^2 - \frac{4 H}{a}\partial_{\eta} + 2(m_2^2-4H^2) - \frac {2k^2}{3a^2}\right ) \Phi^0_{2,2} \,,\label{sigma022eom} \\
-\frac {\delta S_{\text{int}}}{\delta \Phi^{\pm 1}_{1,2}} = \frac {\delta S_2}{\delta \Phi^{\pm 1}_{1,2}} & = 2\left (m_2^2 - 2H^2 + \frac {k^2}{a^2}\right )\Phi^{\pm 1}_{1,2} - \left (\frac {2k}{a^2} \partial_{\eta} - \frac {4 H k}{a}\right )\Phi^{\pm 1}_{2,2}\,, \label{sigma112eom}\\
-\frac {\delta S_{\text{int}}}{\delta \Phi^{\pm 1}_{2,2}} = \frac {\delta S_2}{\delta \Phi^{\pm 1}_{2,2}} & = \frac {2k}{a^2}\partial_{\eta}\Phi^{\pm 1}_{1,2} - \left (\frac 2{a^2} \partial_{\eta}^2 -\frac {4 H}{a} \partial_{\eta} + 2(m_2^2 - 4H^2)\right )\Phi^{\pm 1}_{2,2}\,,\label{sigma122eom}\\
-\frac {\delta S_{\text{int}}}{\delta \Phi^{\pm 2}_{2,2}} = \frac {\delta S_2}{\delta \Phi^{\pm 2}_{2,2}} & = -\left (\frac 2{a^2} \partial_{\eta}^2 -\frac {4 H}{a} \partial_{\eta} + 2(m_2^2-4H^2) + \frac {2k^2}{a^2}\right )\Phi^{\pm 2}_{2,2}\,.\label{sigma222eom}
\end{align}
To solve these equations we consider the different helicity modes separately given that they are decoupled. 

\paragraph{$h=0$ modes} First we focus on the helicity $h=0$ modes. Our aim is to solve algebraically for $\Phi_{0,2}^0$, $\Phi_{1,2}^0$ and $\Phi_{kk,2}^0$ in terms of the interactions and $\Phi_{2,2}^0$. We should then find a dynamical equation for $\Phi_{2,2}^0$ which we can solve formally by introducing bulk-bulk propagators. With this solution at hand, we can plug it into the expressions for the non-dynamical modes to give us the full solutions in terms of propagators and the interacting part of the action. We first note that $\Phi_{0,2}^0$ appears algebraically in \eqref{sigma002eom} and so we can find an expression for $\Phi^0_{0,2}$. We can then plug this solution into \eqref{sigma012eom}, which allows us to solve for $\Phi^0_{1,2}$. We can then substitute both of these solutions into \eqref{sigmakkeom}, and after doing so it becomes algebraic in $\Phi_{kk}$ i.e. all terms with derivatives acting on $\Phi_{kk}$ cancel. This cancellation relies on the precise coefficients that appear in the spin-$2$ free theory and is a consequence of its ghost-free nature. We can therefore solve for $\Phi_{kk}$.  We then substitute all these solutions into the equation of motion for $\Phi^0_{2,2}$ \eqref{sigma022eom}. At the end of this procedure we can express all modes in terms of $\Phi^0_{2,2}$ and the interactions:
\begin{align}
\Phi^0_{0,2} & = \mathcal{O}^{(0)}_2 \Phi^0_{2,2} + \mathcal{O}^{(0)}_{0}\frac {\delta S_{\text{int}}}{\delta \Phi^0_{0,2}} +\mathcal{O}^{(0)}_{1}\frac {\delta S_{\text{int}}}{\delta \Phi^0_{1,2}} + \mathcal{O}^{(0)}_{kk}\frac {\delta S_{\text{int}}}{\delta \Phi_{kk}}\,, \label{sigma002itomsigma022} \\
\Phi^0_{1,2} & = \mathcal{O}^{(1)}_2 \Phi^0_{2,2} + \mathcal{O}^{(1)}_{0}\frac {\delta S_{\text{int}}}{\delta \Phi^0_{0,2}} +\mathcal{O}^{(1)}_{1}\frac {\delta S_{\text{int}}}{\delta \Phi^0_{1,2}} + \mathcal{O}^{(1)}_{kk}\frac {\delta S_{\text{int}}}{\delta \Phi_{kk}}\,, \label{sigma012itosigma022}\\
\Phi_{kk} & = \mathcal{O}^{(kk)}_2 \Phi^0_{2,2} + \mathcal{O}^{(kk)}_{0}\frac {\delta S_{\text{int}}}{\delta \Phi^0_{0,2}} +\mathcal{O}^{(kk)}_{1}\frac {\delta S_{\text{int}}}{\delta \Phi^0_{1,2}} + \mathcal{O}^{(kk)}_{kk}\frac {\delta S_{\text{int}}}{\delta \Phi_{kk}}\,, \label{sigmakkitosigma022}
\end{align}
and the equation of motion for $\Phi^0_{2,2}$ \eqref{sigma022eom} can be written as 
\begin{align} \label{newsigma022eompluggedin}
\mathcal{O}^{(2)}_2 \Phi^0_{2,2} + \mathcal{O}^{(2)}_0\frac {\delta S_{\text{int}}}{\delta \Phi^0_{0,2}} + \mathcal{O}^{(2)}_1 \frac {\delta S_{\text{int}}}{\delta \Phi^0_{1,2}} + \mathcal{O}^{(2)}_{kk} \frac {\delta S_{\text{int}}}{\delta \Phi_{kk}} + \frac {\delta S_{\text{int}}}{\delta \Phi^0_{2,2}} = 0\,.
\end{align}
The various differential operators $\mathcal{O}^{(i)}_{j}$, with $i,j$ denoting $0,1,kk,2$, are somewhat complicated and are given at the end of this appendix. There are a few points to note among these differential operators:
\begin{itemize}
\item The analogous operator in the spin-$1$ case is first-order in time. For this spin-$2$ case, we now encounter second-order operators too. There are four operators that are second-order: $\mathcal{O}^{(0)}_0$, $\mathcal{O}^{(0)}_2$, $\mathcal{O}^{(2)}_0$, and $\mathcal{O}^{(2)}_2$. All other operators are first-order or just a constant. 
\item There are of course various relations between these operators. One of particular interest is 
\begin{align} \label{o02o22okkrelation}
    \mathcal{O}^{(0)}_2 =\frac {2H^2k^2\eta^2}{3\sqrt{3}(m_2^2-4H^2)(m_2^2-2H^2)} \mathcal{O}^{(2)}_2 +\mathcal{O}^{(kk)}_2 .
\end{align}
If we ignore interactions and insert this relation into \eqref{sigma002itomsigma022}, we see that if $\Phi_{2,2}^0$ is on-shell, we have $\Phi^0_{0,2} = \Phi_{kk}$. This is a familiar relation and it ensures that $\Phi_{\mu\nu}$ is traceless on-shell.  
\item There are integration by parts relations between these operators. For any linear differential operator $\mathcal{O}$, we define a corresponding operator $\widetilde{\mathcal{O}}$ such that
\begin{align}
\int d\eta f(\eta) \mathcal{O} g(\eta) = \int d\eta g(\eta) \widetilde{O} f(\eta) + \text{boundary terms}.
\end{align}
For example, if $\mathcal{O} = a(\eta) \partial_{\eta}$, then $\widetilde{\mathcal{O}} = -a(\eta) \partial_{\eta} - a'(\eta)$. Our differential operators $\mathcal{O}^{(i)}_{j}$ satisfy
\begin{equation} \label{DifferentialMapping}
\widetilde{\mathcal{O}}^{(i)}_{j} = \mathcal{O}^{(j)}_i.
\end{equation}
These relations will be useful when we write formal solutions to these equations of motion.
\end{itemize}
We can now go ahead and find formal solutions to these equations of motion. We start with the dynamical mode and solve \eqref{newsigma022eompluggedin}. 
Before we include the interactions, let's first look at the free theory. If we impose Bunch-Davies vacuum conditions, then the solution to $\mathcal{O}^{(2)}_2 \Phi^0_{2,2} = 0$ is given by the mode function \cite{Lee:2016vti}
\begin{align}
    \Phi^0_{2,2} (\eta,k)& = \frac {\sqrt{\pi \alpha_1\alpha_2}H}{24\sqrt{2}} e^{i\pi(\nu_2+1/2)/2} (-\eta)^{-1/2}\times \nonumber \\
    & \quad \left [6k\eta[(2+\nu_2)H^{(1)}_{\nu_2+1}(-k\eta)-(2-\nu_2)H^{(1)}_{\nu_2-1}(-k\eta)] - (9-8k^2\eta^2)H^{(1)}_{\nu_2}(-k\eta)\right ]\,, \label{phi022modefunction}
\end{align}
where 
\begin{align} \label{Spin2Order}
\nu_2 = \sqrt{\frac{9}{4}-\frac{m^2}{H^2}}\,.
\end{align}
We can then construct the bulk-boundary and bulk-bulk propagators associated with this mode. The bulk-boundary propagator is
\begin{equation}
    K^0_{2,2}(\eta,k) = \frac {\Phi^0_{2,2}(\eta,k)}{\Phi^0_{2,2}(\eta_0,k)} \,,
\end{equation}
which satisfies
\begin{equation}
    \mathcal{O}^{(2)}_2K^0_{2,2} = 0, \qquad K^0_{2,2}(\eta_0,k) = 1,\qquad \lim_{\eta \to -\infty(1-i\epsilon)}K^0_{2,2} (\eta,k) = 0 \,.
\end{equation}
The bulk-bulk propagator is
\begin{equation}
    G^0_{2,2}(\eta,\eta',k) = \Phi^0_{2,2}(\eta,k) \Phi^{0^*}_{2,2}(\eta',k) \theta (\eta - \eta') + (\eta \leftrightarrow \eta') - \frac {\Phi^0_{2,2}(\eta_0,k)}{\Phi^{0^*}_{2,2}(\eta_0,k)} \Phi^{0^*}_{2,2}(\eta,k) \Phi^{0^*}_{2,2}(\eta',k) \,,
\end{equation}
and it satisfies
\begin{equation} 
    \mathcal{O}^0_{2,2}(\eta,k) G^0_{2,2}(\eta,\eta',k) = i\delta (\eta - \eta'), \qquad \lim_{\eta,\eta' \to \eta_0} G^0_{2,2}(\eta,\eta',k) = 0, \qquad \lim_{\eta,\eta'\to -\infty(1-i\epsilon)}G^0_{2,2}(\eta,\eta',k) = 0 \,.
\end{equation}
With these propagators at hand, we can formally solve \eqref{newsigma022eompluggedin} and write
\begin{align}
&\Phi^0_{2,2}(\eta,\bfk)  = K^0_{2,2}(\eta,k) \bar{\Phi}^0_{2,2}(\bfk) \nonumber \\
&+ i \int d \eta' G_{2,2}^{0}(\eta, \eta', k)\left[\mathcal{O}^{(2)}_0\frac {\delta S_{\text{int}}}{\delta \Phi^0_{0,2}}(\eta',-\bfk) + \mathcal{O}^{(2)}_1 \frac {\delta S_{\text{int}}}{\delta \Phi^0_{1,2}}(\eta',-\bfk) + \mathcal{O}^{(2)}_{kk} \frac {\delta S_{\text{int}}}{\delta \Phi_{kk}}(\eta',-\bfk) + \frac {\delta S_{\text{int}}}{\delta \Phi^0_{2,2}}(\eta',-\bfk) \right]\,.
\end{align}
In terms of deriving Feynman rules, it is not particularly useful to have differential operators acting on the contributions from the interacting part of the action. As with the spin-$1$ case, we instead integrate by parts such that the differential operators act on the bulk-bulk propagator $G_{2,2}^0$. This then leads us to introduce effective propagators that are not simply built out of the mode functions of the $\Phi_{2,2}^0$, but instead depend on the mode functions of the non-dynamical modes. If we go through this procedure, we can write this solution as  
\begin{align}
&\Phi^0_{2,2}(\eta,\bfk)  = K^0_{2,2}(\eta,k) \bar{\Phi}^0_{2,2}(\bfk)  \nonumber \\
& \quad + i\int d\eta' \frac {\delta S_{\text{int}}}{\delta \Phi^0_{0,2}} (\eta',-\bfk)\left (G^0_{2,2\to 0,2}\right )_{\text{eff}}(\eta,\eta',k)+ i\int d\eta' \frac {\delta S_{\text{int}}}{\delta \Phi^0_{1,2}} (\eta',-\bfk)\left (G^0_{2,2\to 1,2}\right )_{\text{eff}}(\eta,\eta',k) \nonumber \\
& \quad + i\int d\eta' \frac {\delta S_{\text{int}}}{\delta \Phi_{kk}} (\eta',-\bfk)\left (G^0_{2,2\to kk,2}\right )_{\text{eff}}(\eta,\eta',k)+ i\int d\eta' \frac {\delta S_{\text{int}}}{\delta \Phi^0_{2,2}} (\eta',-\bfk) \left (G^0_{2,2\to 2,2}\right )_{\text{eff}}(\eta,\eta',k) \,,\label{sigma022solutionfullygeneral}
\end{align}
where we have dropped all boundary terms since they vanish as $\eta_0 \rightarrow 0$. To write this expression without any explicit differential operators, we have introduced the following {\it mixed propagators}:
\begin{align}
\left (G^0_{2,2\to 0,2}\right )_{\text{eff}}(\eta,\eta',k) & = \mathcal{O}^{(0)}_2(\eta',k) G^0_{2,2}(\eta,\eta',k) \\
& = \Phi^0_{2,2}(\eta,k) \Phi^{0^*}_{0,2}(\eta',k)\theta(\eta - \eta') + \Phi^0_{0,2}(\eta',k) \Phi^{0^*}_{2,2}(\eta,k) \theta(\eta'-\eta)\nonumber \\
& \quad  - \frac {\Phi^0_{2,2}(\eta_0,k)}{\Phi^{0^*}_{2,2}(\eta_0,k)} \Phi^{0^*}_{2,2}(\eta,k) \Phi^{0^*}_{0,2}(\eta',k) + \frac {2iH^2k^2\eta'^2}{3\sqrt{3}(m_2^2-4H^2)(m_2^2-2H^2)}\delta (\eta - \eta')\,, \\
\left (G^0_{2,2\to 1,2}\right )_{\text{eff}}(\eta,\eta',k) & = \mathcal{O}^{(1)}_2 (\eta',k) G^0_{2,2}(\eta,\eta',k) \label{22to12prop} \nonumber \\
& =  \Phi^0_{2,2}(\eta,k) \Phi^{0^*}_{1,2}(\eta',k)\theta(\eta - \eta') +  \Phi^0_{1,2}(\eta',k) \Phi^{0^*}_{2,2}(\eta,k)\theta(\eta' - \eta) \nonumber \\
& \quad - \frac {\Phi^0_{2,2}(\eta_0,k)}{\Phi^{0^*}_{2,2}(\eta_0,k)} \Phi^{0^*}_{2,2}(\eta,k) \Phi^{0^*}_{1,2}(\eta',k) \,, \\
\left (G^0_{2,2\to kk,2}\right )_{\text{eff}}(\eta,\eta',k) & = \mathcal{O}^{(kk)}_2 (\eta',k) G^0_{2,2}(\eta,\eta',k) \nonumber \\
& =  \Phi^0_{2,2}(\eta,k) \Phi^{*}_{kk}(\eta',k)\theta(\eta - \eta') +  \Phi_{kk}(\eta',k) \Phi^{0^*}_{2,2}(\eta,k)\theta(\eta' - \eta) \nonumber \\
& \quad - \frac {\Phi^0_{2,2}(\eta_0,k)}{\Phi^{0^*}_{2,2}(\eta_0,k)} \Phi^{0^*}_{2,2}(\eta,k) \Phi^{*}_{kk}(\eta',k)\,.
\end{align}
To arrive at these expressions we have made use of \eqref{DifferentialMapping}. The extra $\delta (\eta - \eta')$ term for $\left (G^0_{2,2\to 0,2}\right )_{\text{eff}}(\eta,\eta',k)$ comes from the fact that $\mathcal{O}^{(0)}_2$ is a second-order operator whereas the others that are relevant here are first-order. As in the spin-$1$ case, these mixed propagators do not necessarily vanish on the boundary but this is not an issue since again the ``true propagator" is $G_{2,2}^0$ and in any case the boundary terms will drop out when we convert to correlators. These mixed propagators rely on the mode functions of the non-dynamical modes and the explicit expressions for these mode functions are \cite{Lee:2016vti} 
\begin{align}
\Phi_{0,2}^0(\eta, k) = \Phi_{kk}(\eta,k) = \sqrt{\frac{\pi \alpha_1 \alpha_2}{6}} H k^2 e^{i \pi(\nu_2+1/2)/2}(-\eta)^{3/2}H^{(1)}_{\nu_2}(- k \eta) \,, \label{phi002modefunction}   \\
\Phi_{1,2}^0(\eta, k) = \sqrt{\frac {\pi k \alpha_1 \alpha_2}{24}}He^{i \pi(\nu_2+1/2)/2} (-k\eta)^{1/2}\left [k\eta (H^{(1)}_{\nu_2+1} (-k\eta) - H^{(1)}_{\nu_2-1}(-k\eta))-H^{(1)}_{\nu_2}(-k\eta)\right ]\,. \label{phi012modefunction}
\end{align}
We can now go ahead and find expressions for the non-dynamical modes. If we plug \eqref{sigma022solutionfullygeneral} into \eqref{sigma002itomsigma022} we obtain
\begin{align}
& \Phi^0_{0,2}(\eta,\bfk)  = K^0_{0,2}(\eta,k) \left (\frac {\Phi^{0^*}_{0,2}(\eta_0,k)}{\Phi^{0^*}_{2,2}(\eta_0,k)} \right ) \bar{\Phi}^0_{2,2}(\bfk)  \nonumber  \\
& \quad + i\int d\eta' \frac {\delta S_{\text{int}}}{\delta \Phi^0_{0,2}}(\eta',-\bfk) \left (G^0_{0,2\to 0,2}\right )_{\text{eff}}(\eta,\eta',k) + i\int d\eta' \frac {\delta S_{\text{int}}}{\delta \Phi^0_{1,2}}(\eta',-\bfk) \left (G^0_{0,2\to 1,2}\right )_{\text{eff}}(\eta,\eta',k) \nonumber \\
& \quad + i\int d\eta' \frac {\delta S_{\text{int}}}{\delta \Phi_{kk}}(\eta',-\bfk) \left (G^0_{0,2\to kk,2}\right )_{\text{eff}}(\eta,\eta',k) + i\int d\eta' \frac {\delta S_{\text{int}}}{\delta \Phi^0_{2,2}}(\eta',-\bfk) \left (G^0_{0,2\to 2,2}\right )_{\text{eff}}(\eta,\eta',k)\label{sigma002solutionfullygeneral},
\end{align}
where we have defined 
\begin{align}
\left (G^0_{0,2\to 0,2}\right )_{\text{eff}}(\eta,\eta',k) & = \mathcal{O}^{(0)}_2 (\eta,k) \left (G^0_{2,2\to 0,2}\right )_{\text{eff}}(\eta,\eta',k) - i \widetilde{\mathcal{O}}^{(0)}_{0}(\eta',k) \delta (\eta -\eta') \\
& = \Phi^0_{0,2}(\eta,k) \Phi^{0^*}_{0,2}(\eta',k) \theta(\eta-\eta') + (\eta \leftrightarrow \eta') - \frac {\Phi^0_{2,2}(\eta_0,k)}{\Phi^{0^*}_{2,2}(\eta_0,k)} \Phi^{0^*}_{0,2}(\eta,k) \Phi^{0^*}_{0,2}(\eta',k) \nonumber \\
& \quad - \partial_{\eta} \left (\frac {2iH^2\eta^2}{3(m_2^2-4H^2)(m_2^2-2H^2)}\delta' (\eta - \eta')\right )- \frac {2i(m_2^2-4H^2 - H^2k^2\eta^2)}{3(m_2^2-4H^2)(m_2^2-2H^2)} \delta (\eta -\eta') \,, \\
\left (G^0_{0,2\to 1,2}\right )_{\text{eff}}(\eta,\eta',k)& = \mathcal{O}^{(0)}_2(\eta,k)\left (G^0_{2,2\to 1,2}\right )_{\text{eff}}(\eta,\eta',k) - i\widetilde{\mathcal{O}}^{(0)}_1 (\eta',k)\delta (\eta -\eta') \label{02to12prop} \\
& = \Phi^0_{0,2}(\eta,k) \Phi^{0^*}_{1,2}(\eta',k) \theta (\eta - \eta') + \Phi^0_{1,2}(\eta',k) \Phi^{0^*}_{0,2}(\eta,k)\theta (\eta' - \eta) \nonumber \\
& \quad - \frac {\Phi^0_{2,2}(\eta_0,k)}{\Phi^{0^*}_{2,2}(\eta_0,k)} \Phi^{0^*}_{0,2}(\eta,k)\Phi^{0^*}_{1,2}(\eta',k) + \partial_{\eta} \left (\frac {2iH^2k\eta^2}{3(m_2^2-4H^2)(m_2^2-2H^2)}\delta (\eta - \eta')\right ) \,,\\
\left (G^0_{0,2\to kk,2}\right )_{\text{eff}}(\eta,\eta',k) & = \mathcal{O}^{(0)}_2(\eta,k) \left (G^0_{2,2\to kk,2}\right )_{\text{eff}}(\eta,\eta',k) - i\widetilde{O}^{(0)}_{kk}(\eta',k)\delta (\eta - \eta') \\
& = \Phi^0_{0,2}(\eta,k) \Phi^{*}_{kk}(\eta',k) \theta (\eta - \eta') + \Phi_{kk}(\eta',k) \Phi^{0^*}_{0,2}(\eta,k)\theta (\eta' - \eta) \nonumber \\
& \quad -\frac {\Phi^0_{2,2}(\eta_0,k)}{\Phi^{0^*}_{2,2}(\eta_0,k)} \Phi^{0^*}_{0,2}(\eta,k)\Phi^{*}_{kk}(\eta',k) + \frac {2iH^2\eta}{(m_2^2-4H^2)(m_2^2-2H^2)}\delta'(\eta - \eta') \nonumber \\
& \quad -\frac {i[3(m_2^2-6H^2)-2H^2k^2\eta^2]}{3(m_2^2-4H^2)(m_2^2-2H^2)} \delta (\eta - \eta')\,.
\end{align}
Here we see that derivatives of delta functions arise. This is ultimately since the operator $\mathcal{O}^{(0)}_2(\eta,k)$ is second order. In practice, when we encounter such an object while computing a wavefunction coefficient or correlator, we will integrate by parts to remove the derivatives from the delta functions. Moving to the other non-dynamical modes, if we plug \eqref{sigma022solutionfullygeneral} into \eqref{sigma012itosigma022}, we can write the solution of $\Phi^0_{1,2}$ as 
\begin{align}
   & \Phi^0_{1,2}(\eta,\bfk)  = K^0_{1,2}(\eta,k) \left (\frac {\Phi^{0^*}_{1,2}(\eta_0,k)}{\Phi^{0^*}_{2,2}(\eta_0,k)} \right )  \bar{\Phi}^0_{2,2}(\bfk)  \nonumber  \\
& \quad + i\int d\eta' \frac {\delta S_{\text{int}}}{\delta \Phi^0_{0,2}}(\eta',-\bfk)\left (G^0_{1,2\to 0,2}\right )_{\text{eff}}(\eta,\eta',k) + i\int d\eta' \frac {\delta S_{\text{int}}}{\delta \Phi^0_{1,2}}(\eta',-\bfk) \left (G^0_{1,2\to 1,2}\right )_{\text{eff}}(\eta,\eta',k) \nonumber \\
& \quad + i\int d\eta' \frac {\delta S_{\text{int}}}{\delta \Phi_{kk}}(\eta',-\bfk) \left (G^0_{1,2\to kk,2}\right )_{\text{eff}}(\eta,\eta',k) + i\int d\eta' \frac {\delta S_{\text{int}}}{\delta \Phi^0_{2,2}}(\eta',-\bfk) \left (G^0_{1,2\to 2,2}\right )_{\text{eff}}(\eta,\eta',k),\label{sigma012solutionfullygeneral}
\end{align}
where we have defined
\begin{align}
    \left (G^0_{1,2\to 1,2}\right )_{\text{eff}}(\eta,\eta',k) & = \mathcal{O}^{(1)}_2 (\eta,k) \left (G^0_{2,2\to 1,2}\right )_{\text{eff}}(\eta,\eta',k) - i\widetilde{\mathcal{O}}^{(1)}_1 (\eta',k) \delta (\eta - \eta') \\
    & = \Phi^0_{1,2}(\eta,k) \Phi^{0^*}_{1,2}(\eta',k) \theta (\eta - \eta') + (\eta \leftrightarrow \eta' ) - \frac {\Phi^0_{2,2}(\eta_0,k)}{\Phi^{0^*}_{2,2}(\eta_0,k)}\Phi^{0^*}_{1,2}(\eta,k) \Phi^{0^*}_{1,2}(\eta',k) \nonumber \\
    & \quad + \frac {i[3(m_2^2-4H^2)+4H^2k^2\eta^2]}{6(m_2^2-4H^2)(m_2^2-2H^2)} \delta (\eta - \eta')\,, \\
\left (G^0_{1,2\to kk,2}\right )_{\text{eff}}(\eta,\eta',k) & =  \mathcal{O}^{(1)}_{2} (\eta,k) \left (G^0_{2,2\to kk,2}\right )_{\text{eff}}(\eta,\eta',k)- i\widetilde{\mathcal{O}}^{(1)}_{kk} (\eta',k) \delta (\eta - \eta')\\
& = \Phi^0_{1,2}(\eta,k) \Phi^*_{kk}(\eta',k) \theta (\eta - \eta') + \Phi^{0^*}_{1,2}(\eta,k) \Phi_{kk}(\eta',k)  \theta (\eta' - \eta) \nonumber \\
& \quad - \frac {\Phi^0_{2,2}(\eta_0,k)}{\Phi^{0^*}_{2,2}(\eta_0,k)}\Phi^{1^*}_{1,2}(\eta,k) \Phi^{*}_{kk}(\eta',k) + \frac {2iH^2k\eta}{(m_2^2-4H^2)(m_2^2-2H^2)} \delta (\eta - \eta') \,.
\end{align}
The expressions for $\left (G^0_{1,2\to 0,2}\right )_{\text{eff}}(\eta,\eta',k)$ and $\left (G^0_{1,2\to 2,2}\right )_{\text{eff}}(\eta,\eta',k)$ follow from \eqref{02to12prop} and \eqref{22to12prop}. Finally for the $h=0$ modes, we can write the solution of $\Phi_{kk}$ by plugging \eqref{sigma022solutionfullygeneral} into  \eqref{sigmakkitosigma022}:
\begin{align}
   & \Phi_{kk}(\eta,\bfk)  = K_{kk}(\eta,k) \left (\frac {\Phi^{0^*}_{kk}(\eta_0,k)}{\Phi^{0^*}_{2,2}(\eta_0,k)} \right ) \bar{\Phi}^0_{2,2}(\bfk)  \nonumber  \\
& \quad + i\int d\eta' \frac {\delta S_{\text{int}}}{\delta \Phi^0_{0,2}}(\eta',-\bfk) \left (G^0_{kk,2\to 0,2}\right )_{\text{eff}}(\eta,\eta',k) + i\int d\eta' \frac {\delta S_{\text{int}}}{\delta \Phi^0_{1,2}}(\eta',-\bfk) \left (G^0_{kk,2\to 1,2}\right )_{\text{eff}}(\eta,\eta',k)\nonumber \\
& \quad + i\int d\eta' \frac {\delta S_{\text{int}}}{\delta \Phi_{kk}}(\eta',-\bfk) \left (G^0_{kk,2\to kk,2}\right )_{\text{eff}}(\eta,\eta',k) + i\int d\eta' \frac {\delta S_{\text{int}}}{\delta \Phi^0_{2,2}}(\eta',-\bfk) \left (G^0_{kk,2\to 2,2}\right )_{\text{eff}}(\eta,\eta',k)\,,\label{sigma0k2solutionfullygeneral}
\end{align}
where we have defined
\begin{align}
\left (G^0_{kk,2\to kk,2}\right )_{\text{eff}}(\eta,\eta',k) & =  \mathcal{O}^{(kk)}_{2} (\eta,k) \left (G^0_{2,2\to kk,2}\right )_{\text{eff}}(\eta,\eta',k)- i\widetilde{\mathcal{O}}^{(kk)}_{kk} (\eta',k) \delta (\eta - \eta')\\
 & = \Phi_{kk}(\eta,k) \Phi^*_{kk}(\eta',k) \theta (\eta - \eta') + (\eta \leftrightarrow \eta') - \frac {\Phi^0_{2,2}(\eta_0,k)}{\Phi^{0^*}_{2,2}(\eta_0,k)} \Phi^*_{kk}(\eta,k) \Phi^*_{kk}(\eta',k) \nonumber \\
& \quad + \frac {6iH^2}{(m_2^2-4H^2)(m_2^2-2H^2)} \delta (\eta - \eta') \,,
\end{align}
with the other expressions given above. To complete the calculation we need to plug these solutions into the action so that we can read off the rules for computing wavefunction coefficients. We will do this once we have derived the solutions for the modes with $h \neq 0$. 

\paragraph{$h=\pm 1$ modes} The helicity $h=\pm 1$ modes are significantly simpler, and the process is akin to the spin-1 case. There are two modes to consider $\Phi_{1,2}^{\pm 1}$ and $\Phi_{2,2}^{\pm 1}$ with the former non-dynamical and the latter dynamical. There are therefore only two equations \eqref{sigma112eom} and \eqref{sigma122eom} to deal with. We first solve \eqref{sigma112eom} for $\Phi^{\pm 1}_{1,2}$ to obtain
\begin{align}
    \Phi^{\pm 1}_{1,2} & = \frac 1{k^2 + (m_2^2-2H^2)a^2}\left [k\left (\partial_{\eta} + \frac 2{\eta}\right )\Phi^{\pm 1}_{2,2} - \frac {a^2}2 \frac {\delta S_{\text{int}}}{\delta \Phi^{\pm 1}_{1,2}}\right ]\,, \label{sigma112itosigma122}
\end{align}
which can be plugged back into \eqref{sigma122eom} to obtain
\begin{align}
\mathcal{O}^{\pm 1}_{2,2} \Phi^{\pm 1}_{2,2} = -\frac {2k}{a^2} \partial_{\eta}\left (- \frac {a^2}{2(k^2 + (m_2^2-2H^2)a^2)} \frac {\delta S_{\text{int}}}{\delta \Phi^{\pm 1}_{1,2}}\right ) - \frac {\delta S_{\text{int}}}{\delta \Phi^{\pm 1}_{2,2}}\,,\label{sigma122dynamical}
\end{align}
where the operator $\mathcal{O}^{\pm 1}_{2,2} \Phi^{\pm 1}_{2,2}$ is defined as
\begin{align}
\mathcal{O}^{\pm 1}_{2,2} & := -\frac {2(m_2^2-2H^2)}{k^2+(m_2^2-2H^2)a^2} \partial_{\eta}^2 - \frac 4{\eta} \left (\frac {(m_2^2-2H^2)^2a^2}{(k^2+(m_2^2-2H^2)a^2)^2}\right ) \partial_{\eta} \nonumber \\
& \qquad -2\left [\frac {(m_2^2-2H^2)[(m_2^2-4H^2)(m_2^2-2H^2)a^4 + 2k^2(m_2^2-5H^2)a^2 + k^4]}{(k^2+(m_2^2-2H^2)a^2)^2}\right ]\,.
\end{align}
We can now solve \eqref{sigma122dynamical}, but we first turn off the interactions and solve the free theory to obtain the mode function \cite{Lee:2016vti}
\begin{align}
    \Phi^{\pm 1}_{2,2} (\eta,k) & = e^{i \pi(\nu_2+1/2)/2} \sqrt{\frac {\pi \alpha_1}{32}} (-\eta)^{-1/2} \left [k\eta (H^{(1)}_{\nu_2+1}(-k\eta) - H^{(1)}_{\nu_2-1}(-k\eta))- 3H^{(1)}_{\nu_2}(-k\eta)\right ]\,, \label{phi122modefunction}
\end{align}
where we have imposed Bunch-Davies vacuum conditions. We can then construct the bulk-boundary and bulk-bulk propagators for this helicity-$\pm 1$ dynamical mode. The bulk-boundary propagator is
\begin{equation}
    K^{\pm 1}_{2,2}(\eta,k) = \frac {\Phi^{\pm 1}_{2,2}(\eta,k)}{\Phi^{\pm 1}_{2,2}(\eta_0,k)} \,,
\end{equation}
which satisfies
\begin{equation}
    \mathcal{O}^{{\pm 1}}_{2,2}K^{\pm 1}_{2,2} = 0, \qquad K^{\pm 1}_{2,2}(\eta_0,k) = 1,\qquad \lim_{\eta \to -\infty(1-i\epsilon)}K^{\pm 1}_{2,2} (\eta,k) = 0 \,.
\end{equation}
The bulk-bulk propagator is
\begin{equation}
    G^{\pm 1}_{2,2}(\eta,\eta',k) = \Phi^{\pm 1}_{2,2}(\eta,k) \Phi^{{\pm 1}^*}_{2,2}(\eta',k) \theta (\eta - \eta') + (\eta \leftrightarrow \eta') - \frac {\Phi^{\pm 1}_{2,2}(\eta_0,k)}{\Phi^{{\pm 1}^*}_{2,2}(\eta_0,k)} \Phi^{{\pm 1}^*}_{2,2}(\eta,k) \Phi^{{\pm 1}^*}_{2,2}(\eta',k) \,,
\end{equation}
and it satisfies
\begin{equation} 
    \mathcal{O}^{\pm 1}_{2,2}(\eta,k) G^{\pm 1}_{2,2}(\eta,\eta',k) = i\delta (\eta - \eta'), \qquad \lim_{\eta,\eta' \to \eta_0} G^{\pm 1}_{2,2}(\eta,\eta',k) = 0, \qquad \lim_{\eta,\eta'\to -\infty(1-i\epsilon)}G^{\pm 1}_{2,2}(\eta,\eta',k) = 0 \,.
\end{equation}
With these propagators at hand, we can formally solve \eqref{sigma122dynamical} with the interactions turned back on to find
\begin{align}
\Phi^{\pm 1}_{2,2}(\eta,\bfk) & = K^{\pm 1}_{2,2} (\eta,k) \bar{\Phi}^{\pm 1}_{2,2}(\bfk) + i \int d\eta' \frac {\delta S_{\text{int}}}{\delta \Phi^{\pm 1}_{2,2}}(\eta',-\bfk) \left (G^{\pm 1}_{2,2\to 2,2}\right )_{\text{eff}}(\eta,\eta',k) \nonumber \\
&  - i\int d\eta' \frac {2k}{a^2(\eta')} \partial_{\eta'}\left (\frac {a^2(\eta')}{2(k^2 + (m_2^2-2H^2)a^2(\eta'))} \frac {\delta S_{\text{int}}}{\delta \Phi^{\pm 1}_{1,2}}(\eta',-\bfk) \right) \left (G^{\pm 1}_{2,2\to 1,2}\right )_{\text{eff}}(\eta,\eta',k) \,,
\end{align}
where for ease of notation we have defined $\left (G^{\pm 1}_{2,2\to 2,2}\right )_{\text{eff}}(\eta,\eta',k) = G^{\pm 1}_{2,2}(\eta,\eta',k)$. We can now in principle use this solution to write wavefunction coefficients but as we have emphasised before, it is not particularly useful to have differential operators acting on the variation of the interacting action. Instead, we integrate by parts to move the differential operator onto the bulk-bulk propagator, and then define a new propagator. Our solution can then be written as  
\begin{align}
\Phi^{\pm 1}_{2,2}(\eta,\bfk) & = K^{\pm 1}_{2,2} (\eta,k) \bar{\Phi}^{\pm 1}_{2,2}(\bfk) + i \int d\eta' \frac {\delta S_{\text{int}}}{\delta \Phi^{\pm 1}_{2,2}}(\eta',-\bfk) \left (G^{\pm 1}_{2,2\to 2,2}\right )_{\text{eff}}(\eta,\eta',k) \nonumber \\
& \quad + i\int d\eta' \frac {\delta S_{\text{int}}}{\delta \Phi^{\pm 1}_{1,2}}(\eta',-\bfk) \left (G^{\pm 1}_{2,2\to 1,2}\right )_{\text{eff}}(\eta,\eta',k) \,,\label{sigma122solutionfullygeneral}
\end{align}
where we have defined the mixed propagator
\begin{align} \label{Spin2h1Mixed}
\left (G^{\pm 1}_{2,2\to 1,2}\right )_{\text{eff}}(\eta,\eta',k)  & = \frac {k}{k^2+(m_2^2-2H^2)a^2(\eta')} \left (\partial_{\eta'} + \frac 2{\eta'}\right ) G^{\pm 1}_{2,2}(\eta,\eta',k) \\
& = \Phi^{\pm 1}_{2,2}(\eta,k) \Phi^{\pm 1^*}_{1,2}(\eta',k) \theta (\eta - \eta') + \Phi^{\pm 1^*}_{2,2}(\eta,k) \Phi^{\pm 1}_{1,2}(\eta',k) \theta (\eta' - \eta) \nonumber \\
& \quad -\frac {\Phi^{\pm 1}_{2,2}(\eta_0,k)}{\Phi^{\pm 1^*}_{2,2}(\eta_0,k)}\Phi^{\pm 1^*}_{2,2}(\eta) \Phi^{\pm 1^*}_{1,2}(\eta')\,,
\end{align}
and have introduced the mode function of the non-dynamical mode \cite{Lee:2016vti}
\begin{align}
\Phi^{\pm 1}_{1,2} (\eta,k) & = \sqrt{\frac {\pi \alpha_1}{8}}e^{i\pi (\nu_2+1/2)/2} k(-\eta)^{1/2}H^{(1)}_{\nu_2}(-k\eta) \,, \label{phi112modefunction}
\end{align}
which follows from \eqref{sigma112itosigma122} with the interactions turned off. Note that in contrast to many of the $h=0$ cases, this effective propagator does not contain a delta function. Plugging this solution for $\Phi^{\pm 1}_{2,2}$ back into \eqref{sigma112itosigma122}, we obtain
\begin{align}
\Phi^{\pm 1}_{1,2}(\eta,\bfk) & = K^{\pm 1}_{1,2}(\bfk) \left (\frac {\Phi^{\pm 1^*}_{1,2}(\eta_0,k)}{\Phi^{\pm 1 ^*}_{2,2}(\eta_0,k)} \right ) \bar{\Phi}^{\pm 1}_{2,2}(\bfk)  + i\int d\eta' \frac {\delta S_{\text{int}}}{\delta \Phi^{\pm 1}_{2,2}}(\eta'-\bfk) \left (G^{\pm 1}_{1,2\to 2,2}\right )_{\text{eff}}(\eta,\eta',k)  \nonumber \\
& \quad + i\int d\eta' \frac {\delta S_{\text{int}}}{\delta \Phi^{\pm 1}_{1,2}}(\eta',-\bfk) \left (G^{\pm 1}_{1,2\to 1,2}\right )_{\text{eff}}(\eta,\eta',k) \,,\label{sigma112solutionfullygeneral}
\end{align}
where we have defined
\begin{align}
 \left (G^{\pm 1}_{1,2\to 1,2}\right )_{\text{eff}}(\eta,\eta',k) & = \frac {k}{k^2+(m_2^2-2H^2)a^2(\eta)} \left (\partial_{\eta} + \frac 2{\eta}\right )\left (G^{\pm 1}_{2,2\to 1,2}\right )_{\text{eff}}(\eta,\eta',k)  \nonumber \\
 & \quad + \frac {ia^2(\eta)}{2[k^2+(m_2^2-2H^2)a^2(\eta)]}\delta (\eta - \eta') \label{1to1intermediate} \\
 & = \Phi^{\pm 1}_{1,2}(\eta,k) \Phi^{\pm 1^*}_{1,2}(\eta',k) \theta (\eta - \eta') + (\eta \leftrightarrow \eta') - \frac {\Phi^{\pm 1}_{2,2}(\eta_0,k)}{\Phi^{\pm 1^*}_{2,2}(\eta_0,k)} \Phi^{\pm 1^*}_{1,2}(\eta,k) \Phi^{\pm 1^*}_{1,2}(\eta',k) \nonumber \\
 & \quad + \frac i{2(m_2^2-2H^2)} \delta (\eta - \eta') \,.
\end{align}
As we have now seen many times, we see that the non-local delta function that appears in \eqref{1to1intermediate} cancels with the non-local one generated by the differential operator acting on $\left (G^{\pm 1}_{2,2\to 1,2}\right )_{\text{eff}}(\eta,\eta',k)$. We are therefore left with an effective propagator that is constructed out of the mode functions of the non-dynamical mode only, with the addition of a delta function that is local. 
\paragraph{$h=\pm 2$ modes} Finally, we have the $h=\pm 2$ modes in which case there are no non-dynamical modes. The only relevant equation of motion is therefore \eqref{sigma222eom}. In the absence of any interactions, the solution to this equation with Bunch-Davies vacuum conditions is \cite{Lee:2016vti}
\begin{align}
    \Phi^{\pm 2}_{2,2} & = \sqrt{\frac {\pi k}{8 H^2}}e^{i\pi(\nu_2+1/2)/2} (-k\eta)^{-1/2}H^{(1)}_{\nu_2}(-k\eta)\,. \label{phi222modefunction}
\end{align}
Using this mode function we can construct the bulk-boundary and bulk-bulk propagators. The bulk-boundary propagator is
\begin{equation}
    K^{\pm 2}_{2,2}(\eta,k) = \frac {\Phi^{\pm 2}_{2,2}(\eta,k)}{\Phi^{\pm 2}_{2,2}(\eta_0,k)} \,,
\end{equation}
which satisfies
\begin{equation}
    \mathcal{O}^{{\pm 2}}_{2,2}K^{\pm 2}_{2,2} = 0, \qquad K^{\pm 2}_{2,2}(\eta_0,k) = 1,\qquad \lim_{\eta \to -\infty(1-i\epsilon)}K^{\pm 2}_{2,2} (\eta,k) = 0 \,,
\end{equation}
where we defined the operator $\mathcal{O}^{\pm 2}_{2,2}$ as
\begin{align}
  \mathcal{O}^{\pm 2}_{2,2}:= -\left (\frac 2{a^2} \partial_{\eta}^2 -\frac {4 H}{a} \partial_{\eta} + 2(m_2^2-4H^2) + \frac {2k^2}{a^2}\right )\,.
\end{align}
The bulk-bulk propagator is
\begin{equation}
    G^{\pm 2}_{2,2}(\eta,\eta',k) = \Phi^{\pm 2}_{2,2}(\eta,k) \Phi^{{\pm 2}^*}_{2,2}(\eta',k) \theta (\eta - \eta') + (\eta \leftrightarrow \eta') - \frac {\Phi^{\pm 2}_{2,2}(\eta_0,k)}{\Phi^{{\pm 2}^*}_{2,2}(\eta_0,k)} \Phi^{{\pm 2}^*}_{2,2}(\eta,k) \Phi^{{\pm 2}^*}_{2,2}(\eta',k) \,,
\end{equation}
and it satisfies
\begin{equation} 
    \mathcal{O}^{\pm 2}_{2,2}(\eta,k) G^{\pm 2}_{2,2}(\eta,\eta',k) = i\delta (\eta - \eta'), \qquad \lim_{\eta,\eta' \to \eta_0} G^{\pm 2}_{2,2}(\eta,\eta',k) = 0, \qquad \lim_{\eta,\eta'\to -\infty(1-i\epsilon)}G^{\pm 2}_{2,2}(\eta,\eta',k) = 0 \,.
\end{equation}
Now we can write the solution to \eqref{sigma222eom} with interactions turned on as
\begin{align}
\Phi^{\pm 2}_{2,2}(\eta,\bfk) = K^{\pm 2}_{2,2}(\eta,k) \bar{\Phi}^{\pm 2}_{2,2}(\bfk) + i\int d\eta' \frac {\delta S_{\text{int}}}{\delta \Phi^{\pm 2}_{2,2}}(\eta',-\bfk) \left (G^{\pm 2}_{2,2\to 2,2}\right )_{\text{eff}}(\eta,\eta',k)\,, \label{sigma222solutionfullygeneral}
\end{align}
where again for consistency of notation we have defined $\left (G^{\pm 2}_{2,2\to 2,2}\right )_{\text{eff}}(\eta,\eta',k) = G^{\pm 2}_{2,2}(\eta,\eta',k)$. Since there are no non-dynamical modes in this case, we only have the familiar propagator. 

All the solutions we have found in this appendix, namely, \eqref{sigma022solutionfullygeneral}, \eqref{sigma002solutionfullygeneral}, \eqref{sigma012solutionfullygeneral},  \eqref{sigma0k2solutionfullygeneral}, \eqref{sigma122solutionfullygeneral}, \eqref{sigma112solutionfullygeneral}, and \eqref{sigma222solutionfullygeneral} are summarised by \eqref{spin2generalsolution} in the main text.
\paragraph{\ref{CCstep4} Plug the solutions back into action.} The final step to extract wavefunction coefficients is to plug these solutions back into the action. As we did in the spin-$1$ case, we will focus on contributions that involve a single exchange of the spinning field, and therefore we need to plug our solutions into the interacting part of the action and the free theory of the spinning field. Since the free theory is quadratic in the helicity modes $\Phi^h_{n,S}$, and we assume the interactions are linear in the helicity modes, we have, after integration by parts
\begin{align}
    S& = \frac 12 \sum_{h=-2}^2 \sum_{n} \int d\eta \int_{\bfk} \Phi^h_{n,2}(\eta,\bfk) \frac {\delta S_2}{\delta \Phi^h_{n,2}}(\eta, \bfk) + \text{boundary terms} \nonumber \\
    & \quad + \sum_{h=-2}^2\sum_{n}\int d\eta \int_{\bfk}\Phi^h_{n,2} (\eta,\bfk)\frac {\delta S_{\text{int}}}{\delta \Phi^h_{n,2}}(\eta,\bfk)\,,\label{generalspin2linearinteraction}
\end{align}
where the sum over $n$ sums over all modes including any traces. The boundary terms are important for the wavefunction coefficients of the spinning field, but they are irrelevant for the inflaton wavefunction coefficients. When we plug in the solutions, summarised by \eqref{spin2generalsolution} in the main text, we obtain the result presented in \eqref{Feynmanspin2} in the main text.
\paragraph{Explicit details of the operators $\mathcal{O}^{(i)}_j$} 
We denote the differential operators as
\begin{align}
\mathcal{O}^{(i)}_j = \frac 1{A^2}\left (b^{(i)}_j \partial_{\eta}^2 + c^{(i)}_j \partial_{\eta} + d^{(i)}_j\right )\,,
\end{align}
where the various coefficients are 
\begin{align}
    A & = 9(\alpha_1\alpha_2)^{-1} + 12H^2k^2\eta^2 \alpha_2^{-1}+ 4H^4k^4\eta^4 \,, \\
    b^{(0)}_2 & = -4\sqrt 3 H^4k^2\eta^4 A \,, \\
    c^{(0)}_2 & =  -4\sqrt 3 H^4k^2\eta^3 \left [45(\alpha_1\alpha_2)^{-1} + 36H^2k^2\eta^2\alpha_2^{-1} + 4H^4k^4\eta^4\right ]\,, \\
    d^{(0)}_2 & = 2\sqrt 3 H^2k^2 \eta^2 \alpha_1^{-1}\left [9\alpha_2^{-1}(m_2^2-10H^2) + 12H^2k^2\eta^2 \alpha_2^{-1} + 4H^4k^4\eta^4\right ]\,, \\
    b^{(0)}_0 & = 2H^2\eta^2 (3+4H^2k^2\eta^2\alpha_1)A\,, \\
    c^{(0)}_0 & = 12H^2\eta \left [9(\alpha_1\alpha_2)^{-1} + 24H^2k^2\eta^2 \alpha_2^{-1}  + 4H^4k^4\eta^4(3m_2^2-14H^2)\alpha_1\right ]\,,\\
    d^{(0)}_0 & = 54\alpha_1^{-1}\alpha_2^{-2} + 18 H^2k^2\eta^2 \alpha_2^{-1}(5m_2^2-26H^2)+ 48H^4k^4\eta^4 \alpha_1 \alpha_2^{-1} (m_2^2-6H^2) + 8H^6k^6\eta^6\,, \\
    c^{(0)}_1 & = -2H^2k\eta^2 (3+2H^2k^2\eta^2\alpha_1)A\,, \\
    d^{(0)}_1 & = -4H^2k\eta \alpha_1 \left [27\alpha_1^{-2}\alpha_2^{-1}+45H^2k^2\eta^2(\alpha_1\alpha_2)^{-1} + 24H^4k^4\eta^4(m_2^2-5H^2) + 4H^6k^6\eta^6\right ]\,, \\
    c^{(0)}_{kk} & = -6H^2\eta (3+4H^2k^2\eta^2 \alpha_1) A\,, \\
    d^{(0)}_{kk} & = 3\left [27(\alpha_1\alpha_2)^{-1}(m_2^2-6H^2) + 18 H^2k^2\eta^2\alpha_2^{-1}(3m_2^2-22H^2)\right . \nonumber \\
    & \qquad \qquad \qquad \left . + 12 H^4k^4\eta^4 \alpha_1(3m_2^4-28m_2^2H^2+76H^4) + 8H^6k^6\eta^6 \right ]\,, \\
    c^{(1)}_2 & = -2\sqrt 3 H^2k\eta^2 (3\alpha_2^{-1}+2H^2k^2\eta^2)A\,, \\
    d^{(1)}_2 & = -4\sqrt 3 H^2k\eta (3\alpha_2^{-1}+H^2k^2\eta^2)A\,, \\
    c^{(1)}_0 & = 2H^2k\eta^2 (3 + 2H^2k^2\eta^2 \alpha_1) A\,,\\
    d^{(1)}_0 & = -4H^4k^3\eta^3 \alpha_1 A\,, \\
    d^{(1)}_1 & = -\frac {\alpha_1}2 A(A+24H^4k^2\eta^2)\,, \\
    d^{(1)}_{kk} & = -6H^2k\eta (3 + 2H^2k^2\eta^2 \alpha_1)A \,,\\
    c^{(kk)}_2 & = -12\sqrt 3 H^4k^2\eta^3 A\,, \\
    d^{(kk)}_2 & = 2\sqrt 3 H^2k^2\eta^2 [3(m_2^2-6H^2)+2H^2k^2\eta^2]A\,, \\
    c^{(kk)}_0 & = 6H^2\eta (3+4H^2k^2\eta^2 \alpha_1)A\,, \\
    d^{(kk)}_0 & = \alpha_1 A [A - 2H^2k^2\eta^2 (3\alpha_1^{-1}+2H^2k^2\eta^2)]\,,\\
    d^{(kk)}_1 & = -6H^2k\eta (3 + 2H^2k^2\eta^2 \alpha_1)A \,,\\
    d^{(kk)}_{kk} & = -18H^2 (3 + 4H^2k^2\eta^2 \alpha_1) A\,,\\
    b^{(2)}_2 & = -18H^2\eta^2 (\alpha_1\alpha_2)^{-1} A\,, \\
    c^{(2)}_2 & = -36H^2\eta (\alpha_1\alpha_2)^{-1}[9(\alpha_1\alpha_2)^{-1} -4H^4k^4\eta^4]\,, \\
    d^{(2)}_2 & = -18(\alpha_1\alpha_2)^{-1}[9\alpha_1^{-1}\alpha_2^{-2}+3H^2k^2\eta^2 \alpha_2^{-1}(7m_2^2-38H^2) \nonumber \\ & \qquad \qquad \qquad \qquad \qquad + 16H^4k^4\eta^4(m_2^2-5H^2) + 4H^6k^6\eta^6]\,, \\
    b^{(2)}_0& = -4\sqrt 3 H^4k^2\eta^4 A\,, \\
    c^{(2)}_0 & = -4\sqrt 3 H^4k^2\eta^3[27(\alpha_1\alpha_2)^{-1}+ 12H^2k^2\eta^2 \alpha_2^{-1}- 4H^4k^4\eta^4]\,, \\
    d^{(2)}_0 & = 2\sqrt 3 H^2k^2\eta^2\alpha_2^{-1}(A+48H^4k^2\eta^2)\,, \\
    c^{(2)}_1 & = 2\sqrt 3 H^2k\eta^2 (3\alpha_2^{-1} + 2H^2k^2\eta^2)A\,, \\
    d^{(2)}_1 & = -4\sqrt 3 H^4k^3\eta^3(A-72H^2\alpha_2^{-1})\,, \\
    c^{(2)}_{kk} & = 12\sqrt 3 H^4k^2\eta^3A\,, \\
    d^{(2)}_{kk} & = 2\sqrt 3 H^2k^2\eta^2[27m_2^2(\alpha_1\alpha_2)^{-1} + 18H^2k^2\eta^2\alpha_2^{-1}(3m_2^2-10H^2) \nonumber \\ & \qquad \qquad \qquad \qquad \qquad    + 12H^4k^4\eta^4(3m_2^2-16H^2) + 8H^6k^6\eta^6] \,.
\end{align}
When $k^2 \eta^2 \gg 1$, we see that $A > 0$. As we approach $k^2 \eta^2 \rightarrow 0$ we need to maintain $A>0$ otherwise we will encounter singularities in the equations of motion. A necessary and sufficient condition for $A>0$ at all times is $m_2^2 \geq 4 H^2$ which corresponds to $m^2 \geq 2H^2$. This is precisely the Higuchi bound \cite{Higuchi:1986py}. We didn't have such concerns in the spin-$1$ case and indeed in that case there is no Higuchi bound beyond $m^2 > 0$.  

\section{More details on converting to cosmological correlators}\label{WavefunctionToCorrelatorsDetails}
In this appendix we provide further details on how we go from wavefunction coefficients due to CC exchanges to cosmological correlators. In Section \ref{FromWavefunctionToCorrelatorCC} we worked with a concrete example of the trispectrum due to spin-$1$ exchange, while in this appendix we work with general interactions and consider $(p+q)$-point functions while still focusing on single-exchange diagrams. This will be enough to illustrate the main points we want to convey. The extension to more complicated diagrams is straightforward if not a little tedious. We take the momentum-space action to be 
\begin{align}
S_{\text{int}} =& \int d\eta \int _{\bfk, \bfk_1, \ldots, \bfk_p} \sum_{h} f_{(p)} [\pi](\eta,\bfk_1,...,\bfk_p) \Phi^h_{n,S}(\eta,\bfk) \nonumber \\
+&\int d\eta  \int _{\bfk, \bfk_{1}, \ldots, \bfk_{q}} \sum_{h} g_{(q)}[\pi](\eta,\bfk_{1},...,\bfk_{q}) \Phi^h_{m,S}(\eta,\bfk) \,,
\end{align}
where $f_{(p)}[\pi]$ refers to an operator on $\pi$ that is of $p$-th order, similarly $g_{(q)}[\pi]$ of $q$-th order. Any contractions between momenta and polarisations tensors are contained in these objects. We have assumed that the two different vertices contain modes of the spinning field with the same helicity such that these vertices can be glued together in Feynman diagrams, but we have not assumed that they are identical modes i.e. we can have $n \neq m$. As we seen, there are mixed propagators that connect modes with $n \neq m$ enabling us to draw diagrams that connect such vertices. We have integrated by parts to remove any derivatives acting on the spinning field. We assume that any resulting boundary terms vanish which is almost always the case when we build interactions within the EFToI due to the shift symmetry of $\pi$. The correlator is given by the Born rule
\begin{align} 
\langle \pi(\bfk_1) \ldots \pi(\bfk_{p+q})\rangle = \frac{\int \mathcal{D}  \pi \mathcal{D} \Phi  ~ \pi(\bfk_1) \ldots \pi(\bfk_{p+q}) \Psi[\pi, \Phi] \Psi^{*}[\pi, \Phi]}{\int \mathcal{D} \pi \mathcal{D}  \Phi \Psi[\pi, \Phi] \Psi^{*}[\pi, \Phi]} \,,
\end{align}
with the wavefunction parametrised by \eqref{WavefunctionParam}. The wavefunction only depends on the boundary data of the dynamical modes. By expanding around the Gaussian and performing the relevant Gaussian integrals, we arrive at
\begin{align}
\langle \pi(\bfk_1)...\pi(\bfk_{p+q})\rangle & = \frac {\rho_{p+q}^{\pi...\pi}(\bfk_1,...,\bfk_{p+q}) + \sum_{h}\left (\dfrac {\rho_{p+1,h}^{\pi...\pi\Phi}(\bfk_1,...,\bfk_p,\bfk)\rho_{q+1,h}^{\pi...\pi\Phi}(\bfk_{p+1},...,\bfk_{p+q},-\bfk)}{\rho_{2,h}^{\Phi\Phi}(k)}+\text{perms}\right )}{\rho_2^{\pi\pi}(k_1)...\rho_2^{\pi\pi}(k_{p+q})}, \label{correlatorintermsofrho}
\end{align}
where the permutations refer to distinct channels (i.e. distinct exchange momenta) and by momentum conservation $\bfk = -(\bfk_1 + \ldots + \bfk_p) = (\bfk_{p+1} + \ldots + \bfk_{p+q})$. We have established in the main text that, in the case where $p\neq q$,\footnote{There is a nuance when it comes to the case $m=n$ because $\dfrac {\delta S_{\text{int}}}{\delta \Phi^h_{n,S}}$ would have been the sum of $f$ and $g$. However, we are only dealing with the ``cross term" that involves both $f$ and $g$ here. We will address the case when $p=q$ later.}
\begin{align}
\rho_{p+q}^{\pi...\pi} (\bfk_1,...,\bfk_{p+q}) & = -\int d\eta d\eta' \sum_{h} f_{(p)}[K_{\pi}](\eta, \bfk_1,...,\bfk_p)\left (G^h_{n,2\to m,2}\right )_{\text{eff}} (\eta,\eta',k)g_{(q)}[K_{\pi}](\eta', \bfk_{p+1},...,\bfk_{p+q})\nonumber \\
& \quad + \text{perms of }\{\bfk_1,...,\bfk_{p+q}\} + \text{c.c.}\,, \label{rhop+q}
\end{align}
where again the functions $f_{(p)}$ and $g_{(p)}$ contain any contractions between momenta and polarisation tensors and we have summed over the helicities. Recall that $\rho(\{ \bfk \}) = \psi(\{ \bfk \}) + \psi^{*}(- \{ \bfk \})$ so the second term in \eqref{rhop+q} is not directly the complex conjugate of the first, however once we sum over helicities it does indeed become the complex conjugate (we are again assuming parity-even theories). To compute the correlator we also need wavefunction coefficients with the spinning field on the external lines. Such contributions that are leading order in the couplings, come from the interacting part of the action only. We therefore have 
\begin{align}
    \rho_{p+1,h}^{\pi...\pi\Phi}(\bfk_1,...,\bfk_p,\bfk) & = i\int d\eta f_{(p)}[K_{\pi}](\eta,\bfk_1,...,\bfk_p)  \frac {\Phi^{h^*}_{n,S}(\eta,k)}{\Phi^{h^*}_{S,S}(\eta_0,k)} + \text{perms of }\{\bfk_1,...,\bfk_p\} +\text{c.c.} \,, \label{rhop} \\
    \rho_{q+1,h}^{\pi...\pi\Phi} (\bfk_{p+1},...,\bfk_{p+q},\bfk) & = i\int d\eta g_{(q)}[K_{\pi}](\eta,\bfk_{p+1},...,\bfk_{p+q}) \frac {\Phi^{h^*}_{m,S}(\eta,k)}{\Phi^{h^*}_{S,S}(\eta_0,k)} \nonumber \\
    & \quad + \text{perms of }\{\bfk_{p+1},...,\bfk_{p+q}\} +\text{c.c.}\,,\label{rhoq}
\end{align}
where we have used the free theory solution of $\Phi^h_{n,S}(\eta,\bfk)$:
\begin{align}
\Phi^{h,\text{free}}_{n,S}(\eta,\bfk) & = \frac {\Phi^{h^*}_{n,S}(\eta,k)}{\Phi^{h^*}_{S,S}(\eta_0,k)} \bar{\Phi}^h_{S,S}(\bfk)\,.
\end{align}
Recall that there is always a dependence on the boundary data of the dynamical modes $\Phi_{S,S}^h$ since in the Born rule we are integrating over these modes only. Now we substitute \eqref{rhop+q}, \eqref{rhop}, \eqref{rhoq}, and 
\begin{align}
    \rho_2^{\pi\pi}(k) & = [\pi(\eta_0,k) \pi^*(\eta_0,k)]^{-1} \,, \\
    \rho_{2,h}^{\Phi\Phi}(k) & = [\Phi^h_{S,S}(\eta_0,k) \Phi^{h^*}_{S,S}(\eta_0,k)]^{-1} \,,
\end{align}
into \eqref{correlatorintermsofrho} to obtain
\begin{align}
    \langle \pi(\bfk_1)...\pi(\bfk_{p+q})\rangle & = -\prod_{a=1}^{p+q} (\pi(\eta_0,k_a)\pi^*(\eta_0,k_a)) \int d\eta d\eta' \sum_{h} \left (f_{(p)}[K_{\pi}](\eta,\bfk_1,...,\bfk_p) + \text{perms of }\{\bfk_1,...,\bfk_p \}\right ) \times \nonumber \\
   &\quad  \left [\left (G^h_{n,2\to m,2}\right )_{\text{eff}}(\eta,\eta',k) + \frac {\Phi^{h^*}_{n,S}(\eta,k)}{\Phi^{h^*}_{S,S}(\eta_0,k)} \frac {\Phi^{h^*}_{m,S}(\eta',k)}{\Phi^{h^*}_{S,S}(\eta_0,k)} (\Phi^h_{S,S}(\eta_0,k) \Phi^{h^*}_{S,S}(\eta_0,k))\right ] \times \nonumber \\
   &\quad \left (g_{(q)}[K_{\pi}](\eta',\bfk_{p+1},...,\bfk_{p+q}) + \text{perms of }\{\bfk_{p+1},...,\bfk_{p+q}\}\right )\label{++diagram}\\
   & \quad -\prod_{a=1}^{p+q} (\pi(\eta_0,k_a)\pi^*(\eta_0,k_a)) \int d\eta d\eta'  \sum_{h} \left (f^*_{(p)}[K_{\pi}](\eta,\bfk_1,...,\bfk_p) + \text{perms of }\{\bfk_1,...,\bfk_p \}\right ) \times \nonumber \\
   &\quad  \left [\left (G^h_{n,2\to m,2}\right )^*_{\text{eff}}(\eta,\eta',k) + \frac {\Phi^{h}_{n,S}(\eta,k)}{\Phi^{h}_{S,S}(\eta_0,k)}  \frac {\Phi^{h}_{m,S}(\eta',k)}{\Phi^{h}_{S,S}(\eta_0,k)} (\Phi^h_{S,S}(\eta_0,k) \Phi^{h^*}_{S,S}(\eta_0,k))\right ] \times \nonumber \\
   &\quad \left (g^*_{(q)}[K_{\pi}](\eta',\bfk_{p+1},...,\bfk_{p+q}) + \text{perms of }\{\bfk_{p+1},...,\bfk_{p+q}\}\right )\label{--diagram} \\
   & \quad +\prod_{a=1}^{p+q} (\pi(\eta_0,k_a)\pi^*(\eta_0,k_a))\int d\eta d\eta' \sum_{h} \left (f_{(p)}[K_{\pi}](\eta,\bfk_1,...,\bfk_p) + \text{perms of }\{\bfk_1,...,\bfk_p \}\right ) \times \nonumber \nonumber \\
   & \quad \frac {\Phi^{h^*}_{n,S}(\eta,k)}{\Phi^{h^*}_{S,S}(\eta_0,k)} \frac {\Phi^h_{m,S}(\eta',k)}{\Phi^h_{S,S}(\eta_0,k)}\left (\Phi^h_{S,S}(\eta_0,k) \Phi^{h^*}_{S,S}(\eta_0,k) \right ) \times \nonumber \\
   & \quad \left (g^*_{(q)}[K_{\pi}](\eta',\bfk_{p+1},...,\bfk_{p+q}) + \text{perms of }\{\bfk_{p+1},...,\bfk_{p+q}\}\right ) \label{+-diagram} \\
   & \quad +\prod_{a=1}^{p+q} (\pi(\eta_0,k_a)\pi^*(\eta_0,k_a))\int d\eta d\eta' \sum_{h} \left (f^*_{(p)}[K_{\pi}](\eta,\bfk_1,...,\bfk_p) + \text{perms of }\{\bfk_1,...,\bfk_p \}\right ) \times \nonumber \nonumber \\
   & \quad \frac {\Phi^{h}_{n,S}(\eta,k)}{\Phi^{h}_{S,S}(\eta_0,k)} \frac {\Phi^{h^*}_{m,S}(\eta',k)}{\Phi^{h^*}_{S,S}(\eta_0,k)}\left (\Phi^h_{S,S}(\eta_0,k) \Phi^{h^*}_{S,S}(\eta_0,k) \right ) \times \nonumber \\
   & \quad \left (g_{(q)}[K_{\pi}](\eta',\bfk_{p+1},...,\bfk_{p+q}) + \text{perms of }\{\bfk_{p+1},...,\bfk_{p+q}\}\right ) \label{-+diagram}\\
   &\quad + \text{other channels}/,, \nonumber 
\end{align}
where the dependence on the mode functions of the spinning field in \eqref{++diagram}, \eqref{--diagram}, \eqref{+-diagram}, and \eqref{-+diagram} correspond to the $++$, $--$, $+-$, and $-+$ effective Schwinger-Keldysh propagators given in the main text in Section \ref{FromWavefunctionToCorrelatorCC}, respectively. Crucially for the $++$ and $--$ propagators, this procedure only eliminates the factorised part of $\left (G^h_{n,S\to m,S}\right )_{\text{eff}}$ i.e. the parts that depend on the boundary data of the dynamical modes, but it does not cancel any delta function terms that arise due to the non-dynamical nature of some of the modes. For e.g. the trispectrum, we therefore find contact diagram contributions to the correlators in addition to the exchange ones. This final form of the $(p+q)$-point correlator adheres to the generalised rules we outlined in Section \ref{FromWavefunctionToCorrelatorCC}.  

The $p=q$ case works in a similar way, but more subtly. In this case, the number of channels is halved: swapping the momenta sets $\{\bfk_1,...,\bfk_p\}$ and $\{\bfk_{p+1},...,\bfk_{p+q}\}$ is now possible, but it gives the same channel since $\Phi$ is still exchanging the same momenta $\bfk = \pm \left (\bfk_1+\cdots +\bfk_p\right )$. Therefore, the number of terms in the permutations of \eqref{correlatorintermsofrho} is halved. However, now $\rho_{2p}^{\pi...\pi\Phi}$ receives contributions from both $f_{(p)}$ and $g_{(p)}$, thereby doubling the contributions from these wavefunction coefficients. So while the number of terms in the permutations is halved, each term is doubled, yielding the same result as above.

\bibliographystyle{JHEP}
\bibliography{refs}

\end{document}